%% file: dwyatte_dissertation.tex
\documentclass[defaultstyle,12pt]{thesis}
\usepackage{apa} 
\usepackage{times}
\usepackage{graphicx}
\usepackage{siunitx}
\usepackage{url}
\usepackage{amsmath}

\title{What happens next and when ``next'' happens: \\ Mechanisms of spatial and temporal prediction}
\author{Dean R.}{Wyatte}
\otherdegrees{B.S., Indiana University Bloomington, 2007 \\
M.A., University of Colorado Boulder, 2010}
\degree{Doctor of Philosophy}{Ph.D., Cognitive Neuroscience}
\dept{Department of}{Psychology and Neuroscience}
\advisor{Prof.}{Randall C. O'Reilly}
\reader{Prof.~Tim Curran}	
\readerThree{Prof.~Albert Kim}		%  3rd person to sign thesis
\readerFour{Prof.~Joshua Correll}		%  4th person to sign thesis
\readerFive{Prof.~Nikolaus Correll}		%  5th person to sign thesis
\abstract{  \OnePageChapter
The physics of the environment provide a rich spatiotemporal structure for our experience. Objects move in predictable ways and their features and identity remain stable across time and space. How does the brain leverage this structure to make predictions about and learn from the environment? This thesis describes research centered around a mechanistic description of sensory prediction called LeabraTI (TI: Temporal Integration) that explains precisely how predictive processing is accomplished in neocortical microcircuits. The fundamental prediction of LeabraTI is that predictions and sensations are interleaved across the same neural tissue at an overall rate of 10 Hz, corresponding to the widely studied alpha rhythm of posterior cortex. Experiments described herein tested this prediction by manipulating the spatiotemporal properties of three-dimensional object stimuli in a laboratory setting. EEG results indicated that predictions were subserved by $\sim$10 Hz oscillations that reliably tracked the onset of stimuli and differentiated between spatially predictable and unpredictable object sequences. There was a behavioral advantage for combined spatial and temporal predictability for discrimination of unlearned objects, but prolonged study of objects under this combined predictability context impaired discriminability relative to other learning contexts. This counterintuitive pattern of results was accounted for by a neural network model that learned three-dimensional viewpoint invariance with LeabraTI's spatiotemporal prediction rule. Synaptic weight scaling from prolonged learning built viewpoint invariance, but led to confusion between ambiguous views of objects, producing slightly lower performance on average. Overall, this work advances a biological architecture for sensory prediction accompanied by empirical evidence that supports learning of realistic time- and space-varying inputs.
}
\acknowledgements{	\OnePageChapter
Many thanks are in order to:
\begin{itemize}
\item Randy O'Reilly for his guidance throughout my graduate career. His experience and good intuition (not to mention his unabashed optimism) shaped the way in which I now approach problems as a computational modeler and general scientist. 

\item Tim Curran and Al Kim for their knowledge in areas where Randy lacked expertise -- especially experimental work (Randy proudly claims to have never personally run an experiment).

\item Tim Curran's research assistants. A special thanks to Chris Bird and Krystin Corby for coordinating everything and expediting my experiments in the face of other important deadlines.

\item The many friends I have made through mountain biking (Jim, Heather, Matt, Caroline, Greg, Keith, and others). Thanks for joining me on countless adventures and sorry that they always seemed to end with getting lost deep in the woods.

\item My wife, Debra -- a never ending source of emotional (and financial) support. Thanks for keeping me grounded in reality so that I could see the bigger picture of my work.

\end{itemize}
}
	
\emptyLoT	% use this if there is no List of Tables

\begin{document}

\input{chap_intro.tex}
\input{chap_leabrati.tex}
\input{chap_pleast.tex}
\input{chap_bpleast.tex}
\input{chap_sims.tex}
\input{chap_discuss.tex}

\newpage
\bibliographystyle{apa}
\bibliography{ccnlab}

\end{document}

%% file: chap_intro.tex
\chapter{Introduction}
\label{chap:intro}

\sloppy

\section{Sensory predictions and temporal integration}
The brain is often framed as a general purpose ``prediction machine'' \cite{HawkinsBlakeslee04,Clark13}. The fundamental assertion of this framework is that the sole evolved function of the neocortex is to minimize error in its representation of predictions about the physical world. This distillation of function is central to a number of models of neocortical function (e.g., \nopcite{DayanHintonNealEtAl95,RaoBallard99,LeeMumford03,Friston05,GeorgeHawkins09}), but is surprisingly often overlooked in psychology and neuroscience investigations of sensory processing. For example, most experiments are designed to measure evoked responses to a randomly chosen, isolated stimulus under the tacit assumption that response variability is irrelevant noise that averages out across many presentations. Computational models of perceptual processing often operate under similar assumptions in which stimuli are presented as random ``snapshots'' from which some common set of features should be learned to minimize representational variability across presentations (e.g., \nopcite{Fukushima80,RiesenhuberPoggio99,MasquelierThorpe07,OReillyWyatteHerdEtAl13}; although see \nopcite{Foldiak91}, for a notable exception). These experimental and modeling assumptions stand in contrast to the event structure of the physical world, which is highly structured from one moment to the next. It could be the case that response variability does not simply reflect noise, but is actually related to meaningful predictive processing that captures this temporal structure \cite{ArieliSterkinGrinvaldEtAl96,WhiteRolfsCarrasco13,WilderJonesAhmedEtAl13,FischerWhitney14}.

There are a number of important questions that need answered to fully characterize prediction and its role in sensory processing. What are the neural mechanisms responsible for making predictions? Computationally, there is a fundamental tradeoff in making decisions about and generating actions from the constant stream sensory information versus actively generating predictions about what will happen next. Do standard mechanisms balance these tradeoffs or is there special purpose, dissociable machinery specifically for predictive processing? Another line of questioning is concerned with how the brain knows \textit{when} to make predictions. Prediction requires integrating information over some time frame and using the result to drive the actual prediction, but when should integration start? And how long should it last? 

The goal of this thesis is to develop a line of research designed to provide answers to some of these questions and of course, to raise others. The work is largely predicated on a modeling framework referred to as LeabraTI (TI: Temporal Integration), an extension of the standard Leabra cortical learning algorithm \cite{OReillyMunakata00,OReillyMunakataFrankEtAl12} that describes how prediction is accomplished in biological neural circuits. The framework brings together a large number of independent findings from the systems neuroscience literature to describe exactly how multiple interacting mechanisms trade off prediction with sensory processing and learn associations across temporally extended sequences of input. 

The biological details of LeabraTI give rise to a number of testable predictions that can be used to determine the validity of the overall framework. Central to the these testable predictions is the idea that internally generated predictions and sensory events are interleaved through the same neural tissue over intervals of 100 ms. These intervals correspond to individual cycles of the widely observed $\sim$10 Hz alpha rhythm over posterior cortical areas \cite{PalvaPalva07,HanslmayrGrossKlimeschEtAl11,VanRullenBuschDrewesEtAl11}. The temporal interleaving of prediction and sensory processing allows powerful error-driven learning mechanisms to minimize prediction error over multiple episodes, but with the side-effect of discretization artifacts and other temporal oddities due to suppressing sensory processing in favor of prediction for a portion of each 100 ms period.

The empirical work described in this thesis takes advantage of the brain's putative 10 Hz prediction-sensation rate by presenting exogenous stimulation either in phase or out of phase with this endogenous processing. This allows testing of how the spatiotemporal predictability of stimuli influence their encoding for perceptual judgements or prolonged learning. The thesis also describes a neural network model that implements of the broader LeabraTI framework with the goal of accounting for the results of the experimental work.

\section{Organization of the thesis} 
The organization of this thesis is as follows. Chapter \ref{chap:leabrati} contains a comprehensive description of the LeabraTI framework in terms of the low-level biological details of the cortical microcircuitry and response properties required for the temporally interleaved prediction-sensation computation. The chapter also compares LeabraTI with other modeling frameworks and describes a number of testable predictions that differentiate it and can more generally be used to determine its overall validity. 

Chapters \ref{chap:pleast} and \ref{chap:bpleast} describe two experiments designed to test the fundamental predictions of the LeabraTI. The Chapter \ref{chap:pleast} experiment involves manipulating the predictability of time- and space- varying stimulus sequences so that they are either presented in phase or out of phase with the endogenous $\sim$10 Hz alpha rhythm. The stimulus sequences are followed by a perceptual judgement to determine the effect of multiple successful (or unsuccessful) sensory predictions on the encoded representation. EEG is also recorded during the experiment to investigate the effects of these manipulations on endogenous alpha oscillations. The Chapter \ref{chap:bpleast} experiment expands on the basic Chapter \ref{chap:pleast} experimental paradigm to investigate the behavioral effects of predictability on prolonged learning of a subset of the stimuli.

Chapter \ref{chap:sims} describes the neural network model that implements the broader LeabraTI framework which is capable of reproducing the results of the Chapter \ref{chap:pleast} and \ref{chap:bpleast} experiments. The model also provides insight into how predictive learning can alter the representation of the stimuli over prolonged periods of exposure. Finally, Chapter \ref{chap:discuss} discusses the results of the cumulative work accompanied by several open questions as well as directions for future work.

%% file: chap_leabrati.tex
\chapter{The LeabraTI framework: Spatiotemporal prediction with thalamocortical rhythms}
\label{chap:leabrati}

\sloppy
\interfootnotelinepenalty=10000

\section{Introduction}
\begin{figure}[h!]
\begin{center}
\begin{tabular}{ll}
\textbf{A} & \textbf{B} \\
\includegraphics[width=85mm]{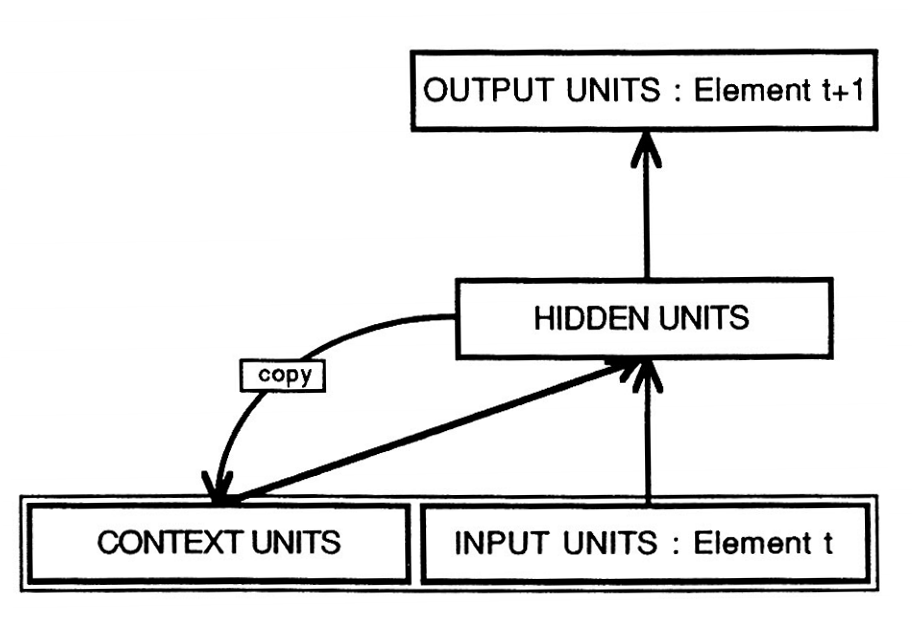} &
\includegraphics[width=75mm]{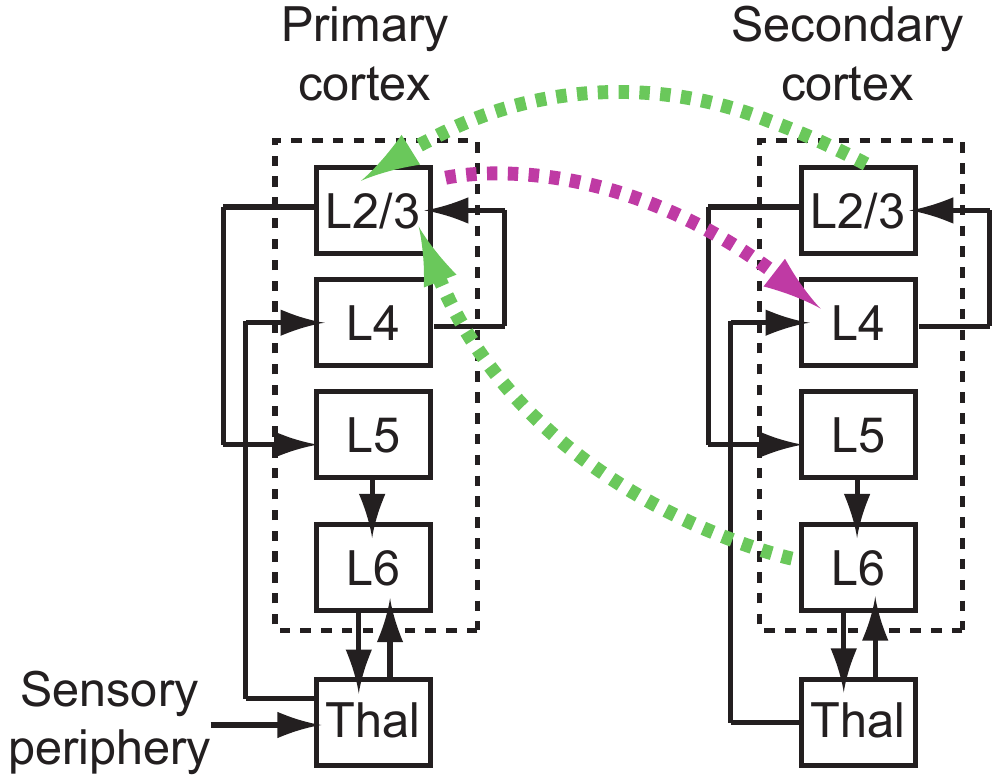} \\
\end{tabular}
\end{center}
\caption{The Simple Recurrent Network (SRN) and microcircuitry of the neocortex}{\textbf{A}: The SRN represents temporal information explicitly using discrete context units that are updated once per time step. Context is integrated with more current inputs to predict information at the subsequent time step. Reproduced from \protect\incite{Servan-SchreiberCleeremansMcClelland91}. \textbf{B}: The neocortex is laminated with canonical circuitry between neurons across layers and between areas. Principal intra-areal connections are shown in black with inter-areal feedforward connections in purple and feedback connections in green. Note that the bifurcation of intra-areal and inter-areal processing streams could allow local context to represented by in a continuous manner.}
\label{fig:srn_circuit}
\end{figure}

This chapter describes the LeabraTI (TI: Temporal Integration) framework, which is a mechanistic description and general model of how prediction and temporal integration works in the brain. It is closely related to the Simple Recurrent Network (SRN) \cite{Elman90,Servan-SchreiberCleeremansMcClelland91} a neural network architecture that explicitly represents temporally lagged information in discrete ``context'' units whose activity gets integrated with more current information to predict what happens in the next time step (Figure \ref{fig:srn_circuit}A). This method of copying a contextual representation from an intermediate representation at discrete intervals was originally shown to be a robust way to leverage powerful error-driven learning to represent latent temporal structure in auditory streams and artificial grammars. More generally, the SRN's explicit representation of temporally lagged context can capture the latent structure of any stimulus that varies systematically over time, making it a good basis for a generic prediction and temporal integration mechanism.

LeabraTI differs in several key ways from the classical SRN architecture as well as other predictive learning frameworks, primarily in the way context is represented and used in predictive processing. These differences are due to biological constraints imposed by the microcircuitry of the neocortex, and thus form a number of testable predictions that can be used to evaluate the validity of the LeabraTI framework. The central prediction of LeabraTI is that temporally lagged context is represented by deep (Layer 6) neurons, which is possible in part to the bifurcation of intra-areal and inter-areal processing streams. As neural processing is a continuous operation, LeabraTI requires a regular interval over which to integrate deep context and make predictions, which is approximately every 100 ms. Predictions are made by driving superficial (Layers 2 and 3) neurons with the state of deep neurons through the intra-areal pathway, which is interleaved with standard peripheral sensory inputs over a total period that is also 100 ms. The strong 100 ms dependency in LeabraTI corresponds to the brain's $\sim$10 Hz alpha rhythm, which has been studied extensively using scalp EEG \cite[see][for comprehensive reviews]{PalvaPalva07,HanslmayrGrossKlimeschEtAl11,VanRullenBuschDrewesEtAl11}.

% srn/microcircuit fig

\section{LeabraTI biological details}

\subsection{Laminar structure and microcircuitry of the neocortex}
A salient feature of the brain, and potential clue in realizing how an SRN-like computation might be carried out in biological neural circuits, is the laminar structure prevalent across the neocortex (Figure \ref{fig:srn_circuit}B). Incoming information from the sensory periphery is transmitted through the thalamus and targets Layer 4 neurons in the primary sensory cortices (e.g., V1). From there, Layer 4 neurons propagate spikes to superficial neurons (Layers 2 and 3) which in turn target Layer 4 neurons of higher-level cortices, forming the prominent corticocortical feedforward pathways that subserve visual and auditory recognition \cite{FellemanVanEssen91}. Corticocortical feedback originates in superficial layers or Layer 6 of the higher-level cortex and generally terminates on superficial neurons of the lower-level cortex \cite{RocklandPandya79}. In addition to these inter-areal pathways, there exists a canonical microcircuit of the form Layer 4 $\rightarrow$ Layer 2/3 $\rightarrow$ Layer 5 $\rightarrow$ Layer 6 that routes spike propagation through the local neuronal structure \cite{DouglasMartin04,ThomsonLamy07,DacostaMartin10}. This microcircuit forms the core computational unit of LeabraTI, as will be described in this and the following sections.

The importance of the local microcircuit was first suggested by Mountcastle in his proposal regarding the gross columnar organization of the neocortex \cite[see][for a comprehensive review]{Mountcastle97}. Mountcastle's proposal states that microcolumns composed of around 80-100 neurons extending vertically through all six lamina with canonical circuitry form the core repeating structure of the neocortex. Neurons within a single micrcolumnnar circuit possess nearly identical receptive field tunings across lamina while neurons in neighboring microcolumns (radial separation greater than 60 \SI{}{\micro\meter}) possess very different receptive field tunings but contribute to the higher-order macrocolumn (i.e., hypercolumn) structure \cite{HubelWiesel77,Jones00}. Microcolumns have been identified in a variety of neural systems with this electrophysiological mapping and are also prominently visible under Nissl staining. Despite this evidence for their structural existence, any function of the microcolumn aside from an organizing principle remains debated \cite{BuxhoevedenCasanova02,HortonAdams05}.

LeabraTI provides a computational role for the microcolumn, by mapping an SRN-like computation onto their Layer 4 $\rightarrow$ Layer 2/3 $\rightarrow$ Layer 5 $\rightarrow$ Layer 6 circuit (Figure \ref{fig:srn_circuit}). In this mapping, superficial neurons continuously integrate feedforward and feedback inter-areal synapses to process current information. Layer 2/3 $\rightarrow$ Layer 5 $\rightarrow$ Layer 6 provides an intra-areal pathway for explicitly representing temporal context in Layer 6 neurons, which are relatively isolated from nonlocal inputs. There is also appropriate circuitry for recirculating this context through the local microcolumn via Layer 4 to drive the learning of temporal associations. This basic idea provides a concise explanation for the strong degree of isotuning throughout a single microcolumn, as Layer 6 neurons need to represent the same overall information as superficial neurons except at a delayed interval.

\subsection{Layer 5 rhythmic bursting and contextual gating}
The laminocolumnar organization of the neocortex provides the dual pathways necessary for continuous information processing and the SRN's explicit temporal context representation. The Layer 4 $\rightarrow$ Layer 2/3 $\rightarrow$ Layer 5 $\rightarrow$ Layer 6 microcircuit only contains four synapses plus the transthalamic re-entrant synapses. Intracolumnar monosynaptic latencies for regular spiking neurons are on the order of 5 ms or faster \cite{Armstrong-JamesFoxDas-Gupta92,LumerEdelmanTononi97} and thus this relatively small amount of tissue, if driven with constant input, would circulate spikes at a rate much too fast to perform substantial temporal integration for making useful predictions. Several studies have noted that a subset of Layer 5 neurons exhibit intrinsic bursting at $\sim$10 Hz when over threshold (\nopcite{ConnorsGutnickPrince82,SilvaAmitaiConnors91}; \abbrevnopcite{FranceschettiGuatteoPanzicaEtAl95}). This rhythmic busting might implement a gating mechanism for updating Layer 6 context information at a regular 100 ms interval.

More specifically, Layer 5 neurons can be roughly divided into 5a and 5b subtypes \cite{ThomsonLamy07}. Layer 5a neurons have relatively small cell bodies and exhibit regular spiking depolarization responses. They collect input from other Layer 5a neurons both within and across columns \cite{SchubertKotterStaiger07} and pass it to 5b neurons and thus, likely play a simple information integration role. Layer 5b neurons, in contrast, have larger cell bodies and exhibit the aforementioned 10 Hz intrinsic bursting response profile. In the context of LeabraTI, the interpretation of this data is that the 5a neurons serve to integrate information from multiple Layer 2/3 neurons, with the 5b neurons gating context to Layer 6 neurons with each 10 Hz burst.

Layer 6 corticothalamic neurons receive strong inputs from Layer 5b neurons and send axons toward the thalamus completing the microcircuit within the local column and allowing the temporally lagged Layer 6 responses to integrate with more current Layer 4 inputs. Information is relayed from the thalamus back up to layer 4 in a focal one-to-one manner that maintains microcolumnar separation \cite{ShermanGuillery06,Thomson10}, which could allow temporal associations to be formed by local Hebbian learning mechanisms that track high probability co-occurences across past and present events \cite{Foldiak91}.

\subsection{Thalamic gating and sensory prediction}
Both the SRN computation and the Leabra algorithm \cite{OReillyMunakata00,OReillyMunakataFrankEtAl12} that are used to implement the LeabraTI framework are predicated on using powerful error-driven learning mechanisms (in addition to more standard Hebbian learning mechanisms) to represent the mapping between sensory inputs and outputs. In the context of temporal integration, error-driven learning would allow computation of error signals based on the difference between what is predicted to happen at a given moment (given the previous moments context as an input) and what actually happens. However, this computation requires that both the prediction and the actual sensation are represented by the same neural tissue so that an error signal can be computed, which is not possible if the sensory periphery is continuously transmitting incoming inputs. 

To resolve this issue, the LeabraTI framework posits that predictions about sensory events and the sensory events themselves are temporally interleaved through the same population of neurons in an alternating manner. This requires a mechanism to periodically downmodulate or even block the transmission of inputs from the sensory periphery. A subset of cells in the thalamus exhibit $\sim$10 Hz intrinsic bursting properties similar to those of Layer 5b neurons (\nopcite{LopesdaSilva91}; \abbrevnopcite{HughesLorinczCopeEtAl04}; \nopcite{LorinczCrunelliHughes08}; \nopcite{LorinczKekesiJuhaszEtAl09}), and thus perhaps perform a similar gating computation of sensory inputs into cortical circuits. In the context of LeabraTI, these bursting neurons might shift the balance of inputs to Layer 4 and superficial neurons between endogenous inputs local to the microcolumn that represent predictions and quick bursts of actual sensory information that conveys the current state of the environment. 

More specifically, during the non-bursting intervals of thalamic intrinsically bursting neurons' response, environmental inputs are downmodulated due to the relative quiescence and Layer 6 neurons provide the dominant driving potential to the microcolumn. Layer 6 corticothalamic neurons exhibit a strong regular spiking depolarization response with facilitating short-term dynamics \cite{Thomson10}, unlike all other pyramidal neurons, which exhibit depressing dynamics. This might suggest a specialized function for Layer 6 neurons, which in the context of LeabraTI is to drive a sustained prediction about upcoming sensory information. These corticothalamic neurons also sustain their drive through reciprocal projections with the thalamic relay cells that they project to. In addition to the Layer 6 $\rightarrow$ Layer 4 transthalamic pathway, Layer 6 neurons also project directly to Layer 4. While these projections are relatively weak \cite{HirschMartinez06b}, they do activate a metabotropic glutamate receptor (mGluR) that produces sustained depolarization similar to Layer 6 corticothalamic neurons \cite{LeeSherman09}. These direct ascending synapses are another possible route for sustained context information to drive Layer 4 neurons. 

The burst response of thalamic intrinsic bursting neurons synchronized with the burst response of Layer 5b neurons destabilizes the Layer 6 sustained prediction through biased competition \cite{WyatteHerdMingusEtAl12,DesimoneDuncan95}. This provides a snapshot of the current state of the sensory periphery as well as an opportunity to integrate the state of current sensory event into a prediction about what will happen during the next moment. At this time, standard error-driven learning mechanisms that compute short timescale firing rate differences \cite{OReillyMunakata00,OReillyMunakataFrankEtAl12} compute an error signal between the previous moment's prediction and the sensory outcome to minimize overall prediction error.

\section{Summary of the LeabraTI computation}
\begin{figure}[h!]
\begin{center}
\includegraphics[width=160mm]{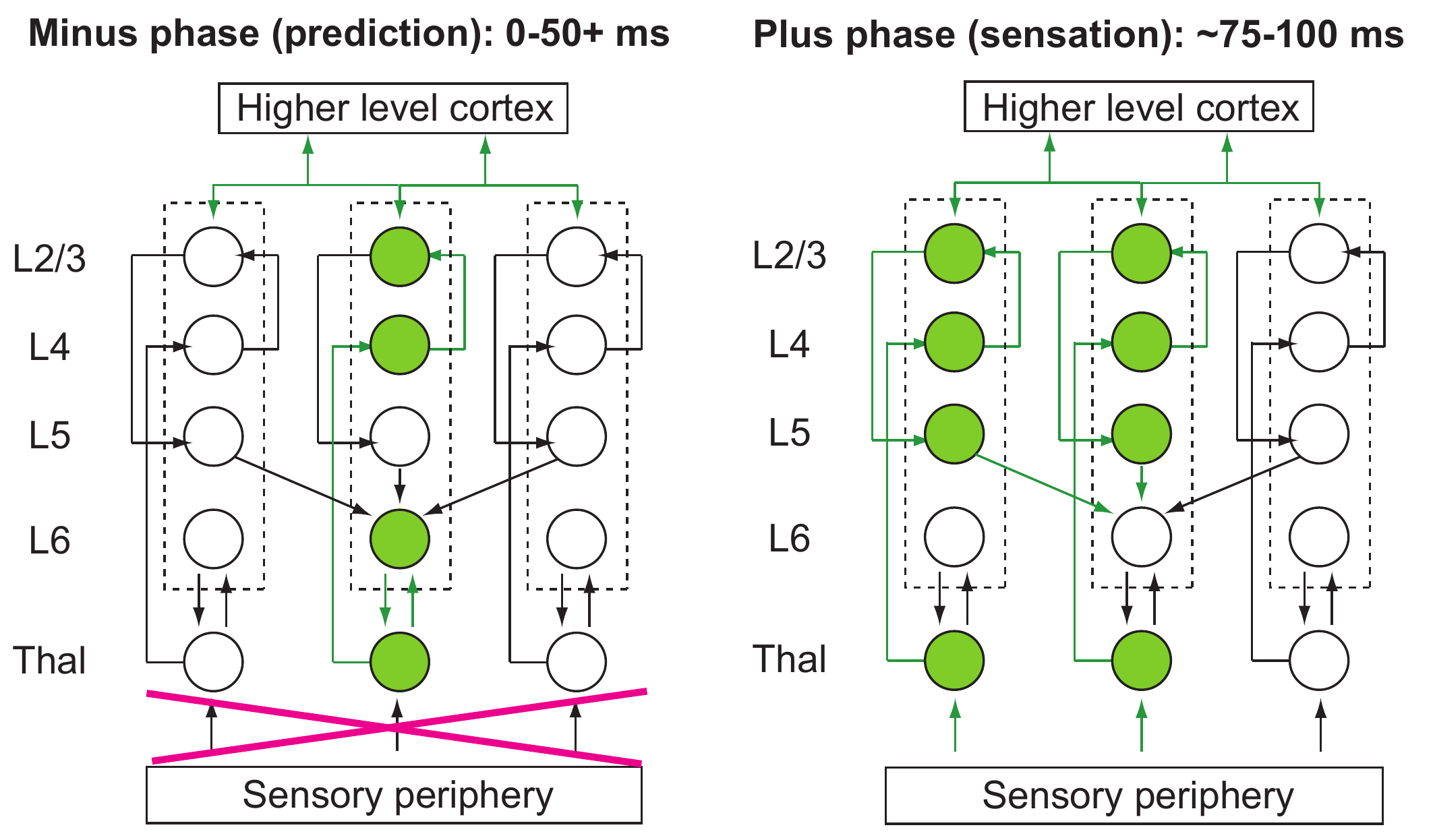}
\end{center}
\caption{The LeabraTI computation}{LeabraTI consists of a \textit{minus} phase prediction about what is about to happen in the following \textit{plus} phase sensory event. The minus phase is characterized by a sustained prediction from deep Layer 6 neurons based on lagged context from the previous moment. Input from the sensory periphery is extremely downmodulated or even blocked during the minus phase due to the relative quiescence of thalamic intrinsically bursting neurons. The plus phase is characterized by a rapid burst of information from the sensory periphery that drives also drives Layer 5 intrinsically bursting neurons to threshold so that they can integrate a snapshot of the current moment as context. Full feedforward and feedback-mediated processing is active in both phases so that predictions are not entirely dependent on deep Layer 6 neurons, but can also be driven by higher-level expectations. Green units and synapses indicate activation during the respective phase.}
\label{fig:leabrati_comp}
\end{figure}

The overall computation of LeabraTI is shown in Figure \ref{fig:leabrati_comp} and summarized here. Standard Leabra processing operates across two phases: a \textit{minus} phase that represents the system's expectation for a given input and a \textit{plus} phase, representing observation of the outcome. Generally, feedforward sensory inputs provide the dominant drive during the minus phase while the plus phase is driven by a combination of feedforward sensory inputs and the target outcome, conveyed via feedback projections. Leabra networks typically only model at the level of superficial (Layers 2 and 3) neurons, since they provide the principal feedforward and feedback projections necessary for error-driven learning.

% leabrati computation

In LeabraTI, deep Layer 6 neurons provide the dominant input to superficial Layer 2/3 neurons during the minus phase, relayed via Layer 4 transthalamic and direct ascending projections (direct ascending projections not pictured in Figure \ref{fig:leabrati_comp}). This input represents a sensory prediction about what is about to happen during the subsequent plus phase, using temporally lagged context from the previous plus phase. To be computationally useful, the prediction needs to last at least 50 ms in order to allow contributions from both the feedforward drive and slightly longer latency feedback from higher-order cortical areas. The short-term facilitating dynamics unique to Layer 6 corticothalamic neurons or mGluR activation from the direct ascending projections provide the sustained spiking response throughout the minus phase. Reciprocal thalamocortical synapses back to Layer 6 might also assist in maintaining minus phase input drive. During this period, input from the sensory periphery is extremely downmodulated due to the relative quiescence of thalamic intrinsically bursting neurons. Downstream areas basically repeat this predictive process driven by their respective deep Layer 6 context, although they do also receive direct Layer 4 drive from the previous cortical area. 

The plus phase is characterized by a shift of input from the local column to the sensory periphery, driven by the destabilizing response from thalamic intrinsically bursting cells. The rapid influx of synchronous drive from the thalamus serves two purposes. First, it rapidly drives the propagation of responses through downstream areas by quickly driving neurons along the Layer 4 $\rightarrow$ Layer 2/3 $\rightarrow$ Layer 4 feedforward pathway \cite{FellemanVanEssen91} to threshold, similar to the concept of ``synfire chains'' that simultaneously ignite entire pools of neurons to propagate waves of spikes back and forth through a network \cite{BrunoSakmann06,WangSpencerFellousEtAl10,TiesingaFellousSejnowski08}. Second, it drives responses down through the local Layer 2/3 $\rightarrow$ Layer 5 $\rightarrow$ Layer 6 pathway, which also requires a large influx of thalamic drive \cite{BeierleinFallRinzelEtAl02}. This rapid transmission between and within areas causes Layer 5b neurons across areas to burst with relatively little cross-area delay so that the context that the bursting captures represents information from a single moment in time \cite{BollimuntaChenSchroederEtAl08}.

Once thalamic bursting quiets, the input to the microcolumn shifts back to the newly integrated Layer 6 context. This prediction-sensation process repeats approximately every 100 ms corresponding to the brain's $\sim$10 Hz alpha rhythm, which is suggested to have strong thalamocortical and deep laminar sources \cite{LopesdaSilva91,KlimeschSausengHanslmayr07,PalvaPalva07,LorinczKekesiJuhaszEtAl09,BollimuntaMoSchroederEtAl11,HanslmayrGrossKlimeschEtAl11}. 

\section{Relation to other frameworks}

\subsection{The Simple Recurrent Network (SRN)}
\begin{figure}[h!]
\begin{center}
\includegraphics[width=140mm]{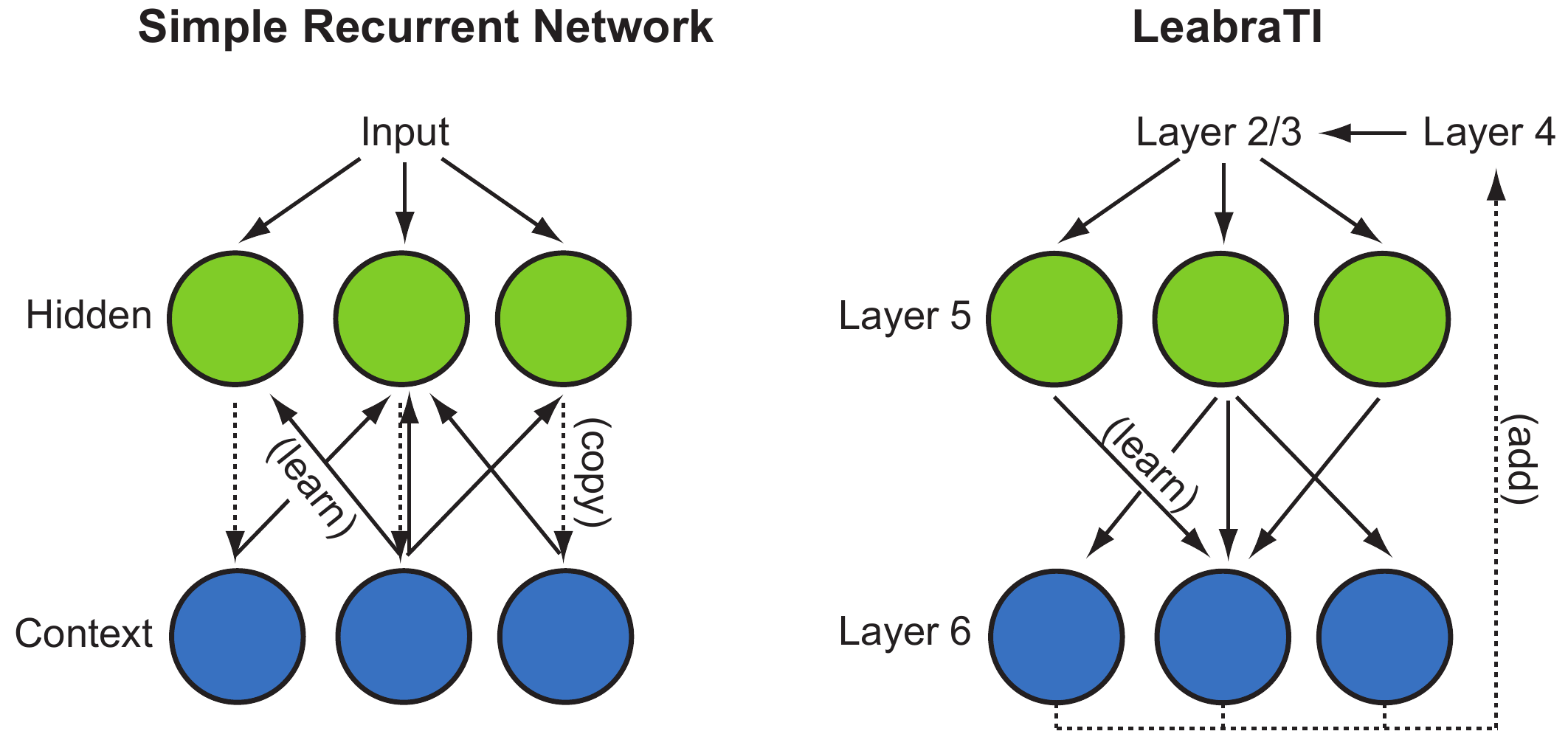}
\end{center}
\caption{Relationship between the Simple Recurrent Network (SRN) and LeabraTI}{The Simple Recurrent Network (SRN) projects a direct copy of Hidden units' state to context units that is then integrated at the next time step with the current input. LeabraTI, in contrast, projects first through Layer 5 $\rightarrow$ Layer 6 plastic synapses triggering an immediate prediction that is then used additively in computing the input at the next time step. These operations are mathematically equivalent, but LeabraTI provides a better fit with the circuitry of the microcolumn.}
\label{fig:srn_invert}
\end{figure}

LeabraTI is perhaps most closely related to the Simple Recurrent Network (SRN) \cite{Elman90,Servan-SchreiberCleeremansMcClelland91}, but with several key differences that are necessary due to the implementation using cortical circuits. The SRN is capable of learning temporal sequences by virtue of a ``copy'' operation (Figure \ref{fig:srn_invert}) that represents an exact replica of the previous time step's intermediate units in separate group of context units. Thus, the SRN integrates two inputs at each time step -- the previous time step's processed inputs and the current time step's unprocessed inputs. Predictions are made in a separate group of output units and prediction errors are backpropagated to learn the weights that allow the best prediction over items in the input sequence. 

Like the SRN, LeabraTI uses discrete context units (deep Layer 6 neurons), but it does not implement a direct copy of sending units at each time step that is then used for predictive learning. LeabraTI is actually an inversion of the SRN formulation, where context is immediately sent through plastic integrative synapses before being sent through an additional input channel when computing the activation states of superficial (Layer 2 and 3) units (Figure \ref{fig:srn_invert}). The LeabraTI formulation is mathematically equivalent to the SRN, but provides a better fit with the circuitry of the microcolumn, which contains plastic synapses between Layer 5 and Layer 6 (primarily the 5a $\rightarrow$ 5b integrative synapses). The inversion also allows a sustained, uncorrupted minus phase prediction that would not be possible if the directionality of the projections were reversed due to the massive numbers of afferent and efferent synapses on superficial neurons. 

% srn/microcircuit fig

One other difference between LeabraTI and the SRN is that the latter makes predictions using a completely separate group of output units. LeabraTI, in contrast, makes predictions using the very same neural tissue that is used to represent sensory outcomes. This is a much more natural mechanism for predictive learning as it does not require a separate populations of neurons for predictions and sensations -- it does, however, tradeoff the neural space savings with temporal interleaving of predictions and sensory outcomes, which could cause discretization artifacts and other temporal oddities in perception. There is increasing evidence of discretization at the 10 Hz rate proposed by LeabraTI \cite{VanRullenKoch03b,VanRullenDubois11,VanRullenBuschDrewesEtAl11}, which is discussed in the section following this one. 

% columns are essentially bio SRN's, shifts computational unit from neuron to microcolumn

\subsection{Predictive coding models}
Another framework related to LeabraTI is the predictive coding model \cite[e.g.,]{RaoBallard99,Friston05,DenOudenKokDeLange12}. The fundamental claim of predictive coding models is that sensory events themselves are not explicitly encoded by neurons. Instead, prediction errors -- the difference between predictions and sensory events -- are what is encoded. The prediction error computation is generally repeated at each stage of processing with only the residual error being propagated downstream for further processing. Predictive coding models have gained traction due to this efficient ``delta'' code and due to their positing a distinct role for feedback projections in generating predictions. A recent description of predictive coding makes connections with the microcolumnar circuitry in the same way as LeabraTI. In their proposal, \abbrevincite{BastosUsreyAdamsEtAl12} suggested that deep infragranular neurons code prediction errors which ascend back to Layer 4 neurons or down to lower-level areas as feedback predictions. 

The primary problem with predictive coding models is in their computation of prediction error. This computation requires feedback predictions to be inhibitory in nature so that they can be subtracted from the excitatory input, leaving only the residual error. This computation is not possible in cortical circuits since feedback via pyramidal axons is fundamentally excitatory. This issue could be resolved by inverting the excitatory signal, but feedback axons predominantly target other excitatory cells and the net postsynaptic potentials generated by their activation are also excitatory \cite{Budd98,JohnsonBurkhalter96,JohnsonBurkhalter97}.

LeabraTI resolves this issue by computing predictions and sensory events explicitly in discrete phases, based on excitatory input with inhibitory competition to suppress spurious errors in coding \cite{WyatteHerdMingusEtAl12,DesimoneDuncan95}. Thus predictions are not fundamentally inhibitory, but excitatory due to being driven through Layer 4 neurons from deep context. This assumption is consistent with the predictive remapping phenomenon in cortex \cite{DuhamelColbyGoldberg92,NakamuraColby02,CavanaughHuntAfrazEtAl10} during which excitatory neurons fire in anticipation of a stimulus that falls outside of their receptive field at one moment, but will fall inside their receptive field upon making a saccade.

\section{Testable predictions}
\subsection{Electrophysiological predictions}
LeabraTI describes how prediction and temporal integration work at the relatively low level of individual microcolumns and their lamina. This level of description creates a number of testable predictions for electrophysiological investigations that have received initial support in the literature. For example, recent studies that have employed depth electrodes to simultaneously record from multiple layers of neurons within a small patch of cortex have indicated large differences in the spectral coherence of superficial and deep neurons \cite{MaierAdamsAuraEtAl10} as would be predicted by LeabraTI.

For example, \incite{BuffaloFriesLandmanEtAl11} employed single-unit recordings in visual sites V1, V2, and V4 in awake, behaving monkeys during a simple directed attention task, finding a dissociation in spike coherence frequency in superficial (gamma spectrum, peak $\sim$50 Hz) and deep layers (alpha spectrum, peak $\sim$10 Hz). A similar experimental paradigm expanded on these findings by demonstrating cross-frequency coupling between gamma and alpha spectra localized to superficial and deep layers, respectively \cite{SpaakBonnefondMaierEtAl12}. The cross-frequency coupling was characterized by a clear nesting of gamma activity within alpha cycles, suggesting that deep neurons' alpha coherence might subserve a task-independent 10 Hz duty cycle \cite{JensenMazaheri10,JensenBonnefondVanRullen12,JensenGipsBergmannEtAl14} for continuous integration by superficial neurons.

The strong 10 Hz coherence of deep neurons persists even with constant sensory stimulation \textit{in vivo}. \incite{MaierAuraLeopold11} found Layer 5 potentials that were not phase-locked to visual stimulation with a strong $\sim$10 Hz component as long as the stimulus was present. Another study that recorded individual Layer 5 neurons found that they exhibited 50-100 ms ``packeted'' responses that were somewhat irregular during a sustained stimulus, but generally adhered to a $\sim$10 Hz pattern \cite{LuczakBarthoHarris13}. 

Altogether, these results are consistent with LeabraTI's 10 Hz context updating by deep Layer 5b $\rightarrow$ Layer 6 bursts. Several studies have noted similar $\sim$10 Hz bursting in subpopulations of thalamic neurons (\abbrevnopcite{HughesLorinczCopeEtAl04}; \nopcite{LorinczCrunelliHughes08}; \nopcite{LorinczKekesiJuhaszEtAl09}). However, the focus of this thesis is to investigate the overall role of 10 Hz predictive processing during perception, which is most plausibly studied using the relatively coarser alpha rhythm observable from the scalp, discussed next.\footnote{Scalp observable alpha is commonly assumed to correspond to the underlying $\sim$10 Hz thalamocortical generators \cite{LopesdaSilva91,KlimeschSausengHanslmayr07,PalvaPalva07,LorinczKekesiJuhaszEtAl09,BollimuntaMoSchroederEtAl11,HanslmayrGrossKlimeschEtAl11}, opposed to physiologically distinct oscillators.}

\subsection{The alpha rhythm and sensory prediction}
LeabraTI's temporal interleaving of predictions and sensations across the same neural tissue is likely to cause some discretization artifacts due to the suppression of information from the sensory periphery during prediction. The question of whether perception is discrete or continuous has occupied the literature for over 30 years \cite[see][for a review]{VanRullenKoch03b}. Our everyday experience suggests that perception is undeniably continuous, but a number of phenomena reviewed here support the discretization of perception at $\sim$10 Hz, as predicted by LeabraTI.

The wagon wheel illusion occurs when a rotating wagon wheel appears to switch direction at certain speeds. The effect was first noted in cinematography, due to the temporal aliasing effect of a camera's shutter. VanRullen and colleagues have demonstrated the wagon wheel effect under continuous illumination and have noted that it is maximal at $\sim$10 Hz, suggesting the alpha rhythm implements a similar temporal aliasing effect \cite{VanRullenReddyKoch05,VanRullenReddyKoch06}. A recent investigation indicated that a static wagon wheel-like stimulus also flickers at rates estimated at $\sim$10 Hz when viewed in the visual periphery outside the scope of overt attention \cite{SokoliukVanRullen13}. Illusory jitter of high-contrast edges has been shown to occur at $\sim$10 Hz \cite{AmanoArnoldTakedaEtAl08}. All of these effects have been correlated with increased alpha-band power over the visual cortices.

The phase of ongoing alpha has activity has also been related to sensory processing, but in terms of efficacy. Errors made processing at-threshold stimuli have been suggested to arise from alpha phase at stimulus onset \cite{MathewsonGrattonFabianiEtAl09,BuschDuboisVanRullen09}. Specifically, analyses that group data based on whether stimuli were successfully detected have indicated opposing phases for successful versus unsuccessful detection. For undetected stimuli, the alpha cycle was at approximately in phase at target onset and 180 degrees out of phase 50 ms later when processing begins in primary visual cortex \cite{NowakBullier97,FoxeSimpson02}. This low point in the alpha cycle is suggested to be the point of maximal thalamocortical suppression \abbrevcite{MathewsonLlerasBeckEtAl11} so that an internal prediction can be successfully driven by Layer 6 context. 

Given the importance of alpha phase in determining successful perception, it seems that there should be a mechanism in place to ensure that environmental stimuli are not processed during the brain's prediction phase. As such, endogenous oscillations including alpha have been shown to phase-lock to exogenous rhythmic visual and auditory stimulation (\nopcite{WillBerg07,FujiokaTrainorLargeEtAl09}; \abbrevnopcite{StefanicsHangyaHernadiEtAl10}; \nopcite{SpaakdeLangeJensen14,SchroederLakatosKajikawaEtAl08,CalderoneLakatosButlerEtAl14}). This phase-locking ensures that environmental events don't overlap with internal predictions that attenuate sensory efficacy. \abbrevincite{MathewsonPrudhommeFabianiEtAl12} presented subjects with a train of stimuli that either were rhythmic, and thus reliably predicted the temporal onset of a masked probe, or were arrhythmic and unpredictable. Rhythmic stimulus trains entrained endogenous alpha oscillations and as a result, probes that occurred in either 100 ms or 200 ms after the probe were less susceptible to the effects of masking, since they were out of phase with the prediction interval of the alpha rhythm.

Phase resetting is thought to underly the alpha rhythm's entrainment properties \cite{CalderoneLakatosButlerEtAl14}. Phase resetting also provides flexibility for the rhythm  so that it can align to unexpected salient stimuli that capture attention. For example, salient flashes can cause fluctuations in perceptual efficacy that oscillate at $\sim$10 Hz after the flash occurs \cite{LandauFries12}. Salient sounds can also cross-modally reset alpha in the visual cortices with a similar fluctuations in accuracy after the event (\abbrevnopcite{FiebelkornFoxeButlerEtAl11}; \nopcite{RomeiGrossThut12}). After phase reset, sensory efficacy is typically impaired for 50-100 ms as the brain captures the newly established in state of the environment as Layer 6 context so that it can resume generating reliable predictions.

To summarize the data reviewed here, there are a number of visual phenomena that show periodicities in perceptual processing. In the context of LeabraTI, these periodicities are due to the 10 Hz interleaving of predictions and sensory events across the same neural tissue which necessarily suppresses information from the sensory periphery during prediction. As such, there is a strong correlation between these periodicities and alpha power as well as the phase of the ongoing alpha oscillations. Furthermore, mechanisms exist to reset and lock the phase of alpha to important environmental stimuli. 

%% file: chap_pleast.tex
\chapter{Spatial and temporal prediction during novel object recognition: Temporal and spectral signatures}
\label{chap:pleast}

\sloppy

\section{Introduction}
% write without referencing LeabraTI too much, this chapter should stand by itself
How does the brain integrate information from one moment to the next and use it it to drive predictions about what will happen? A number of models of general neocortical function have highlighted the centrality of prediction in neural processing (e.g., \nopcite{DayanHintonNealEtAl95,RaoBallard99,LeeMumford03,Friston05,GeorgeHawkins09}), but it is surprisingly often overlooked in psychology and neuroscience investigations of sensory processing. Predictability might be useful in perceptual processes like object recognition, for example, by allowing better integration across features extracted over the course of several samples \cite{Foldiak91,StringerPerryRollsEtAl06,WallisBaddeley97,IsikLeiboPoggio12}. The timing of predictions might also be important. The LeabraTI model (Chapter \ref{chap:leabrati}) as well as several other theories of sensory prediction \cite{ArnalGiraud12,GiraudPoeppel12} emphasize that predictions occur in a pacemaker manner at regular intervals and thus temporal properties of endogenous processes as well as exogenous stimulation might also affect perceptual processing.

Previous work has implicated 7-13 Hz alpha oscillations in the anticipation of the temporal onset of stimuli at a spatially attended location (\nopcite{GouldRushworthNobre11}; \abbrevnopcite{BelyusarSnyderFreyEtAl13}; \nopcite{RohenkohlNobre11}). Specifically, alpha oscillations desynchronize in the hemisphere contralateral to the attended region of visual space as well as in anticipation of the onset of a time-varying stimulus. Pre-stimulus alpha oscillations have also been related to the detection rate of at-threshold point-light targets \cite{MathewsonGrattonFabianiEtAl09,BuschDuboisVanRullen09}, suggesting that spontaneous fluctuations in alpha oscillatory properties could also be related to the ability to properly anticipate a stimulus. Together, these results implicate the role of alpha oscillations in hemifield-based spatial attention tasks and for prediction of relatively simple stimuli. 

One issue with relating the extant literature on alpha oscillations to prediction is that is unclear whether the experiments described therein measure actual predictive processing about \text{what} would happen or comparatively simple anticipatory attention mechanisms about \textit{where} a stimulus might appear. For example, the alpha rhythm might simply correspond to shifts of spatial attention, which can be oriented approximately 10 times per second \cite{VanRullenDubois11} and studies that implement a Posner spatial cueing task \cite{Posner80} support this idea \cite{CapotostoBabiloniRomaniEtAl09,BuschVanRullen10}. Other studies, however, suggest that prediction and attention are separable mechanisms with dissociable neural and behavioral effects \cite{KokRahnevJeheeEtAl12,WyartNobreSummerfield12,HorschigJensenVanSchouwenburgEtAl13}. Experiments have generally not examined the role of alpha oscillations in complex perceptual tasks that require actual predictive processing, although it was recently demonstrated that an alpha-band presentation rate (12.5 Hz) is optimal for maximal fMRI BOLD response in a dynamic face recognition experiment \cite{SchultzBrockhausBulthoffEtAl13}.

The work described in this chapter investigated the role of predictive processing during a novel object recognition task. The experiment made use of novel three-dimensional stimuli that required integration over multiple sequential views to extract their three-dimensional structure. The stimuli were presented at central fixation and thus, manipulating the ordering of the views could be used to test the effect of their predictability without confounding it with spatial attention. This is henceforth referred to as ``spatially'' predictable, although it is likely a combination of spatially- and featurally- predictive processing (i.e., the prediction of specific features at specific locations of a spatial map). To test the relationship between alpha oscillations and predictive processing, the experiment took advantage of the entrainability of endogenous alpha oscillations by exogenous rhythmic stimulation \cite{SchroederLakatosKajikawaEtAl08,CalderoneLakatosButlerEtAl14}. Whereas previous experiments related pre-stimulus alpha to  anticipation via a post-hoc grouping of detected and missed stimuli \cite{MathewsonGrattonFabianiEtAl09,BuschDuboisVanRullen09}, the current experiment entrained alpha oscillations by presenting subsequent views of the stimulus at a regular 10 Hz rate to determine the causal effect of alpha oscillations on prediction (henceforth referred to as ``temporally'' predictable).

The results of the experiment indicated that spatial and temporal predictability of an entraining sequence enhanced discriminability of a subsequently presented probe stimulus. Temporal predictability also speeded response times for the probe judgement. EEG amplitude averaging indicated separable time courses for spatial and temporal prediction over posterior sites with temporal predictability always preceding the onset of stimuli and spatial predictability manifesting at the onset of stimuli and persisting for over 100 ms after in the case of the probe. Oscillatory analyses indicated strong bilateral alpha power and phase coherence modulation as a function of stimulus predictability. In addition to these main effects, right hemisphere sites exhibited synergistic effects of combined spatial and temporal probe predictability on EEG amplitude and 10 Hz phase coherence approximately 200 ms after probe onset. Exploratory analyses indicated that oscillatory effects in the lower frequency delta-theta (5 Hz) band, which has also been associated with predictive processing \cite{ArnalGiraud12,GiraudPoeppel12} were similar to those in the alpha band, if not more prominent. 

\section{Methods}

\subsection{Participants}
% 29 D464B EEG 
% 29 Behavioral only
% ===
% 58 total, 27 female (ages 18-28 mean=21)
A total of 58 students from the University of Colorado Boulder participated in the experiment (ages 18-28 years, mean=21; 31 male, 27 female). EEG was recorded from 29 of the participants while they completed the experiment. The remaining 29 participants completed a solely behavioral experiment without EEG recording. All participants were right-handed and reported normal or corrected-to-normal vision. Participants either received course credit or payment of \$15 per hour as compensation for their participation. Informed consent was obtained from each participant prior to the experiment in accordance with Institutional Review Board policy at the University of Colorado.

\subsection{Stimuli}
\begin{figure}[h!]
\begin{center}
\begin{tabular}{ll}
\textbf{A} \\
\includegraphics[width=160mm]{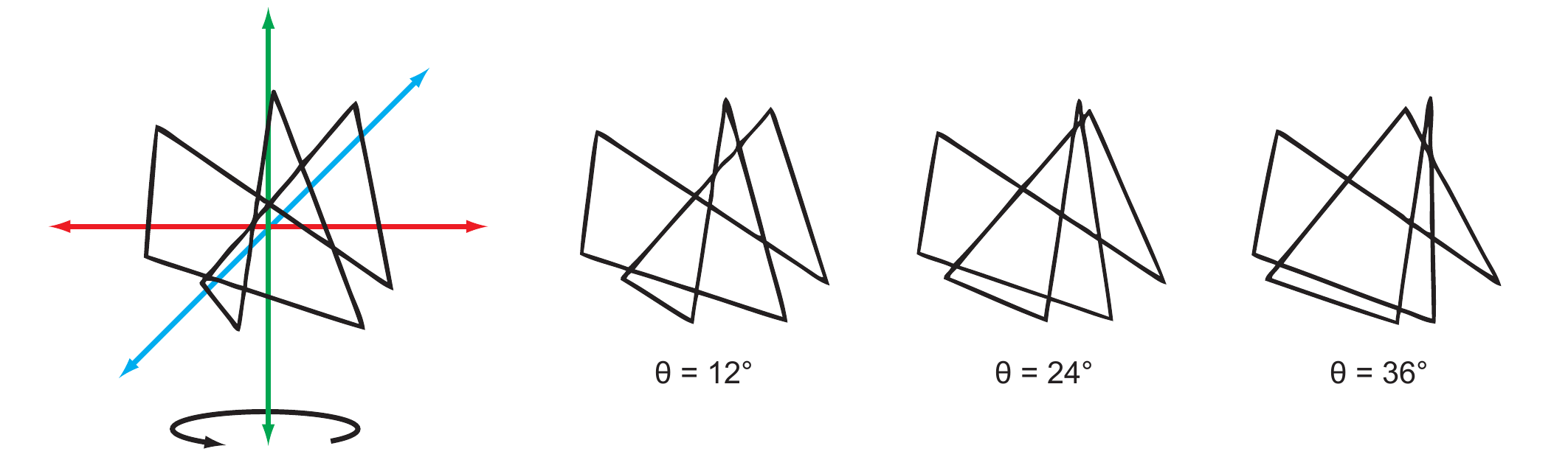} \\
\textbf{B} \\
\includegraphics[width=160mm]{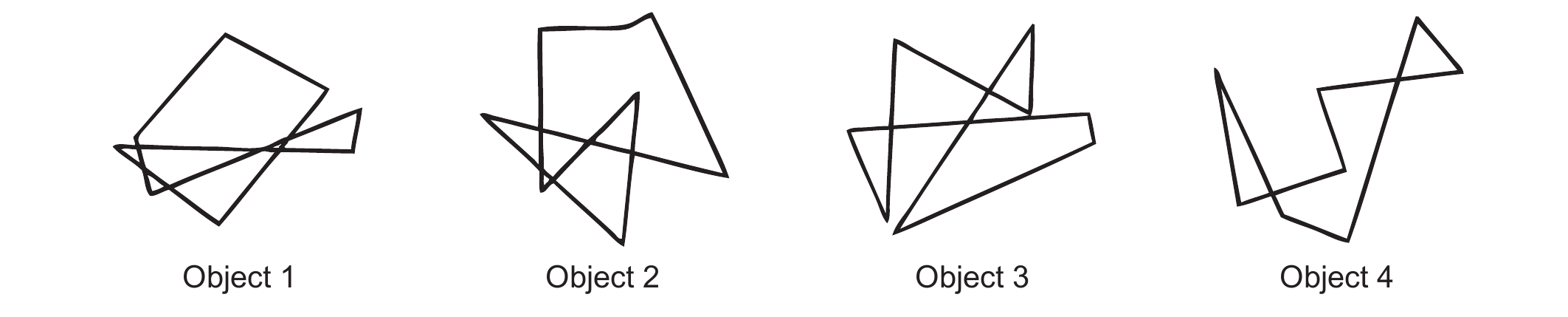} \\
\end{tabular}
\end{center}
\caption{Novel ``paper clip'' objects}{\textbf{A:} Objects were composed of eight three-dimensional vertices joined together with line segments. To render the objects to bitmap images, each object was rotated completely about its vertical axis in steps of 12 degrees and reduced to an orthographic projection. \textbf{B:} Four of the 16 objects used in the experiment.}
\label{fig:pleast_objs}
\end{figure}

Novel ``paper clip'' objects similar to those used in previous investigations of three-dimensional object recognition \cite{BulthoffEdelman92,EdelmanBulthoff92,LogothetisPaulsBulthoffEtAl94,LogothetisPaulsPoggio95,SinhaPoggio96} were created using MATLAB. Eight vertices were placed randomly on the surface of a sphere of unit radius and then joined together with line segments. The last and first vertex were also joined to form a closed loop so that line segment terminations were not a salient feature \cite{BalasSinha09b}. Objects were constrained to exclude extremely acute angles between successive segments (less than 20 degrees) and were approximately rotationally balanced (center of mass within 10\% of the origin). Objects were were rotated completely about their vertical axis in steps of 12 degrees and rendered to bitmap images under an orthographic projection. A total of 16 objects were created using this procedure, yielding 480 images (30 images per object). Object examples are shown in Figure \ref{fig:pleast_objs}.

% paperclip fig
% rotations for one obj, several objs

\subsection{Procedure}
\begin{figure}[h!]
\begin{center}
\begin{tabular}{ll}
\textbf{A} \\
\includegraphics[width=160mm]{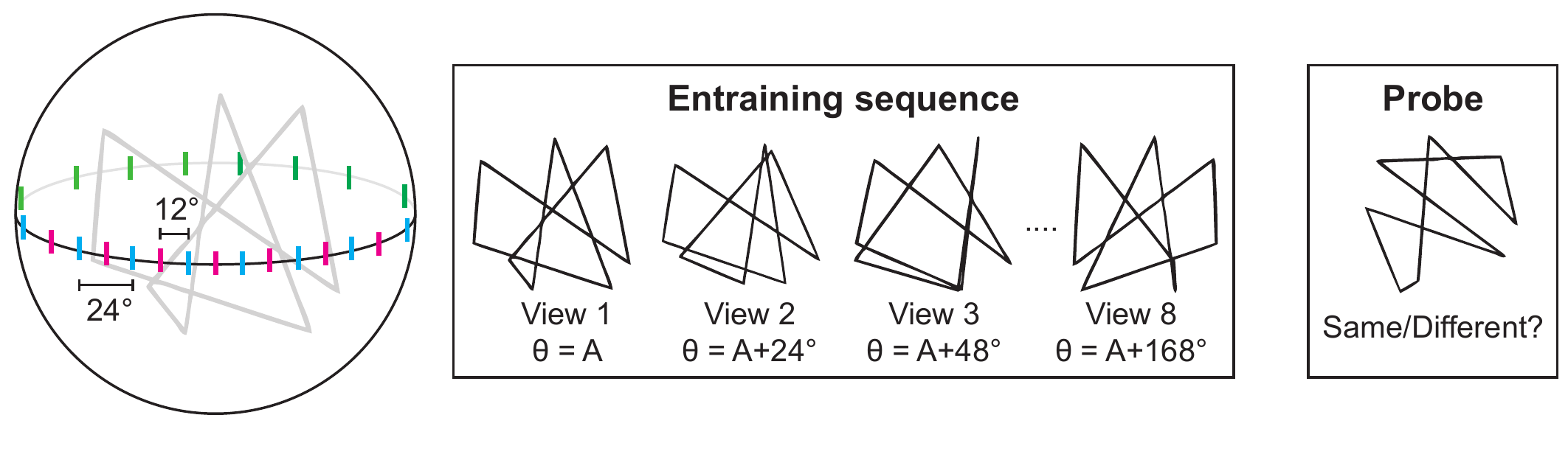} \\
\textbf{B} \hspace{90mm} \textbf{C} \\
\includegraphics[width=160mm]{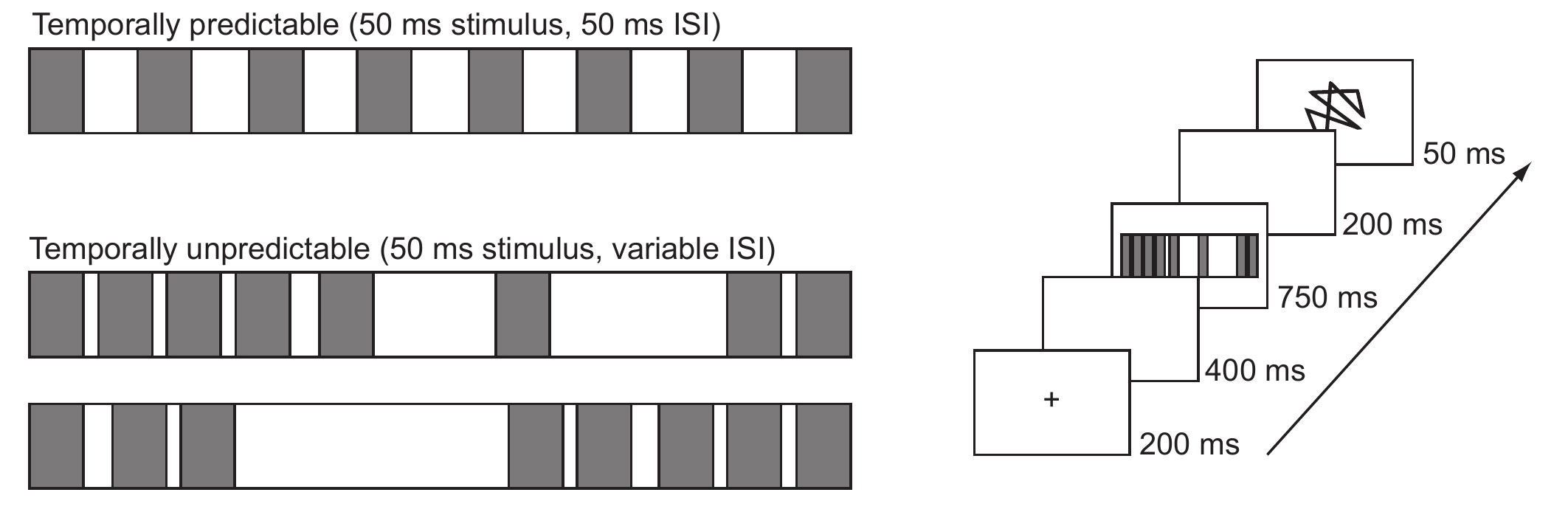} \\
\end{tabular}
\end{center}
\caption{Experimental procedure}{\textbf{A:} Experimental trials consisted of an entraining sequence containing eight views of a single object, followed by a probe stimulus. Entraining views were spaced 24 degrees apart (blue tick marks). The probe depicted an unseen view from the 12 degree interpolations between views of the entraining sequence (magenta tick marks) or from outside the span of the entraining sequence in increments of 24 degrees (green tick marks). \textbf{B:} Entraining views were either presented at 10 Hz with a 50 ms on time and 50 ms off time or in a temporally unpredictable manner with a 50 ms on time (gray segments) and variable off time (white segments). In both cases, the duration of the total entraining sequence was held constant at 750 ms. \textbf{C:} Order and timing  of events within a single trial.}
\label{fig:pleast_task}
\end{figure}

Participants observed an entraining sequence of rotated views of a random object and performed a same-different judgement about a probe stimulus. On each trial, a view was randomly selected as the initial view of the sequence followed by seven additional views spaced 24 degrees apart (Figure \ref{fig:pleast_task}A, blue tick marks). Thus, the entire eight view entraining sequence spanned 168 degrees of the object. The entraining sequence was either presented in order (i.e., spatially predictable) or randomized. Following the entraining sequence after a 200 ms blank was a probe stimulus consisting of either an unseen view from the entraining object or a novel distractor. Unseen views were randomly sampled from the 12 degree interpolations between views of the entraining sequence (Figure \ref{fig:pleast_task}A, magenta tick marks) and from outside of the span of the entraining sequence in increments of 24 degrees (Figure \ref{fig:pleast_task}A, green tick marks).

Distractors were created from the original target objects by randomly selecting new spherical coordinates for six of the eight vertices and re-rendering them to bitmap images using the same method as the original target objects (12 degree steps about the vertical axis). Distractors conformed to the same constraints as the original target objects (no extremely acute angles, approximately rotationally balanced). Participants were instructed to respond ``same'' if they believed the probe depicted the same object as the entraining sequence or ``different'' if it depicted a distractor object. Participants received feedback after each trial according to whether their response was correct or incorrect. % Responses were collected via a millisecond-accurate response box connected through the display computer's serial port. 

During the entraining sequence, object views were presented for 50 ms at either 10 Hz (i.e., temporally predictable) or at a variable rate by manipulating the interstimulus interval (ISI) between subsequent views. Temporally predictable ISIs were 50 ms, totaling 350 ms across the entraining sequence. Variable ISIs were selected by randomly generating seven ISIs that also summed to 350 ms (Figure \ref{fig:pleast_task}B). Variable ISIs were in the range of 17 ms (minimum) to 217 ms (maximum) in increments of 17 ms. Temporal unpredictability was maximized by generating 400 such ISI sequences, calculating the summed squared error (SSE) across subsequent ISIs in a sequence, and selecting the 100 sequences with the highest SSE for use during the experiment.

% task fig

The experiment was displayed on an LCD monitor at native resolution operating at 60 Hz using the Psychophysics Toolbox Version 3 \cite{Brainard97,Pelli97}. Stimuli were presented at central fixation on an isoluminant 50\% gray background and subtended approximately 5 degrees of visual angle. Trials began with a fixation cross (200 ms) followed by a blank (400 ms), the entraining sequence (750 ms total), a second blank (200 ms), and ended with the probe stimulus (50 ms) (Figure \ref{fig:pleast_task}C). Participants were required to respond within 2000 ms or the trial was aborted. Trials were separated by a variable intertrial interval of 2000-2400 ms. The full experiment consisted of 500 trials with an additional 20 practice trials that were characterized by a longer blank (1000 ms) between the entraining sequence and the probe to familiarize participants with the order of events during trials. Participants completed the 20 practice trials (which were discarded from analysis) prior to performing the 500 experimental trials. % random sampling of all variables with replacement

\subsection{EEG recording and preprocessing}
% Net Station 4.3.1
The EEG was recorded using an Electrical Geodesics, Inc.~(EGI) system composed of a 128 channel net (HCGSN 130) amplified through 200 M\SI{}{\ohm} amplifiers (Net Amps 200). The signal was sampled at 250 Hz with impedances for each electrode adjusted to less than 40 k\SI{}{\ohm} before and during the recording. Stimulus and response trigger onsets were measured via the Psychophysics Toolbox using a high precision realtime clock that was synchronized within 2.5 ms of the EEG system's clock before every trial during the experiment.

EEG data were preprocessed using the FieldTrip toolbox \cite{OostenveldFriesMarisEtAl11}. Raw data were first band-pass filtered between 1 Hz and 100 Hz with a 59-61 Hz band-stop and then epoched into 2350 ms segments that spanned the start of the pre-trial blank to 1000 ms after the probe stimulus. Individual segments were visually inspected and rejected if found to contain muscle artifacts or atypical noise. Bad channels were also identified and temporarily removed from the data before performing ICA decomposition \cite{DelormeMakeig04} to remove ocular artifacts. Components related to ocular artifacts were identified based on their topographical distribution across electrodes. The data were reconstructed without the ocular components and any bad channels were replaced using spherical spline interpolation \cite{PerrinPernierBertrandEtAl89}. The resulting segments were re-referenced using an average reference.

\subsection{Event-related averaging}
Event-related averaging was performed separately for the entraining sequence and the subsequent probe. For the entraining sequence, data were aligned to the onset of entraining views 2 through 8 and averaged from the period beginning 50 ms before each entrainer and ending 50 ms after. Baseline correction was performed using the first 50 ms of this period. For the probe, data were aligned to the probe onset and averaged from the period beginning 200 ms before the probe and ending 400 ms after. This allowed detection of predictability effects during the blank period due to differences in phase elicited by the entraining sequence as well as probe-evoked predictability effects.

\begin{figure}[h!]
\begin{center}
\includegraphics[width=80mm]{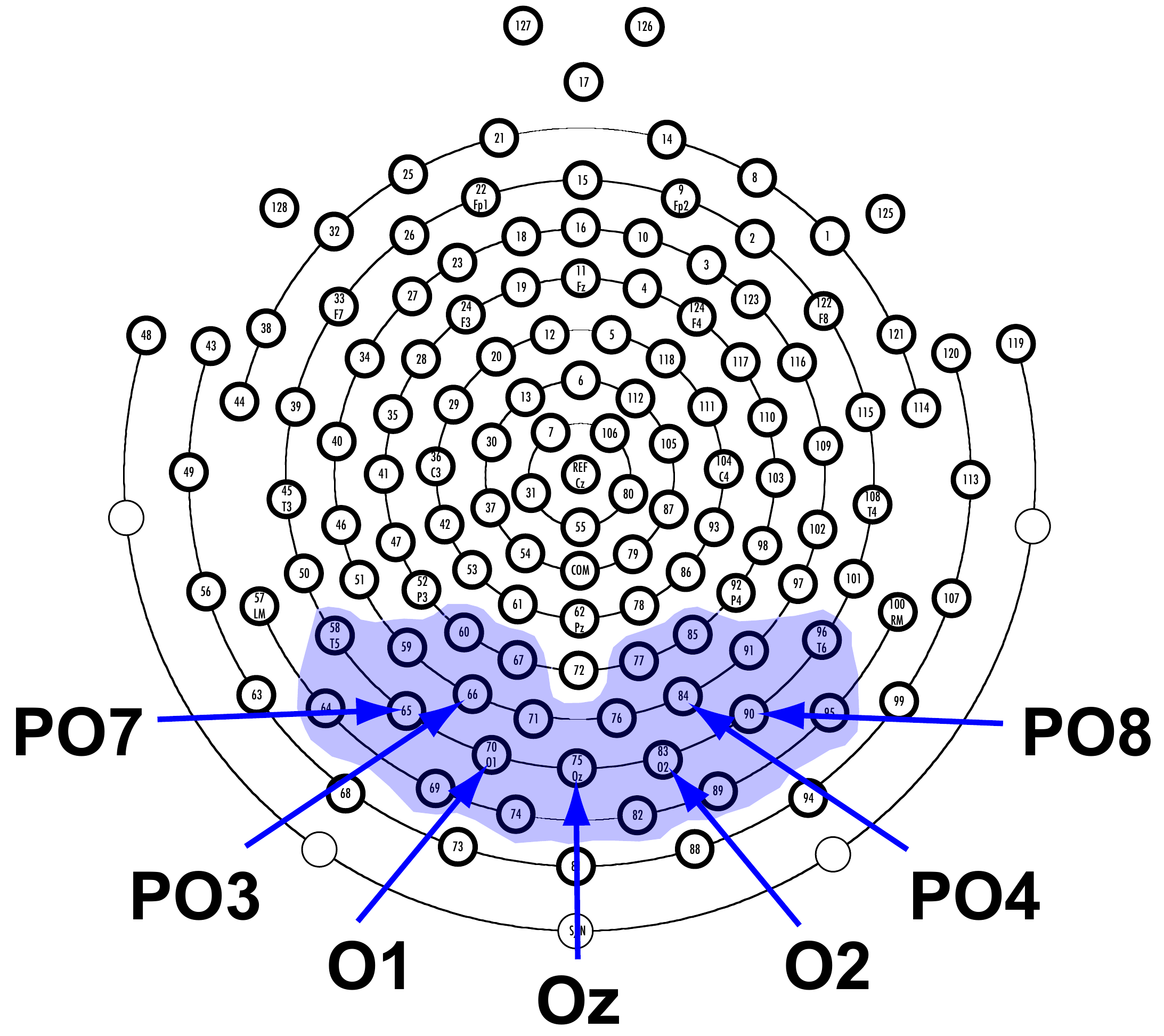}
\end{center}
\caption{Electrode pooling for EEG analyses}{Blue shaded region denotes pooled electrodes. Locations from the 10-10 system are indicated.}
\label{fig:pleast_channels}
\end{figure}

All waveforms were averaged over a montage of 23 electrodes that covered the occipital and parietal cortices (Figure \ref{fig:pleast_channels}). The montage included locations from the 10-10 system that are commonly associated with perceptual processing (Oz, O1/O2, PO3/PO4, and PO7/PO8) \cite[e.g.,]{DohertyRaoMesulamEtAl05,RohenkohlNobre11,FahrenfortScholteLamme07}.

% electrode poolings fig

\subsection{Time-frequency analysis}
Segmented data were used to compute time-frequency data for each trial. Data were first downsampled to 125 Hz and then used to compute the instantaneous Fourier coefficients at each time bin using a multi-taper approach. Hanning tapers were generated at 5-20 Hz and convolved with the data using a sliding time window (four cycles per frequency per time window). The relatively long time window required for low-frequency bands prohibits computation of time-frequency data for short time segments, such as the 200 ms blank before the probe. To address this issue, time-frequency data were computed over the entire 2350 ms trial epoch and then cropped to investigate temporal regions of interest.

Power was computed from the magnitude of the instantaneous Fourier coefficients at each frequency (\textit{f}) and time bin (\textit{t}):
\begin{align*}
Power(f,t) = \frac{1}{n} \sum_{k=1}^{n}{|F_k(f,t)|}^2
\end{align*}

Phase, which is normally defined as the angle of the complex Fourier coefficients, cannot be averaged due to its circularity and thus standard statistical models cannot be applied to assess its significance. One solution to this problem is to instead compute inter-trial coherence (ITC) \cite{LachauxRodriguezMartinerieEtAl99}. ITC is averaged in the complex domain by first normalizing phase information to unit length by dividing off power and then computing the magnitude:
\begin{align*}
ITC(f,t) = \bigg{|}\frac{1}{n} \sum_{k=1}^{n}{\frac{F_k(f,t)}{|F_k(f,t)|}}\bigg{|}
\end{align*}

ITC ranges between 0 and 1 and represents how systematic phase angles are across trials. A value of 0 indicates that phase information is essentially uniformly random across trials while a value of 1 indicates a high degree of phase-locking at a particular frequency across trials.

All time-frequency analyses were averaged over the same montage of 23 occipitoparietal electrodes that were used to compute event-related averages (Figure \ref{fig:pleast_channels}). 

\section{Results}

\subsection{Behavioral measures of spatial and temporal predictability}
\begin{figure}[h!]
\begin{center}
\begin{tabular}{ll}
\includegraphics[width=80mm]{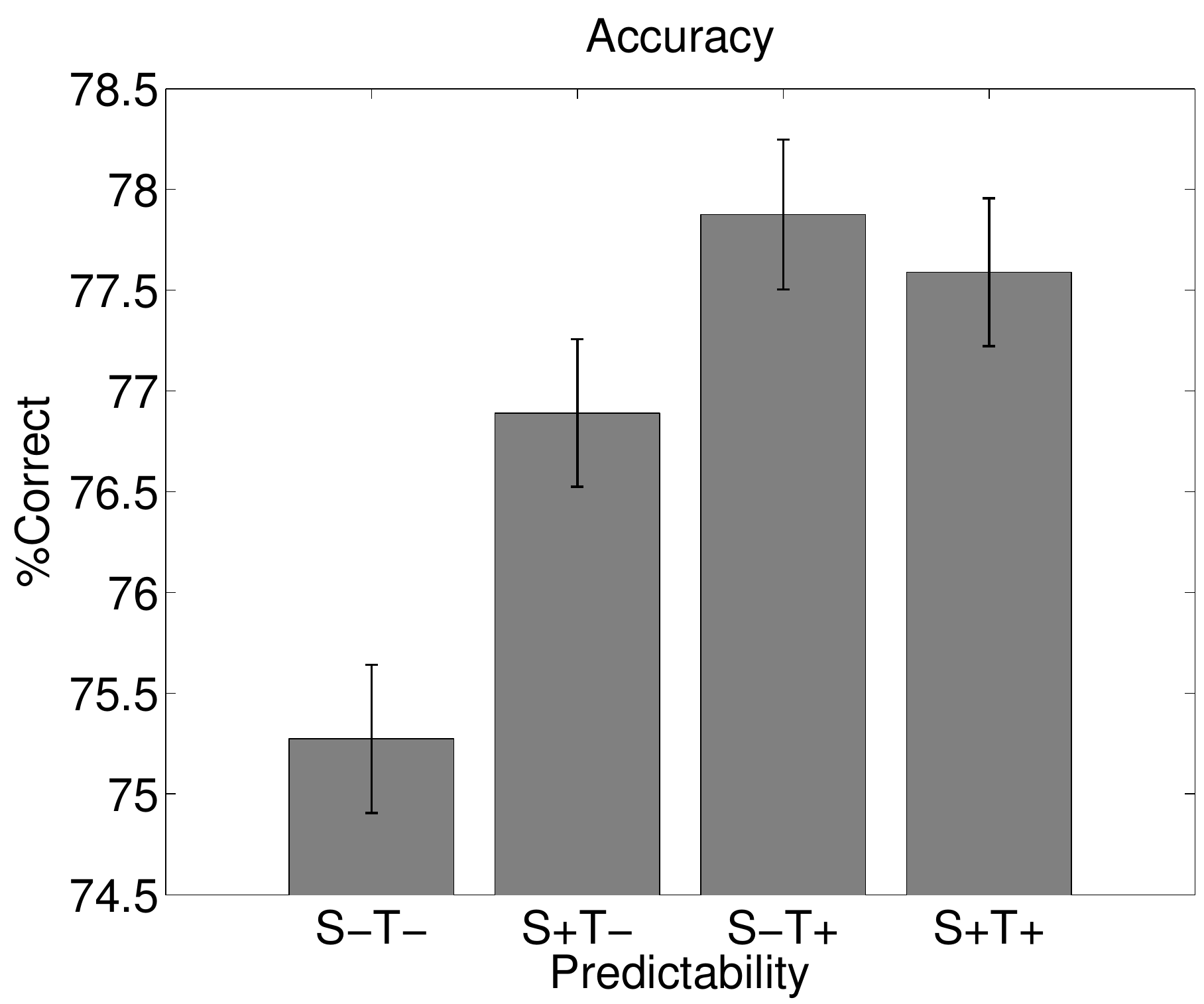} & 
\includegraphics[width=80mm]{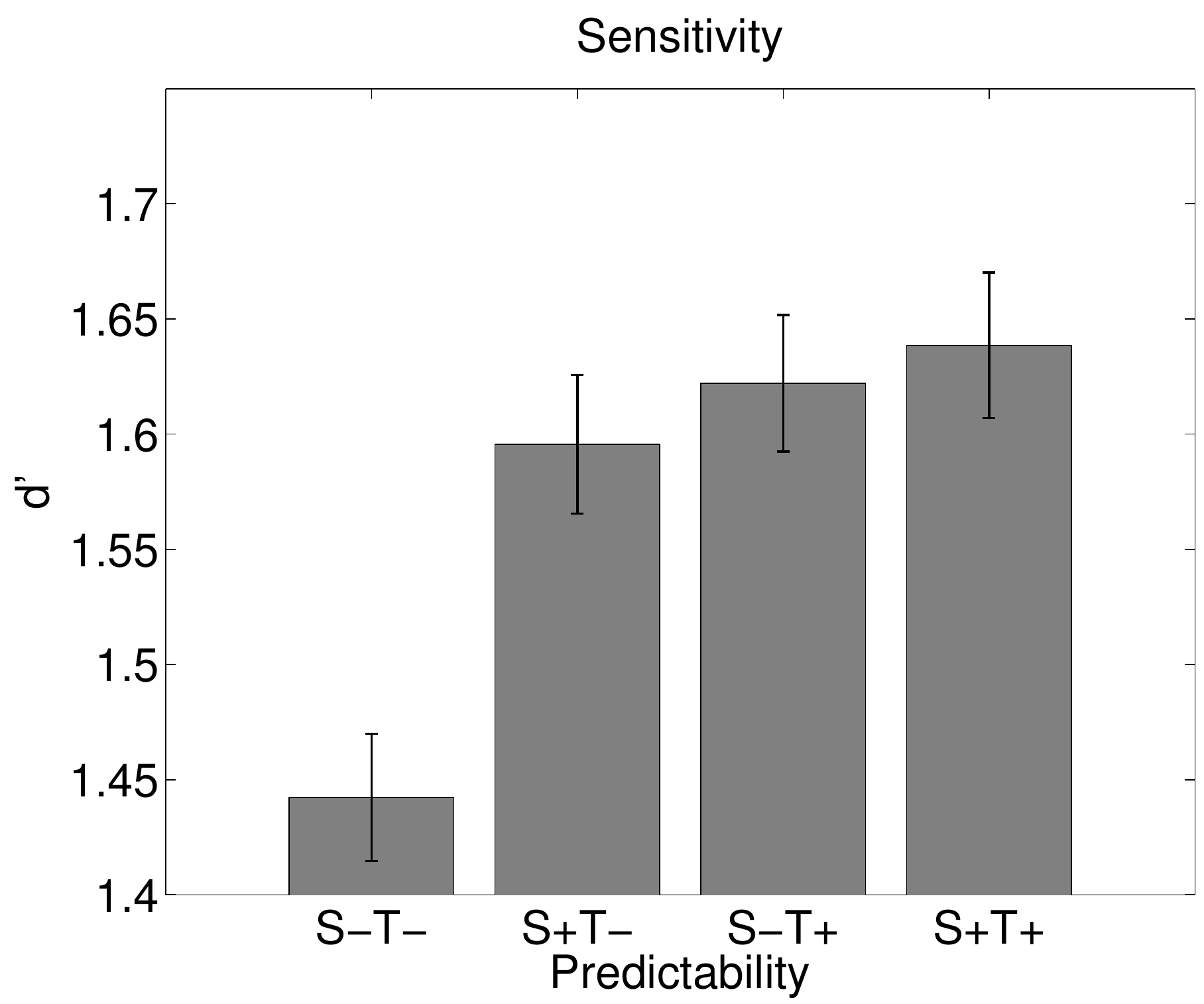} \\
\includegraphics[width=80mm]{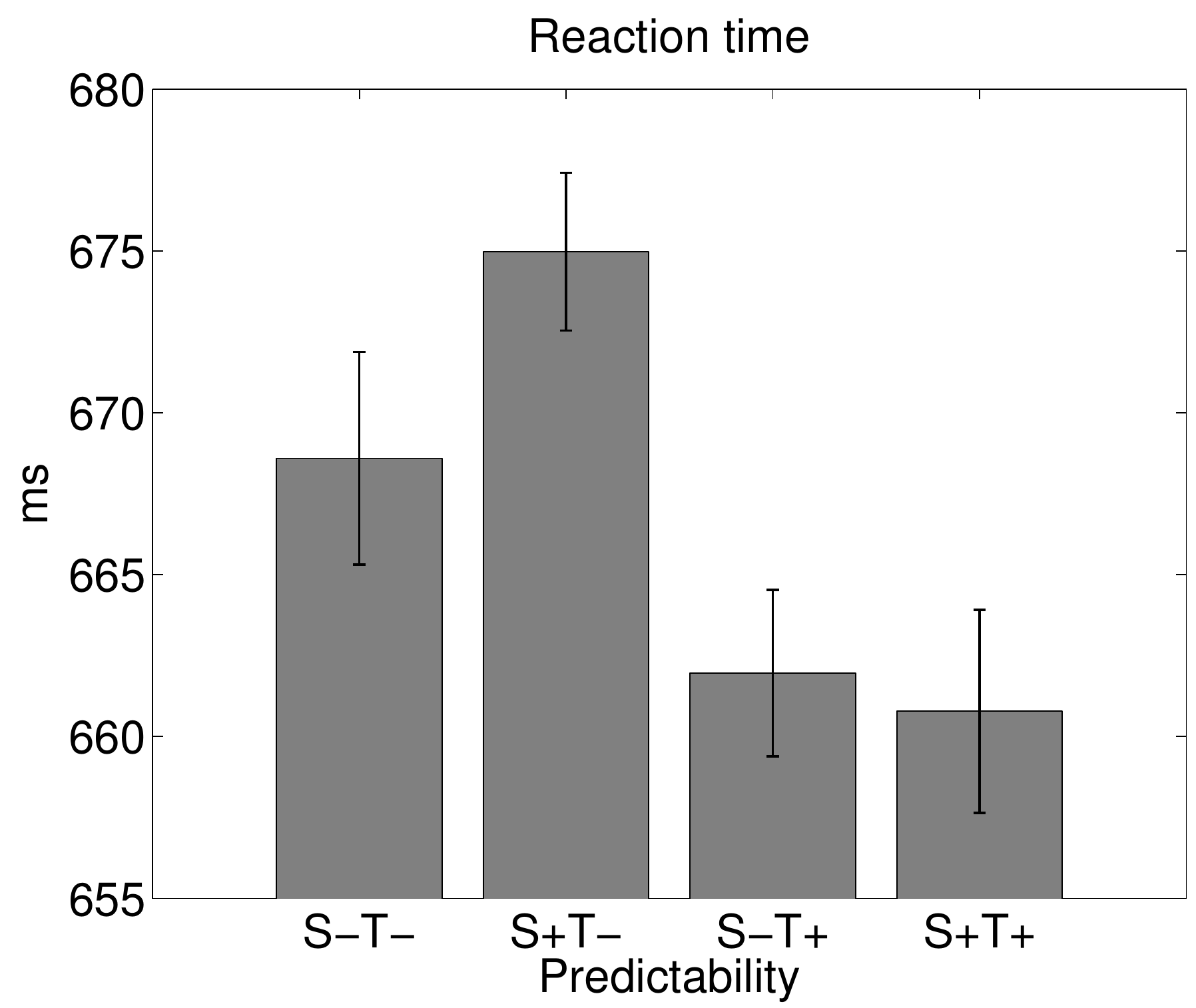} &
\includegraphics[width=80mm]{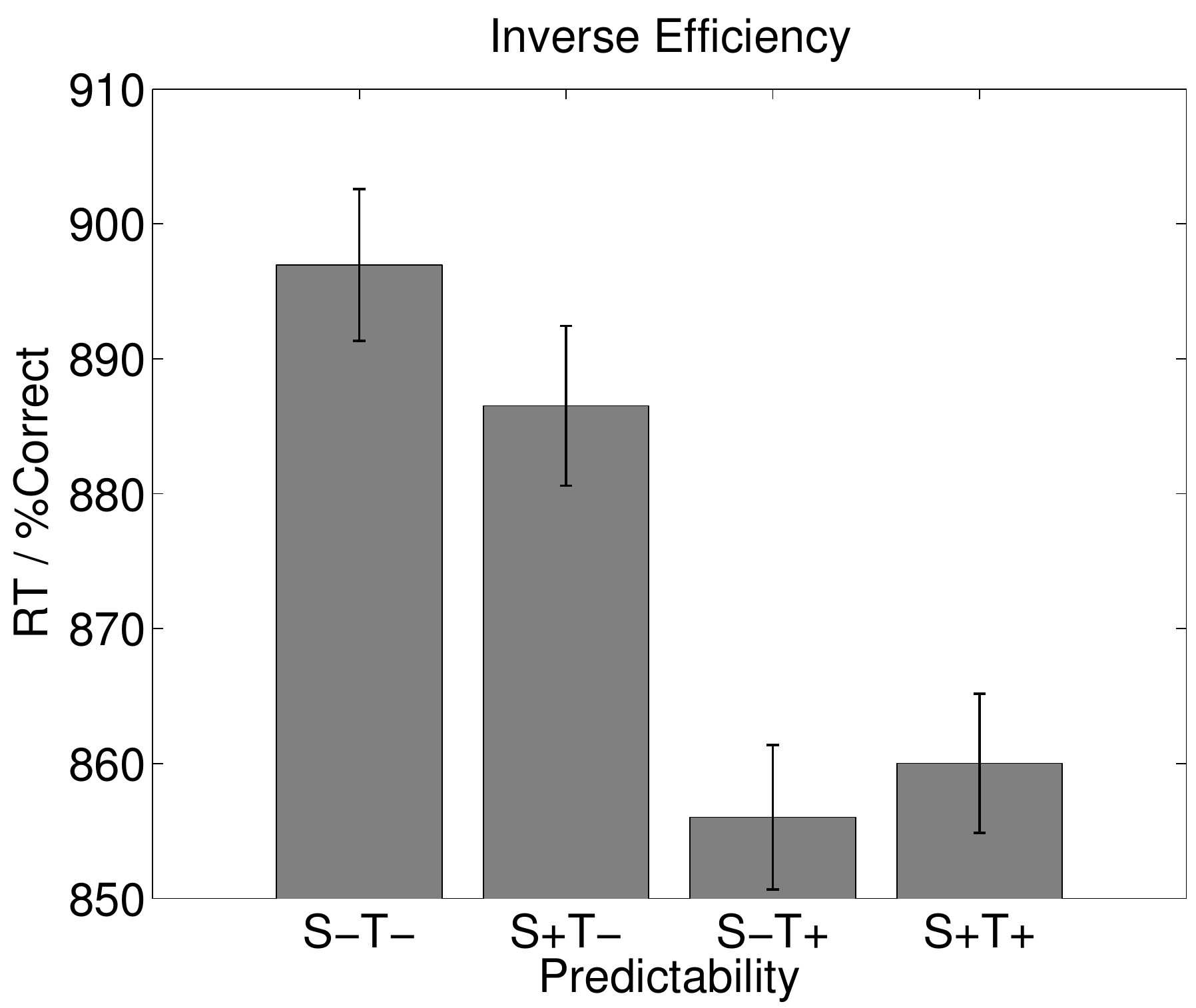} \\
\end{tabular}
\end{center}
\caption{Behavioral measures of spatial and temporal predictability}{Accuracy, \textit{d'} (sensitivity), reaction time, and inverse efficiency (reaction time divided by percent correct) as a function of entrainment condition. S-/+ refers to spatially unpredictable and predictable, T-/+ to temporally unpredictable and predictable. Error bars depict within-subjects error using the method described in \protect\incite{Cousineau05} adapted for standard error.}
\label{fig:pleast_behave}
\end{figure}

Five subjects were excluded from behavioral analysis for accuracy 2.7$\sigma$ (or further) below mean accuracy across subjects. The remaining 53 subjects were submitted to a 2x2 ANOVA with spatial and temporal predictability as within-subjects factors. Experiment type (EEG or behavioral only) was included as an additional between-subjects factor to ensure that it did not interact with any of the within-subjects factors. Accuracy and reaction times were collected during the experiment and were used to compute \textit{d'}, a measure of sensitivity that takes into account response bias, and inverse efficiency, a measure that combines accuracy and reaction times \cite{TownshendAshby78}. These behavioral measures are plotted in Figure \ref{fig:pleast_behave}.

% oh, behave

% D464B subs acc=75, rt=624.72
% E029 subs acc=0.79 rt=742.1212
Subjects that completed the full EEG experiment were on average less accurate (\textit{F}(1, 51) = 4.80, \textit{p} = 0.033) but responded more quickly (\textit{F}(1, 51) = 9.63, \textit{p} = 0.003) than subjects that completed the solely behavioral experiment. These differences reflect a speed-accuracy tradeoff, likely due to differences in instructions given to subjects by experimenters or motivational differences between subject groups based on whether EEG was recorded. Importantly, experiment type did not interact with any within-subjects factors (all \textit{p}'s $>$ 0.05) indicating that the behavioral measures of interest were not dependent on which type of experiment subjects completed.

Overall, subjects were more accurate when the entraining sequence was temporally predictable (\textit{F}(1, 51) = 17.84, \textit{p} $<$ 0.001). A similar effect for spatial predictability failed to reach significance (\textit{F}(1, 51) = 1.85, \textit{p} = 0.18). The interaction between spatial and temporal predictability, however, was significant (\textit{F}(1, 51) = 6.13, \textit{p} = 0.017). The LeabraTI model (Chapter \ref{chap:leabrati}) as well as investigations of predictability on attentional allocation \cite{DohertyRaoMesulamEtAl05,RohenkohlGouldPessoaEtAl14} suggest that combined spatial and temporal predictability should have a superadditive effect on behavioral outcomes. However, the combined spatial and temporal predictability condition here (denoted S+T+ in Figure \ref{fig:pleast_behave}) was subadditive. Although not significantly different from spatial predictability alone (\textit{t}(52) = 1.29, \textit{p} = 0.204) or from temporal predictability alone (\textit{t}(52) = 0.45, \textit{p} = 0.652), this result merits further investigation. 

When responses are transformed into \textit{d'}, there is a significant effect of both spatial (\textit{F}(1, 51) = 4.71, \textit{p} = 0.035) and temporal predictability (\textit{F}(1, 51) = 11.99, \textit{p} $<$ 0.001). This result suggests that response bias can at least partially explain why spatial predictability failed to reach significance for raw accuracy. The interaction between spatial and temporal predictability remained significant for \textit{d'} (\textit{F}(1, 51) = 4.49, \textit{p} = 0.039). The interaction was additive, but was driven primarily by the strong effect of introducing spatial or temporal predictability over complete unpredictability (S-T- versus S+T-, \textit{t}(52) = 3.19, \textit{p} = 0.002; S-T- versus S+T+, \textit{t}(52) = 4.26, \textit{p} $<$ 0.001) opposed to any synergistic effect of combined spatial and temporal predictability (S+T+ versus S+T-, \textit{t}(52) = 0.90; S+T+ versus S-T+, \textit{t}(52) = 0.31; both \textit{p}'s $>$ 0.05).

Reaction times for correct trials were significantly faster when the entraining sequence was temporally predictable (\textit{F}(1, 51) = 9.92, \textit{p} $=$ 0.003). A similar effect for spatial predictability failed to reach significance (\textit{F}(1, 51) = 0.45, \textit{p} = 0.504) nor did the interaction term (\textit{F}(1, 51) = 1.83, \textit{p} = 0.182).

Inverse efficiency, which considers reaction time as a function of accuracy (defined as reaction time divided by percent correct) can be thought of as the amount of energy consumed by the system to produce a behavioral outcome \cite{TownshendAshby83}. It is often used to remove non-monotonicities present in accuracy or reaction times alone, although that effect is not observed here. Nevertheless, it provides another lens under which to inspect the data, and thus it is considered here. Inverse efficiency was significantly lower when the entraining sequence was temporally predictable (\textit{F}(1, 51) = 25.00, \textit{p} $<$ 0.001), but not when it was spatially predictability (\textit{F}(1, 51) = 0.22, \textit{p} = 0.644). The interaction between spatial and temporal predictability failed to reach significance for inverse efficiency (\textit{F}(1, 51) = 1.94, \textit{p} = 0.170).

% below only true when considering correct AND incorrect trials
%
%Inverse efficiency was characterized by a significant cross-over interaction (\textit{F}(1, 51) = 5.85, \textit{p} = 0.019). Spatial predictability of the entraining sequence produced lower inverse efficiency over complete unpredictability. Inverse efficiency was lowest on average when stimuli were temporally predictable, but the addition of spatial predictability caused an increase in inverse efficiency.

\begin{figure}[h!]
\begin{center}
\begin{tabular}{ll}
\includegraphics[width=80mm]{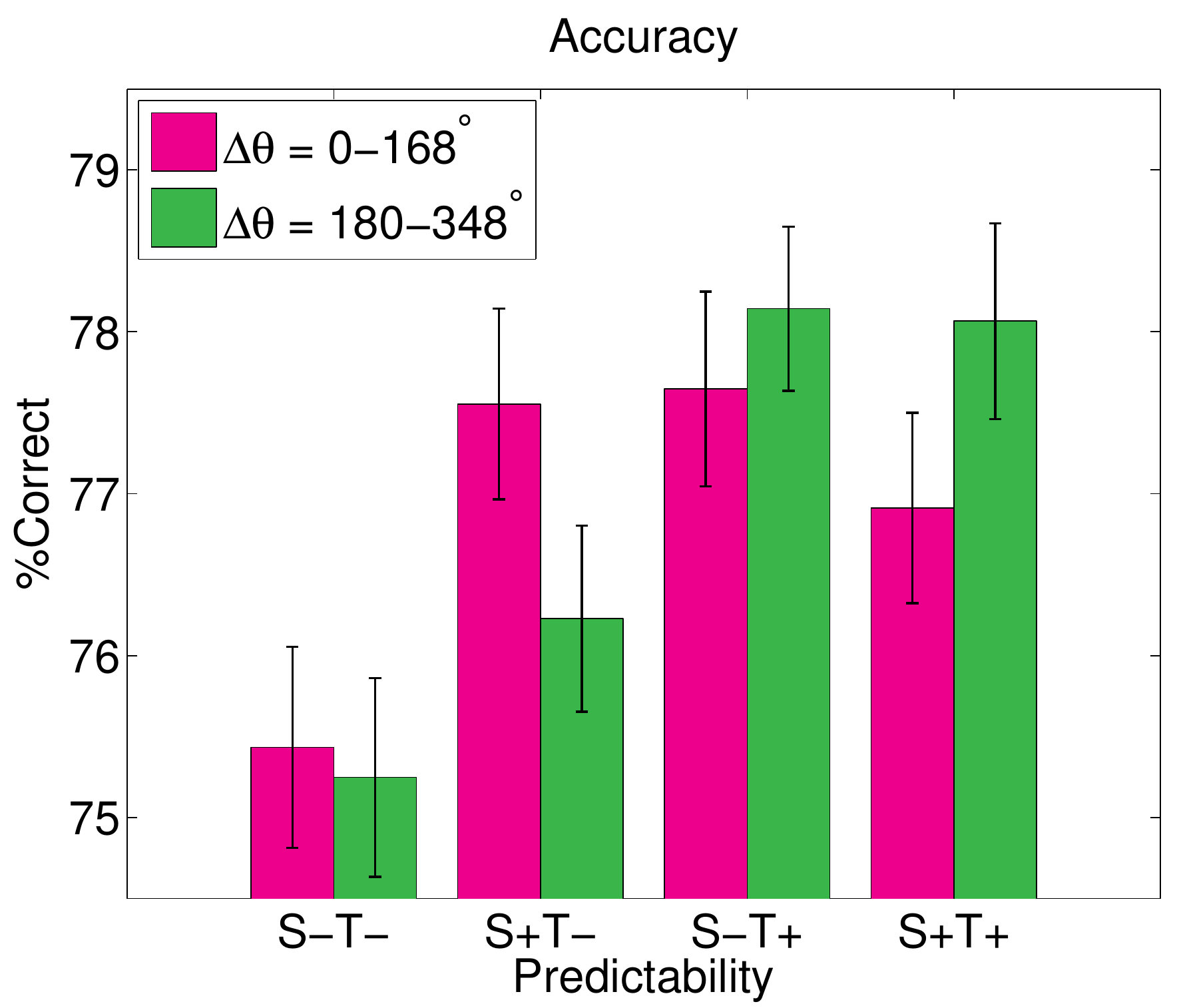} & 
\includegraphics[width=80mm]{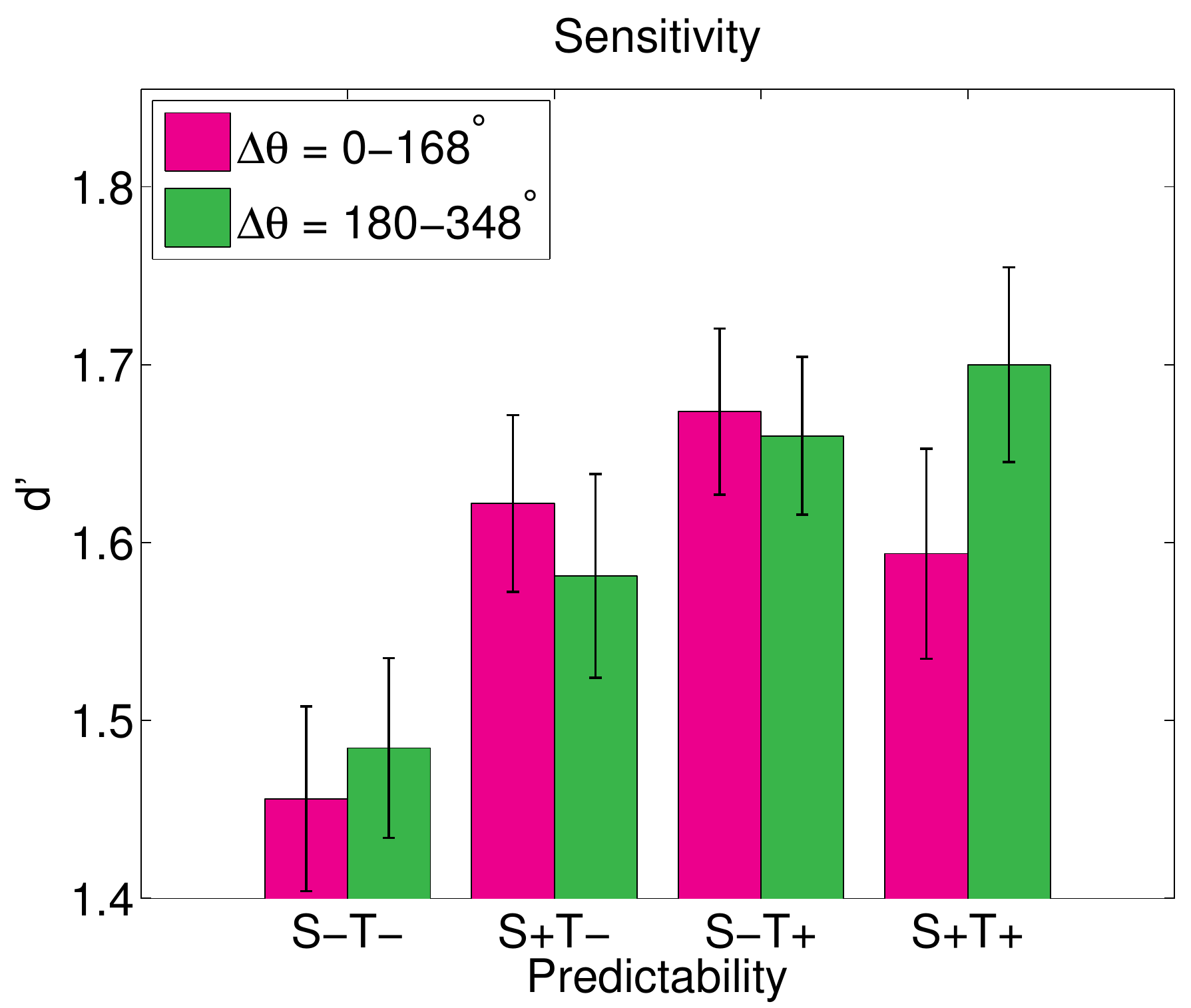} \\
\includegraphics[width=80mm]{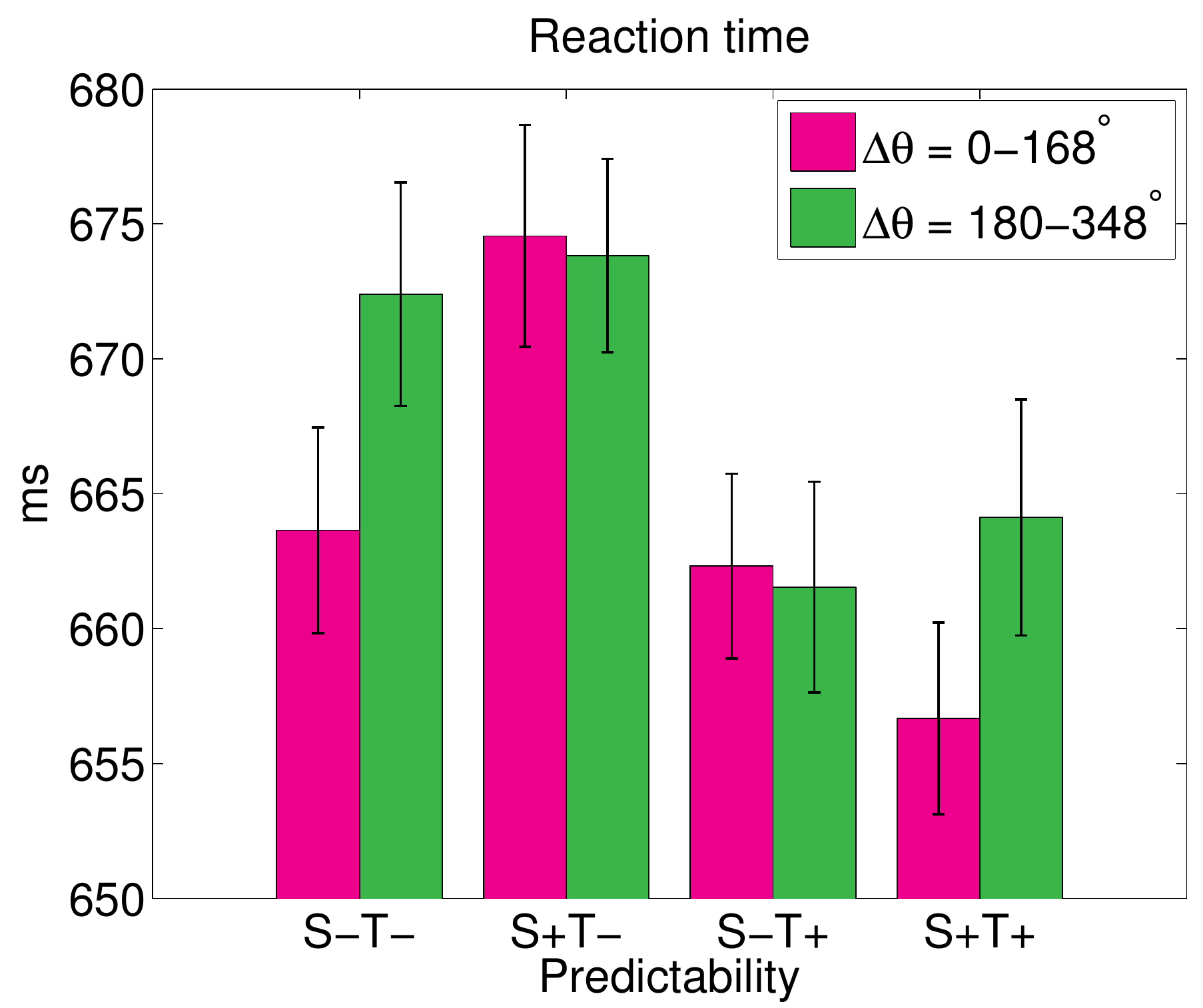} & 
\includegraphics[width=80mm]{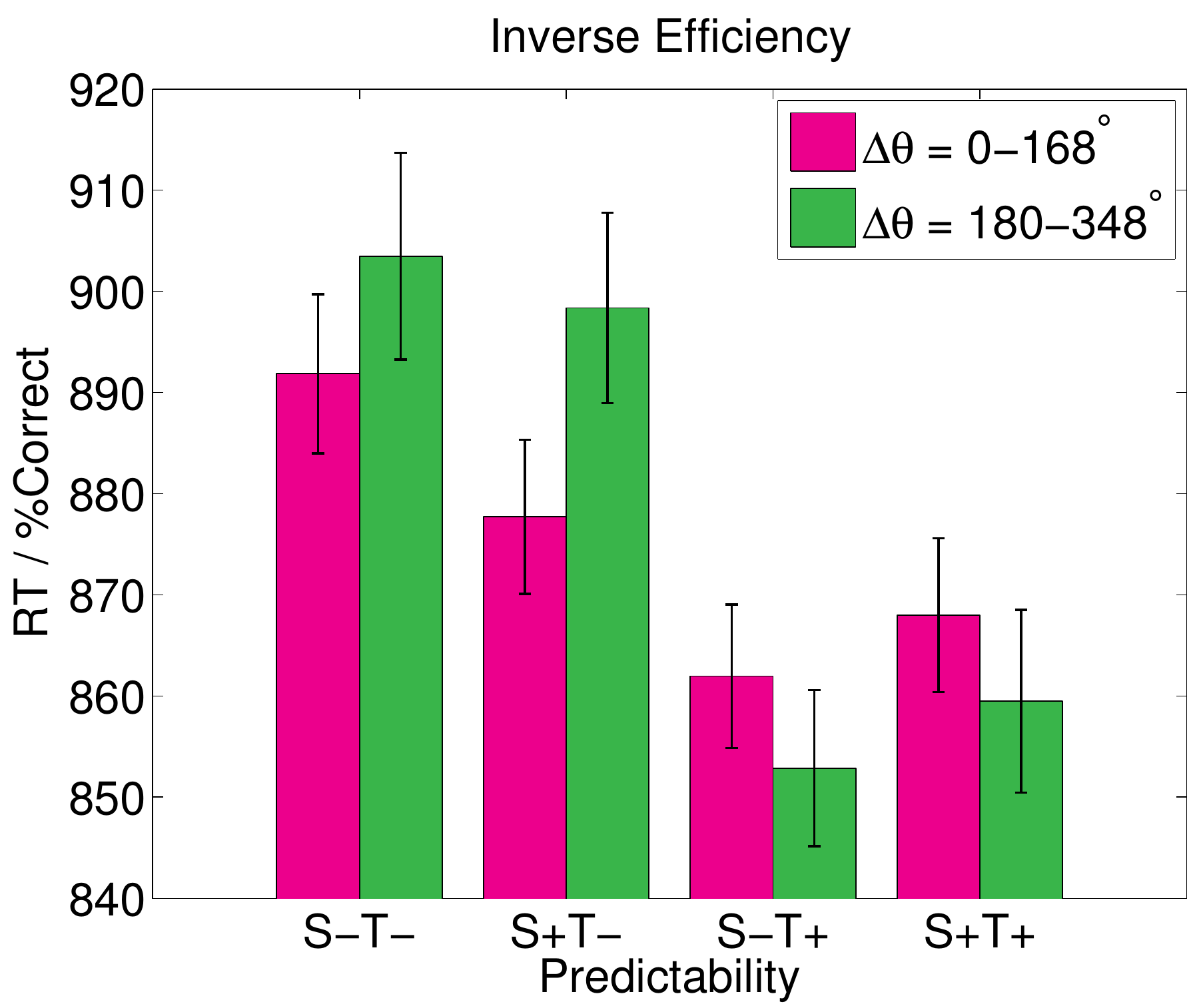} \\
\end{tabular}
\end{center}
\caption{Behavioral effects of predictability on view interpolation and extrapolation}{Same behavioral measures as described in Figure \ref{fig:pleast_behave}, but split according to whether the probe judgement required view interpolation or extrapolation. Magenta bars depict view interpolation (i.e., angular difference between initial view and probe less than 168 degrees), green bars view extrapolation (i.e., angular difference between initial view and probe greater than 168 degrees). Error bars depict within-subjects error using the method described in \protect\incite{Cousineau05} adapted for standard error.}
\label{fig:pleast_behave_angle}
\end{figure}

Responses were also split according to whether the probe depicted a view interpolated from within the entraining sequence or was extrapolated from outside the span of the entraining sequence. This analysis allowed inference about whether either of these cases interacted meaningfully with predictability (Figure \ref{fig:pleast_behave_angle}). For example, prediction might   only be possible for probes that are a logical extension of the entraining sequence. Both ``same'' and ``different'' probe judgements were included in these groupings since distractors shared some features with targets, even across viewing angles. Statistical comparisons indicated that view interpolation/extrapolation did not interact with spatial or temporal predictability for any of the behavioral measures (all \textit{p}'s $>$ 0.05).

% oh, behave pt 2 -- no significant effects

\subsection{Time course of spatial and temporal predictability}
A total of five subjects were excluded from EEG analysis -- three for an overabundance of artifacts in the EEG recording resulting in low trial counts after rejection and two for accuracy 2.7$\sigma$ (or further) below mean accuracy across subjects (these two subjects were also excluded from behavioral analyses, see preceding section). The remaining 24 subjects were included in all EEG analyses. 

A 2x2 ANOVA with spatial and temporal predictability as within-subjects factors was used to assess statistical significance at each time bin of event-related averages. \textit{p}-values were corrected for a maximum false discovery rate (FDR) of 5\% using the method described in \incite{BenjaminiYekutieli01}. Additionally, effects were only considered significant if they persisted for at least 16 ms.

\begin{figure}[h!]
\begin{center}
\includegraphics[width=160mm]{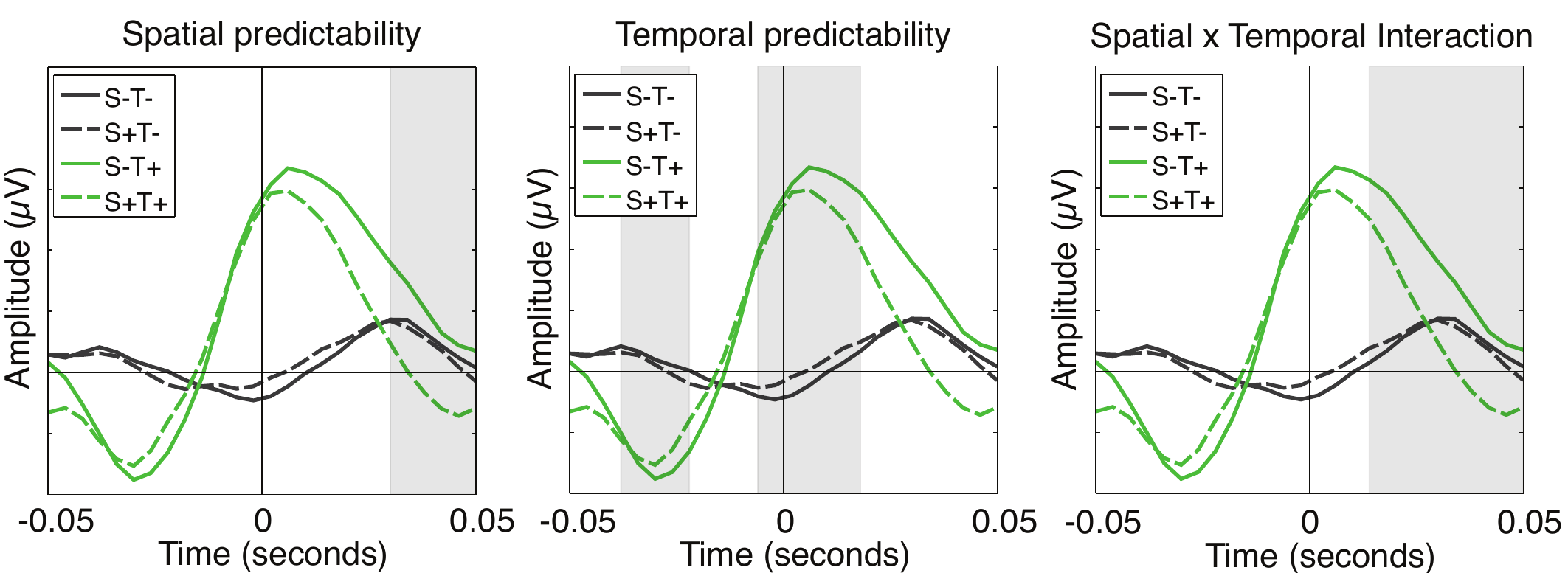}
\end{center}
\caption{Entrainer-evoked activity}{Grand averages for entrainers 2 through 8 as a function of entrainment condition. S-/+ refers to spatially unpredictable and predictable, T-/+ to temporally unpredictable and predictable. All plots depict the grand average with gray shaded regions denoting significant effects of spatial predictability (left), temporal predictability (center), and the interaction between these terms controlling for a maximum false discovery rate (FDR) of 5\%.}
\label{fig:pleast_entrain_tla}
\end{figure}

To investigate the build-up of spatial and temporal predictability over the entraining sequence, activity from the second through final entraining views was averaged for each condition (the first entraining view was always unpredictable, so it was omitted from the average). The results of this analysis are plotted in Figure \ref{fig:pleast_entrain_tla}. The first thing worth noting is that a large 10 Hz periodicity was present for the temporally predictable conditions (S-T+ and S+T+), phase-aligned approximately to the onset of each entrainer. Temporally unpredictable entrainers (S-T- and S+T-) were also approximately periodic, but with less amplitude change from baseline. The reason for the 10 Hz periodicity in these conditions despite being temporally unpredictable is likely due to the 750 ms constant duration of the entraining sequence regardless of condition (Figure \ref{fig:pleast_task}B). Presenting eight stimuli over 750 ms with a variable ISI is still a 10 Hz presentation rate on average. Still, the temporally unpredictable entrainers exhibited markedly weaker amplitude and were approximately 180 degrees out of phase with the the temporally predictable entrainers.

% erp - entrain
% revision todo: recreate these figs with y-axis numbers

The effect of spatial predictability manifested 26 ms after the onset of each entrainer and persisted for at least another 24 ms (one fourth of the 10 Hz period). Temporal predictability, in contrast, manifested prior to (-38 through -22 ms pre-stimulus) and at the onset of each entrainer (-6 ms pre-stimulus through 18 ms post-stimulus). The effect of temporal predictability appears to be driven primarily by the antiphase relationship between T- and T+ conditions at these time points. Together, these effects demonstrate differential time courses for spatial and temporal predictability.

Spatial predictability was enhanced when stimuli were temporally predictable. This effect is characterized by the significant interaction between spatial and temporal predictability starting 14 ms after the onset of the entrainer and persisting for at least another 36 ms. This result indicates that the brain is more capable of differentiating between spatially coherent and random sequences of stimuli when it can properly anticipate the presentation of each stimulus (S-T+ versus S+T+) compared to when the onset is unpredictable.

\begin{figure}[h!]
\begin{center}
\includegraphics[width=160mm]{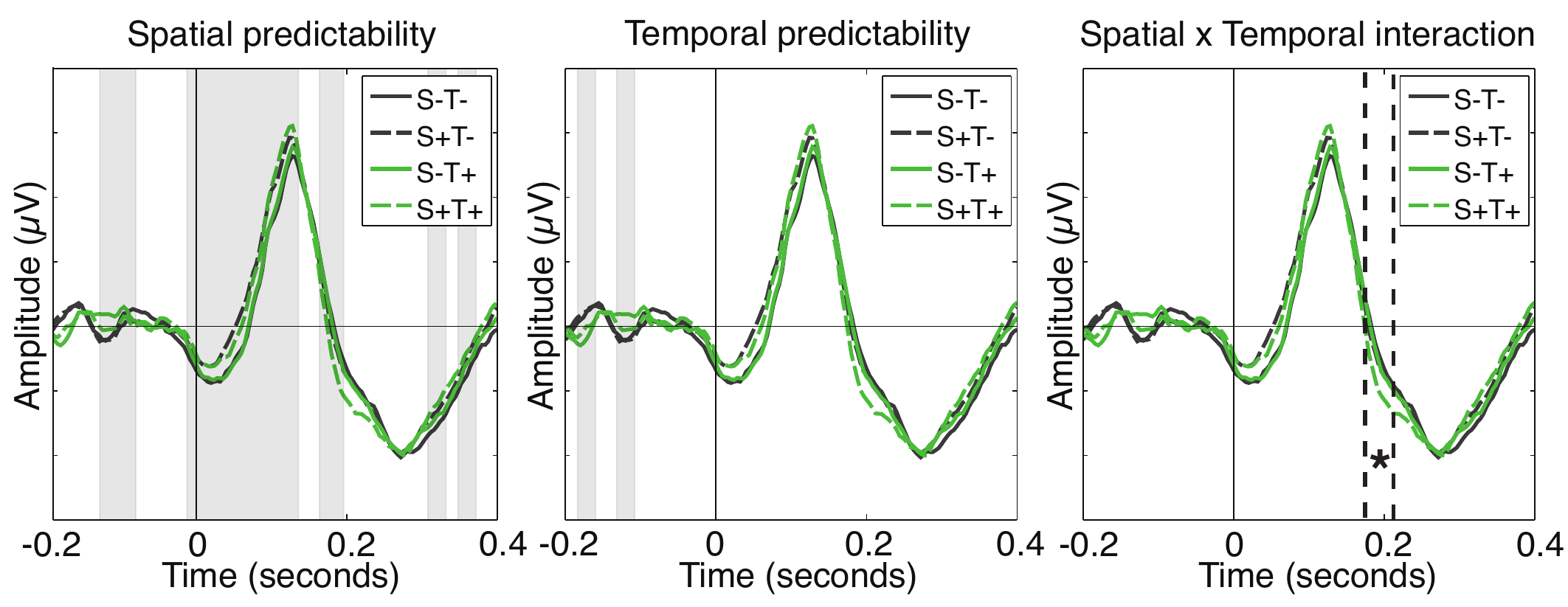}
\end{center}
\caption{Probe-evoked activity}{Grand averages for the probe stimulus, including the preceding 200 ms blank. Asterisk in the interaction plot indicates trending significance at the 5\% level when averaging amplitude within the window defined by the dotted lines.}
\label{fig:pleast_probe_tla}
\end{figure}

To investigate the effect of spatial and temporal predictability on perception of the probe, waveforms were aligned to the probe onset averaged from 200 ms before its presentation through 400 ms after (Figure \ref{fig:pleast_probe_tla}). The results of this analysis essentially mirror the entrainer-evoked effects albeit with a few key differences. Temporal predictability was again a purely anticipatory process, manifesting within the pre-stimulus period (-184 ms through -160 ms and -132 ms through -108 ms pre-stimulus). Spatial predictability also manifested briefly within the pre-stimulus period (-128 ms through -80 ms pre-stimulus), which was not seen for spatially predictable entrainers. The post-stimulus effects of spatial predictability occurred much earlier than for entrainers and persisted much longer, beginning approximately at the onset of the probe (-12 ms pre-stimulus) and lasting 136 ms through the P1 response with several transient effects after.

% erp - probe
% revision todo: recreate these figs with y-axis numbers

Unlike the significant interaction between spatial and temporal predictability for entrainers, the interaction for the probe failed to reach significance given the constraints of the statistical test (maximum FDR of 5\%, 16 ms consecutive significance). However, averaging the amplitude within a window defined by exploratory analysis indicated a trending interaction from 170 ms to 210 ms (\textit{F}(1, 22) = 3.97, \textit{p} = 0.059). This effect was more pronounced and reached significance if only the right hemisphere channels were considered (\textit{F}(1, 22) = 5.61, \textit{p} = 0.027), but failed to reach significance in the left hemisphere (\textit{F}(1, 22) = 2.24, \textit{p} = 0.148), explaining the trending effect when both hemispheres are considered jointly.

\subsection{Predictability entrains alpha oscillations}
The same 24 subjects used in event-related analyses (see preceding section) were used in time-frequency analyses described here. Statistical methods were also identical, consisting of a 2x2 ANOVA with spatial and temporal predictability as within-subjects factors and \textit{p}-values corrected for a 5\% maximum FDR \cite{BenjaminiYekutieli01}. Effects were only considered significant if they persisted for at least 16 ms.

\begin{figure}[h!]
\begin{center}
\includegraphics[width=160mm]{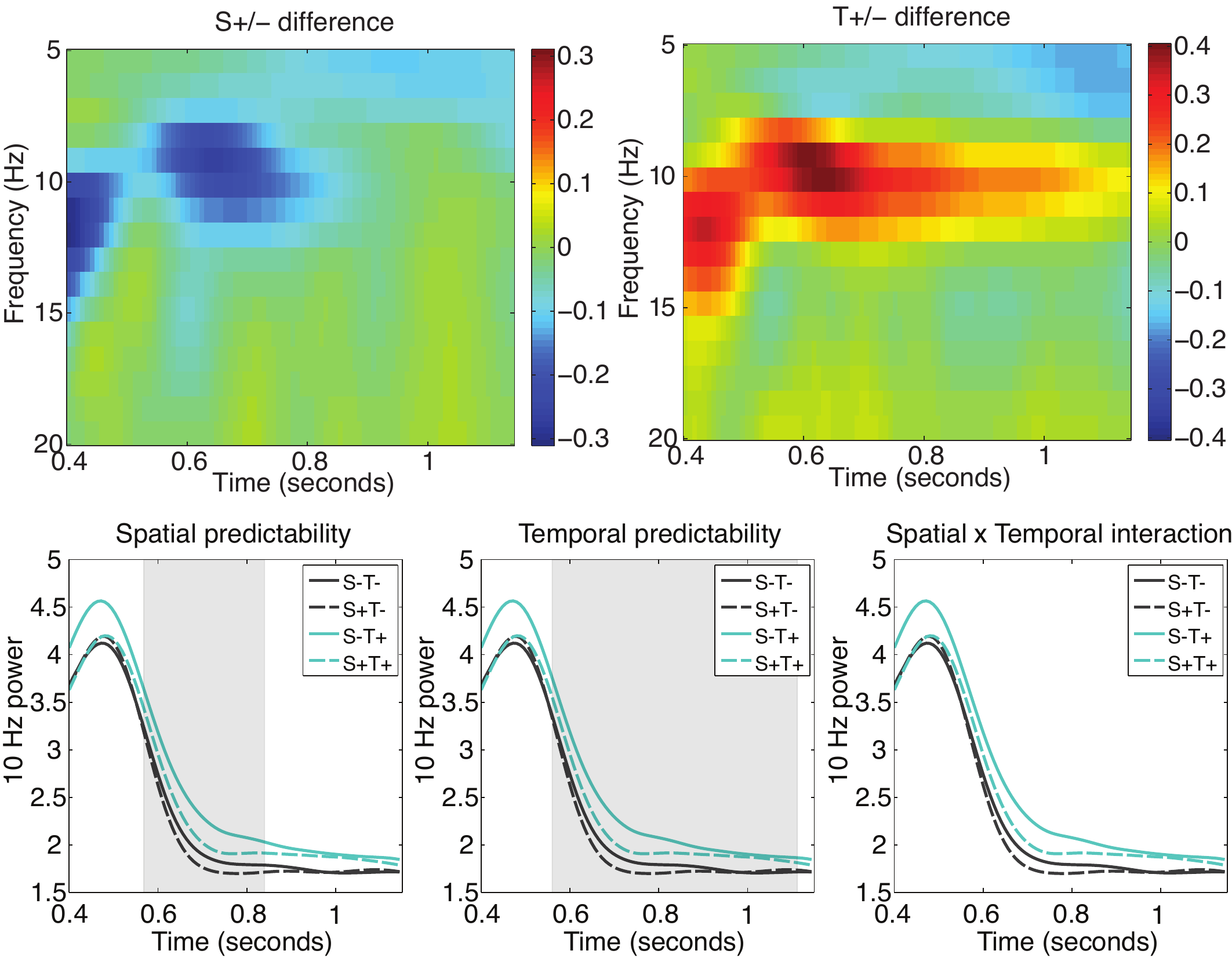}
\end{center}
\caption{Effect of entrainer predictability on alpha power}{Alpha-band power over the entraining sequence. \textbf{Top}: Main effects of spatial and temporal predictability on oscillatory power in the 5-20 Hz frequency range. \textbf{Bottom}: 10 Hz only effects of spatial predictability (left), temporal predictability (center), and the interaction between these terms with gray shaded ranges indicating significance while controlling for a maximum false discovery rate (FDR) of 5\%. S-/+ refers to spatially unpredictable and predictable, T-/+ to temporally unpredictable and predictable. Time axes indicate total trial time after the initial fixation cross with 0.4 seconds corresponding to the first entrainer.}
\label{fig:pleast_entrain_pow_10Hz}
\end{figure}

\begin{figure}[h!]
\begin{center}
\includegraphics[width=160mm]{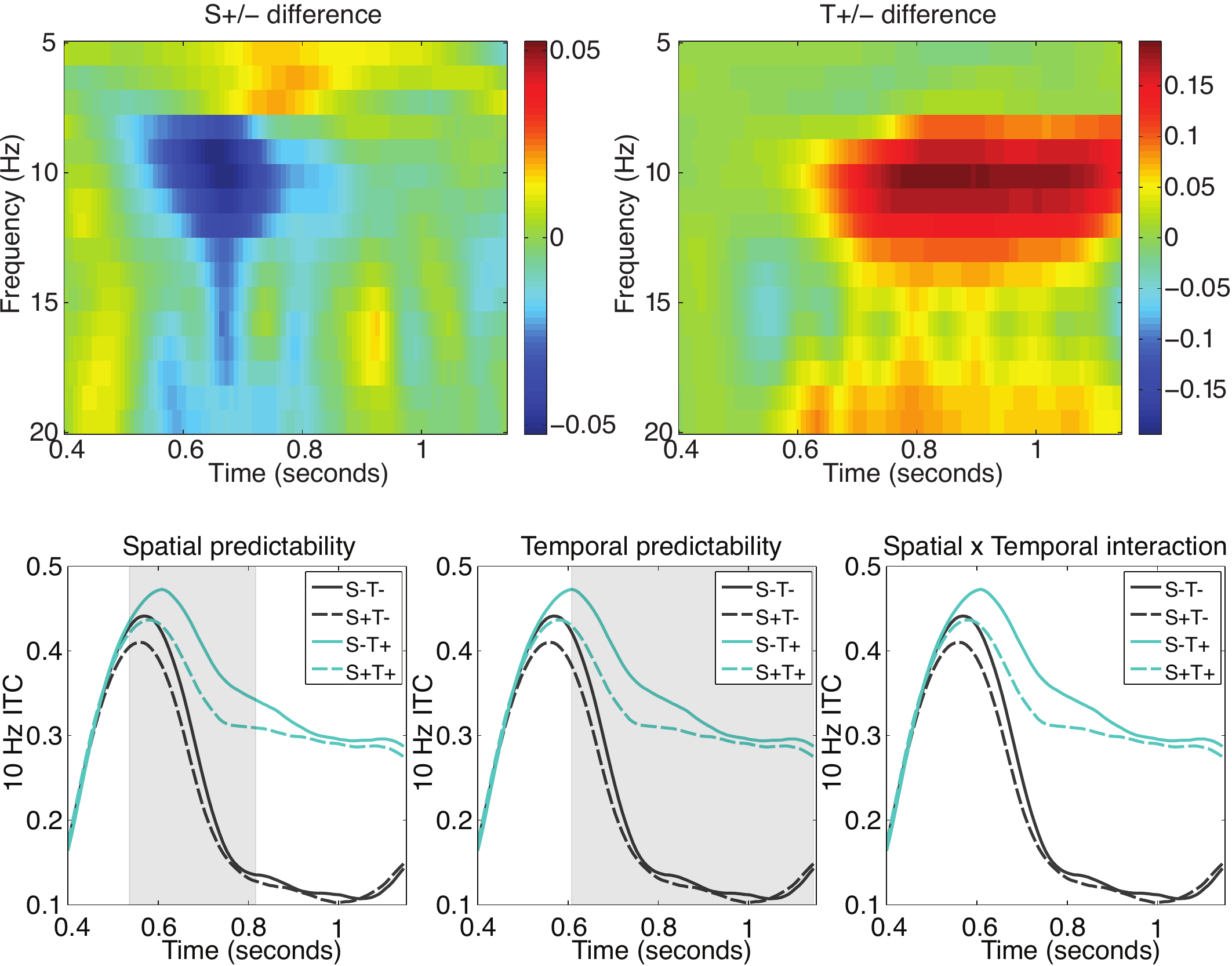}
\end{center}
\caption{Effect of entrainer predictability on alpha phase coherence}{Alpha-band inter-trial coherence (ITC) over the entraining sequence. Axes, legends, and shading for significant regions are the same as those described in Figure \ref{fig:pleast_entrain_pow_10Hz}.}
\label{fig:pleast_entrain_ITC_10Hz}
\end{figure}

Power and inter-trial coherence (ITC), a measure of phase angle consistency across trials \cite{LachauxRodriguezMartinerieEtAl99}, were computed over a 5-20 Hz frequency range to investigate the relationship between entrainer predictability and alpha oscillatory properties (Figures \ref{fig:pleast_entrain_pow_10Hz}-\ref{fig:pleast_entrain_ITC_10Hz}). Both spatial and temporal predictability had a significant effect on 10 Hz power and ITC beginning around 200 ms after the onset of the first entrainer (power: 160-168 ms after entrainer 1; ITC: 136-208 ms after entrainer 1). Together, these results indicate that 10 Hz entrainment affects both power and phase alignment and takes around 2-3 events to establish. Spatial predictability had a suppressive effect on 10 Hz power and phase alignment whereas temporal predictability had an enhancement effect. Any interactions between spatial and temporal predictability failed to reach significant levels altogether.

% 10 hz entrain pow

% 10 Hz entrain ITC

\begin{figure}[h!]
\begin{center}
\includegraphics[width=160mm]{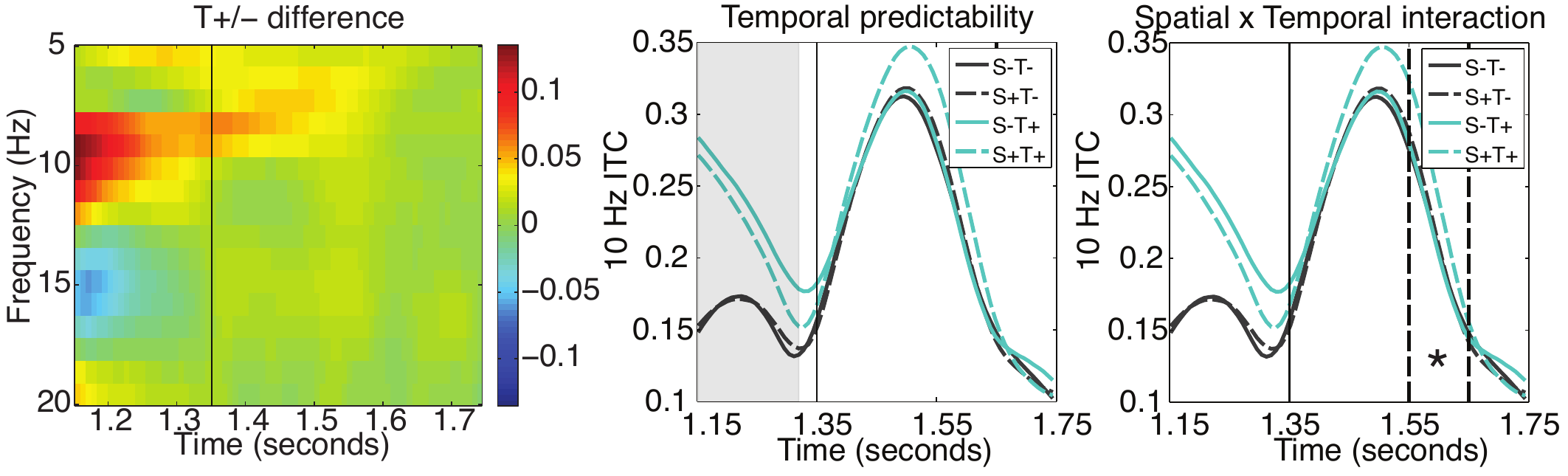}
\end{center}
\caption{Alpha phase coherence before and after probe}{Alpha-band inter-trial coherence (ITC) 200 ms preceding the probe and 400 ms following. Solid vertical line indicates probe onset. Asterisk indicates trending significance at the 5\% level when averaging 10 Hz ITC within the window defined by the dotted lines. Time axes indicate total trial time after the initial fixation cross.}
\label{fig:pleast_probe_ITC_10Hz}
\end{figure}

The effect of spatial predictability only persisted for three entrainers on average (duration ranges for 10 Hz power and ITC: 272-280 ms). Temporal predictability, in contrast, exhibited a sustained effect on 10 Hz power lasting nearly the entire entraining sequence. In the case of ITC, the effect of temporal predictability persisted throughout the blank period that separated the entraining sequence and the probe (Figure \ref{fig:pleast_probe_ITC_10Hz}). This result indicates that alpha phase remained more aligned for temporally predictable stimuli even without exogenous entrainment. 

Predictability effects on 10 Hz ITC failed to reach significance during the presentation of the probe or the following 400 ms period given the constraints of the statistical test (maximum FDR of 5\%, 16 ms consecutive significance). An enhancement in ITC could be observed for the combined predictability (S+T+) condition and trended toward significance when averaged over a 100 ms period beginning 155 ms after the onset of the probe (\textit{F}(1, 22) = 4.11, \textit{p} = 0.055). This effect was more pronounced and reached significance if only the right hemisphere channels were considered (\textit{F}(1, 22) = 7.45, \textit{p} = 0.012), but failed to reach significance in the left hemisphere (\textit{F}(1, 22) = 1.47, \textit{p} = 0.237), explaining the trending effect when both hemispheres are considered jointly. 

% 10 Hz probe ITC

\subsection{Predictability effects in delta-theta bands}
\begin{figure}[h!]
\begin{center}
\includegraphics[width=160mm]{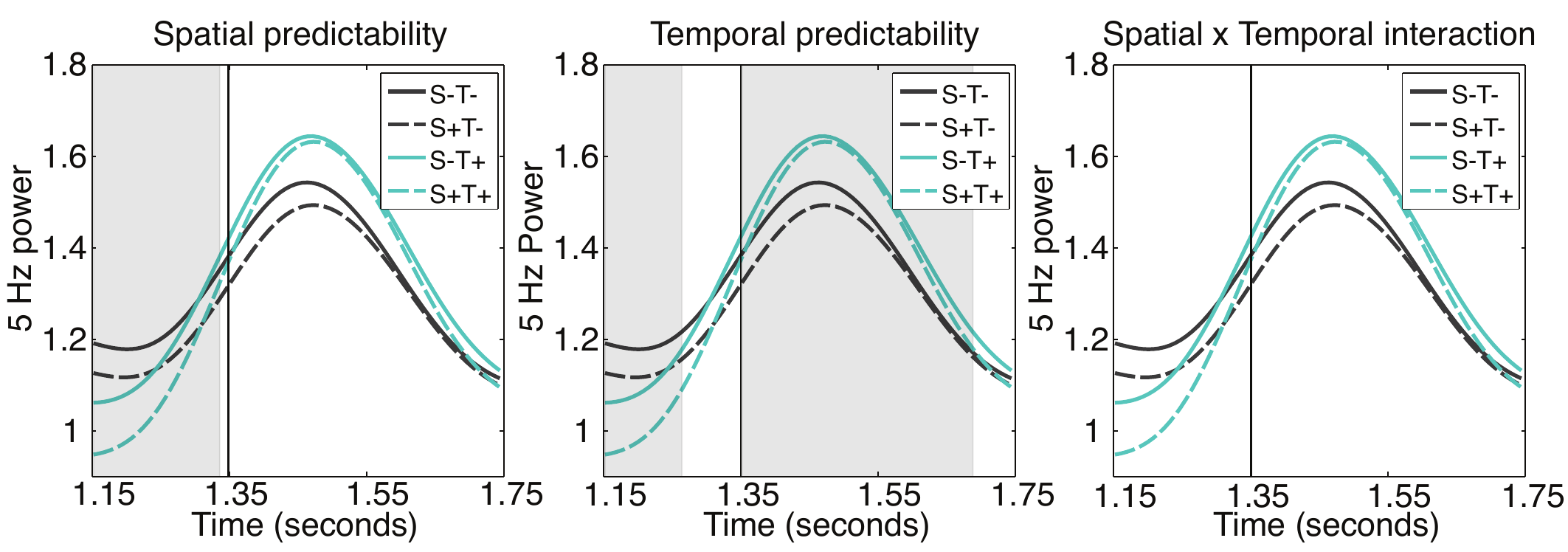}
\end{center}
\caption{Delta-theta power before and after probe}{Delta-theta power 200 ms preceding the probe and 400 ms following. Axes, legends, and shading for significant regions are the same as those described in Figure \ref{fig:pleast_probe_ITC_10Hz}.}
\label{fig:pleast_probe_pow_5Hz}
\end{figure}

\begin{figure}[h!]
\begin{center}
\includegraphics[width=160mm]{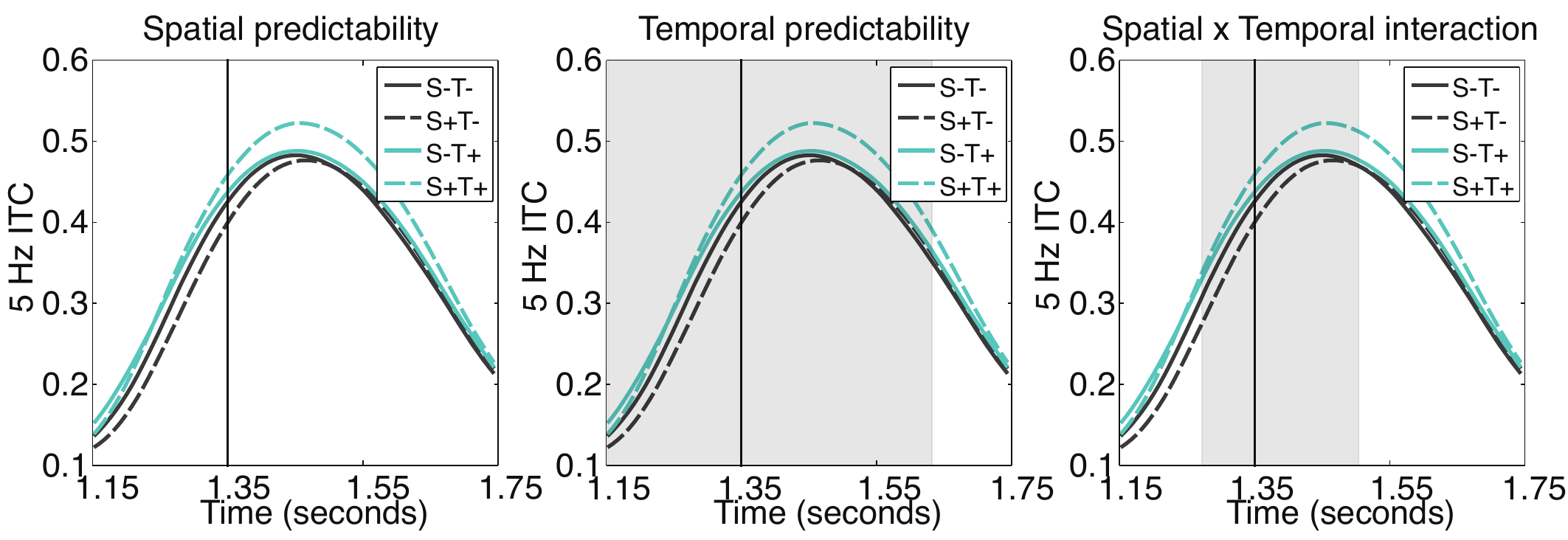}
\end{center}
\caption{Delta-theta phase coherence before and after probe}{Delta-theta inter-trial coherence (ITC) 200 ms preceding the probe and 400 ms following. Axes, legends, and shading for significant regions are the same as those described in Figure \ref{fig:pleast_probe_ITC_10Hz}.}
\label{fig:pleast_probe_ITC_5Hz}
\end{figure}

The effects of spatial and temporal predictability on oscillatory properties during the probe period (-200 ms pre-stimulus through 400 ms after) were also investigated in the delta-theta bands, centered around 5 Hz. This frequency was identified based on exploratory analysis and was also motivated by alternative models of sensory prediction \cite{ArnalGiraud12,GiraudPoeppel12}. Power and ITC at 5 Hz are plotted in Figures \ref{fig:pleast_probe_pow_5Hz}-\ref{fig:pleast_probe_ITC_5Hz}.

% 5 Hz probe pow

% 5 Hz probe ITC

Both spatial and temporal predictability had a significant effect on 5 Hz power during the 200 ms blank period preceding the probe. Temporal predictability had a  suppressive effect on 5 Hz power, in contrast to the positive modulation found for 10 Hz power. This effect reversed following the presentation of the probe and persisted for over 300 ms. The interaction between spatial and temporal predictability failed to reach significance for 5 Hz power at any time points.

Temporal predictability also had a significant effect on 5 Hz ITC beginning during the 200 ms blank period preceding the probe and lasting nearly 300 ms after the presentation of the probe. Whereas 10 Hz ITC decreased during the blank period (yet remained significantly higher for temporally predictable stimuli), 5 Hz ITC increased, and continued to increase until approximately 100 ms after the onset of the probe. 5 Hz ITC was highest for the combined spatial and temporal predictability condition (S+T+), indexed by a significant interaction beginning before the probe onset (-78 ms pre-stimulus) and persisting 154 ms after.

% \subsection{Predictability effects in gamma oscillations}
% 35 Hz int ???

%\subsection{Alpha phase at probe onset}

\section{Discussion}

\subsection{Summary of results}
The work described in this chapter investigated how the brain integrates information from 100 ms samples and uses it to drive predictions about what will happen. The experimental paradigm used to address this question involved entraining alpha oscillatory activity to determine the effects of spatial and temporal predictability on a novel object recognition task. Behaviorally, spatial and temporal predictability increased probe discriminability on a same-different judgement as well as speeded response times. %Inverse efficiency, which combines accuracy and reaction times and can be thought of as the amount of energy consumed by the system to produce a behavioral outcome \cite{TownshendAshby78,TownshendAshby83}, was lower for temporally predictable stimuli on average, but exhibited an increase for combined spatial and temporal predictability.

Event-related analysis of EEG data indicated that temporal predictability caused a strong periodicity (in this case, 10 Hz) phase aligned approximately to the onset of each temporally predictable entraining stimulus. This alignment was approximately 180 degrees out of phase for temporally unpredictable stimuli and these differences in waveform alignment caused amplitude differences that preceded both entraining stimuli and the probe. The effects of spatial predictability generally manifested during the presentation of stimuli. There was an early divergence of probe-evoked activity caused by spatially predictable ordering of entrainers that persisted for over 100 ms through the P1 response with several transient effects after. There was evidence that spatial predictability was only effective when stimuli were also temporally predictable. This was evident for entrainers, and over right hemisphere channels approximately 200 ms after the probe onset. This latter effect failed to reach significance for left hemisphere channels and only trended toward significant levels when both hemispheres were considered jointly. It is likely that this is simply an issue of insufficient power, but other explanations such as hemispheric specialization \cite[e.g.,]{Marsolek99,Dien09b} cannot be completely ruled out.

Both spatial and temporal predictability had effects on the power and phase coherence (indexed by inter-trial coherence, ITC) of neural oscillations. Spatially predictable entrainment caused a suppression of 10 Hz power with a lower degree of phase alignment than spatially random stimuli. Temporally predictable entrainment had the opposite effect, with increased 10 Hz power and phase alignment. Phase alignment due to temporal predictability remained elevated compared to temporally unpredictable stimuli during a 200 ms blank period between the entraining sequence and probe, indicating that the effects of temporal predictability could persist without exogenous entrainment. There was evidence of a selective enhancement in phase alignment for the combined spatial and temporal predictability case approximately 200 ms after the probe onset. As was the case in event-related analyses, this effect was significant in the right hemisphere, but not the left, and trended toward significance when both hemispheres were considered jointly. Power and phase coherence effects were also examined in the delta-theta bands (5 Hz) during the probe judgement on the basis of alternative theories of sensory prediction \cite{ArnalGiraud12,GiraudPoeppel12}. Results were similar to those found for 10 Hz, but more robust with significant effects for both power and phase alignment. 

\subsection{Separate time courses for spatial and temporal prediction}
Spatial and temporal predictability were characterized by distinct and generally non-overlapping time courses. Temporal predictability manifested solely before (or at) the onset of each stimulus and appeared to be driven by an approximate antiphase relationship between temporally predictable and unpredictable stimuli. Spatial predictability generally manifested during the presentation of stimuli, although in the case of the probe, also elicited transient differences after its disappearance. Predictability effects were similar for entraining stimuli and the probe except for the post-stimulus effects in the case of the probe. The similarity between entrainers and the probe suggests that the brain might treat the probe as a continuation of the entraining sequence and process it in the same manner. 

From these results, it can be concluded that temporal prediction is an anticipatory process, occurring during the absence of exogenous stimulation (in between entrainers or before the probe). The effect of temporal predictability extends 16 ms after the onset of each entrainer, but latencies for the first wave of responses in primary visual cortex (V1) are approximately 40-60 ms \cite{NowakBullier97,FoxeSimpson02} so there is no exogenous stimulation \textit{per se} during the duration of the effect despite the stimulus being onscreen. Spatial prediction begins shortly before exogenous stimulation, but in the case of the probe, persists through the initial V1 responses. Spatial prediction, thus, might better be characterized as a post-stimulus process opposed to truly anticipatory process. The computation might involve a comparison between what is expected and what is actually coded by incoming spikes, consistent with the LeabraTI model (Chapter \ref{chap:leabrati}) as well as predictive coding models \cite[e.g.,]{RaoBallard99,Friston05,DenOudenKokDeLange12}.

\subsection{Oscillatory mechanisms of spatial and temporal prediction}
Spatial and temporal predictability had effects on both power and phase coherence of neural oscillations. In both cases, predictability took 2-3 entraining stimuli to establish, consistent with previous investigations (\nopcite{MathewsonFabianiGrattonEtAl10}; \abbrevnopcite{MathewsonPrudhommeFabianiEtAl12}). Spatial predictability was then characterized by a suppression of 10 Hz power and phase angle variability causing a lower degree of alignment than spatially random stimuli. This effect is opposite than that of temporal predictability, which was characterized by increased 10 Hz power and phase angle alignment. Previous investigations have generally not simultaneously manipulated temporal rhythmicity and spatial coherence \cite[although see]{DohertyRaoMesulamEtAl05,RohenkohlGouldPessoaEtAl14} and thus, the relative suppression and decreased phase coherence during spatially predictable entrainment were unexpected results.

Successful oscillatory entrainment is thought to be a result of repeated phase resetting of endogenous oscillations causing phase to move into alignment with the frequency of exogenous stimulation \cite{SchroederLakatosKajikawaEtAl08,CalderoneLakatosButlerEtAl14}. Oscillatory phase resetting has been shown to be caused by salient, unexpected events (\abbrevnopcite{FiebelkornFoxeButlerEtAl11}; \nopcite{LandauFries12,RomeiGrossThut12}) and thus these unexpected results might be accounted for by considering successive views in terms of the amount of ``surprise'' \cite{IttiBaldi09,MeyerOlson11} they evoke. Subsequent views of the entraining sequence have significant feature overlap, characterized by the same populations of neurons spiking from one view to the next. Repeated spiking, especially as a function of expectation, has been shown to evoke rate suppression mechanisms \cite{SummerfieldTrittschuhMontiEtAl08}. When entraining views are presented out of order, feature overlap is minimized and each view is ``surprising'', causing an initial fast spiking burst response, which could led to a higher degree of phase resetting.

The entrainment effects of spatial predictability were transient and dissipated before the end of the entraining sequence whereas the effects of temporal predictability persisted much longer. This might account for the null effects of spatial predictability on most behavioral measures. Temporal predictability effects persisted through the 200 ms blank period between the entraining sequence and probe and had robust effects on all behavioral measures. Thus, the present experiment could be modified to use a shorter entrainment sequence which would likely elicit successful spatial predictability for probe judgements.

The enhancement of 10 Hz phase angle alignment after the probe was presented did not reach significant levels assuming the FDR-corrected significance test at each time bin but did for 5 Hz phase alignment. Furthermore, temporal and spatial predictability main effects around the probe onset were more pronounced for 5 Hz oscillatory properties. One potential explanation for these effects is that the 200 ms blank period between the entraining sequence and the probe corresponded to two cycles at 10 Hz, but only one cycle at 5 Hz. Phase angles at 10 Hz exhibited significant dealignment over this period, and actually increased in alignment at 5 Hz. Thus, it is possible that this increase in phase alignment lead to a more pronounced selective enhancement for combined spatial predictability at 5 Hz, consistent with recent data \cite{CravoRohenkohlWyartEtAl13,ArnalDoellingPoeppel14}. A more robust effect might be found for 10 Hz if the blank period between the entraining sequence and probe was only 100 ms in duration.

Another limitation of the present experimental paradigm is that it is unclear whether exogenous entrainment simply created new oscillations akin to steady state visually evoked potentials (SSVEP), or actually entrained existing endogenous oscillations. Thus, it might not be particularly surprising that predictable 10 Hz stimulation caused increases in 10 Hz power and phase alignment. However, entrained alpha-band periodicity has been shown to correlate with individual resting alpha oscillation frequency \abbrevcite{DeGraafGrossPatersonEtAl13} so it is likely that the paradigm recruited existing oscillations and caused them to align to the exogenous entrainment frequency opposed to creating new oscillations. The fact that phase alignment continued through the 200 ms blank period without exogenous entrainment also supports this claim.

\subsection{Synergistic effects of combined spatiotemporal prediction}
Combined spatial and temporal predictability further potentiated differences in spatial predictability for probe-evoked activity as well as 10 Hz phase angle alignment approximately 200 ms after probe offset. A similar synergistic effect has been described in the literature on attentional allocation with temporal anticipation causing further potentiation of spatial attention's effect on the P1 response (peak latency of 110-130 ms) \cite{DohertyRaoMesulamEtAl05}. The synergistic effect observed in the current experiment occurred later, closest to the N2 response, and further resulted in increased 10 Hz phase alignment with the same latency. 

The effects of spatial attention have been shown to manifest early, typically affecting the P1 response \cite{LuckHeinzeMangunEtAl90,HillyardAnllo-Vento98,HillyardVogelLuck98}, whereas the posterior N2 component has been associated with more complex processing such as appearance of anticipated or relevant stimulus features \cite{FolsteinVanPetten08}. Thus, the early synergistic effect of combined spatiotemporal anticipation reported in \incite{DohertyRaoMesulamEtAl05} likely reflected a recent shift of spatial attention, which has been shown to modulate EEG amplitude and behavioral measures \cite{BuschVanRullen10}. In contrast, the relatively late synergistic effect of combined spatiotemporal predictability observed in the current experiment likely indexes true stimulus-predictive processing. Consistent with this idea, a recent study that used MEG to investigate changes in facial expression identified a component that was sensitive to predictability of the expression change, peaking between approximately 165-200 ms after the change \abbrevcite{FurlRijsbergenKiebelEtAl10}.

\subsection{Alpha oscillations index stimulus-predictive processing}
Alpha oscillations have previously been implicated in the allocation of hemifield-based spatial attention and anticipation of the temporal onset of relatively simple stimuli (\nopcite{GouldRushworthNobre11}; \abbrevnopcite{BelyusarSnyderFreyEtAl13}; \nopcite{RohenkohlNobre11,MathewsonGrattonFabianiEtAl09,BuschDuboisVanRullen09}). It was unclear whether these results engaged actual predictive processing about \text{what} would happen or comparatively simple anticipatory attention mechanisms about \textit{where} a stimulus might appear. 

The current experiment consisted of a relatively complex perceptual task that that required integration across features extracted over the course of a sequence of stimulus presentations to perform a subsequent probe judgment. Both spatial and temporal predictability benefited probe discriminability and thus, it can reasonably be concluded that the experiment engaged actual predictive processing. Alpha oscillations indexed both the spatial and temporal predictability of the entraining stimuli as well as tracked the onset of the probe through phase alignment during the 200 ms blank period separating the entraining sequence and probe.

Cumulatively, these results suggest that alpha oscillations serve at least two roles in prediction that can roughly be characterized as ``spatial'' and ``temporal'' in nature. The first is a prediction about the spatio-featural content of upcoming sensory events. This prediction could simply be about where a stimulus might show up, irrespective of what features it contains. Importantly however, the prediction can also carry content about the probability of specific features showing up at a specific location in spatial map \cite{KokRahnevJeheeEtAl12,WyartNobreSummerfield12,HorschigJensenVanSchouwenburgEtAl13}. The second function of alpha oscillations is a pacemaker function that allows predictions to be made at regular intervals. This pacemaker function can phase align to the onset of stimuli \cite{SchroederLakatosKajikawaEtAl08,CalderoneLakatosButlerEtAl14} so that predictions can reliably be made in advance of actual sensory events. Overall, these suggested roles of alpha oscillations are compatible with the LeabraTI model (Chapter \ref{chap:leabrati}) and other models of sensory predictions that highlight the role of oscillations in making about making predictions about incoming sensory information \cite{ArnalGiraud12,GiraudPoeppel12}.

%% file: chap_bpleast.tex
\chapter{Effects of spatial and temporal prediction during prolonged learning of novel objects}
\label{chap:bpleast}

\sloppy

\section{Introduction}
The core challenge of object recognition is concerned with solving the invariance problem \cite{DicarloZoccolanRust12}. Essentially, object identity must remain invariant across large changes in a an object's visual position, scale, rotation, and viewpoint to generate successful behavior. Understanding how exactly the brain solves this problem has been a major focus of the object recognition literature with the bulk of data and models suggesting that it is solved gradually by a hierarchy of neural processing mechanisms from V1 through inferior temporal (IT) cortex that extract increasingly complex features at each stage with increasing tolerance to transformations \cite{Fukushima80,RiesenhuberPoggio99,WallisRolls97,MasquelierThorpe07,OReillyWyatteHerdEtAl13}.

One question that remains to be fully answered is how the synaptic projections that support invariance are learned in the first place. One intriguing hypothesis is that a temporal association rule might form associations between multiple samples of a single object as it undergoes transformations \cite{StringerPerryRollsEtAl06,WallisBaddeley97,IsikLeiboPoggio12}. It has been demonstrated that some neurons can form temporal associations between arbitrary pairs of stimuli \cite{SakaiMiyashita91}, including a population in monkey IT cortex \cite{MeyerOlson11}. Experiments by DiCarlo and colleagues have indicated that these temporal associations can build new invariance for specific object transformations including changes in position and size \cite{CoxMeierOerteltEtAl05,LiDiCarlo08,LiDiCarlo10}. This new invariance can be learned without supervised reward, suggesting that it could be a natural consequence of generic neural processing mechanisms given the spatiotemporal statistics of the physical world \cite{LiDiCarlo12}.

Evidence of invariance due to temporal association has yet to be demonstrated in IT neurons for three-dimensional changes in viewpoint \cite[although see][for relevant human behavioral work]{WallisBulthoff01,WallisBackusLangerEtAl09}. IT neurons typically have a tuning curve of approximately 90 degrees for three-dimensional objects \cite{LogothetisPaulsBulthoffEtAl94,LogothetisPaulsPoggio95}, but these objects can be recognized irrespective of viewpoint after prolonged exposure \cite{WallisBulthoff99,EdelmanBulthoff92,TarrGauthier98}. Intuitively, predictable motion from one moment to the next could be considered important for encoding three-dimensional structure \cite{LawsonHumphreysWatson94,Stone98,VuongTarr04,BalasSinha09b,BalasSinha09c,ChuangVuongBulthoff12}, and thus a temporal association rule could plausibly be used to group together multiple viewpoints from naturalistic spatial structure of objects.

The work described in this chapter investigated the role of predictable spatiotemporal information during a novel object learning task. In the context of the LeabraTI model (Chapter \ref{chap:leabrati}) as well as several other theories of sensory prediction \cite{ArnalGiraud12,GiraudPoeppel12}, spatial structure might be learned from predictions about incoming sensory information made at regular temporal intervals. To test this hypothesis, both the spatial and temporal predictability of changes in objects' viewpoint were manipulated during a training period followed by a series of same-different judgements over static test views.

Somewhat surprisingly, the results of the experiment indicated that accuracy was lowest when stimuli were learned in a combined spatially and temporally predictable context and highest when learned in a completely unpredictable context. Reaction times were also slower when objects were learned with spatial predictability.

\section{Methods}

\subsection{Participants}
A total of 62 students from the University of Colorado Boulder participated in the experiment (ages 18-22, mean=19.11 years; 30 male, 32 female). All participants reported normal or corrected-to-normal vision and received course credit as	compensation for their participation. Informed consent was obtained from each participant prior to the experiment in accordance with Institutional Review Board policy at the University of Colorado.

\subsection{Stimuli}
\begin{figure}[h!]
\textbf{A} \\
\includegraphics[width=160mm]{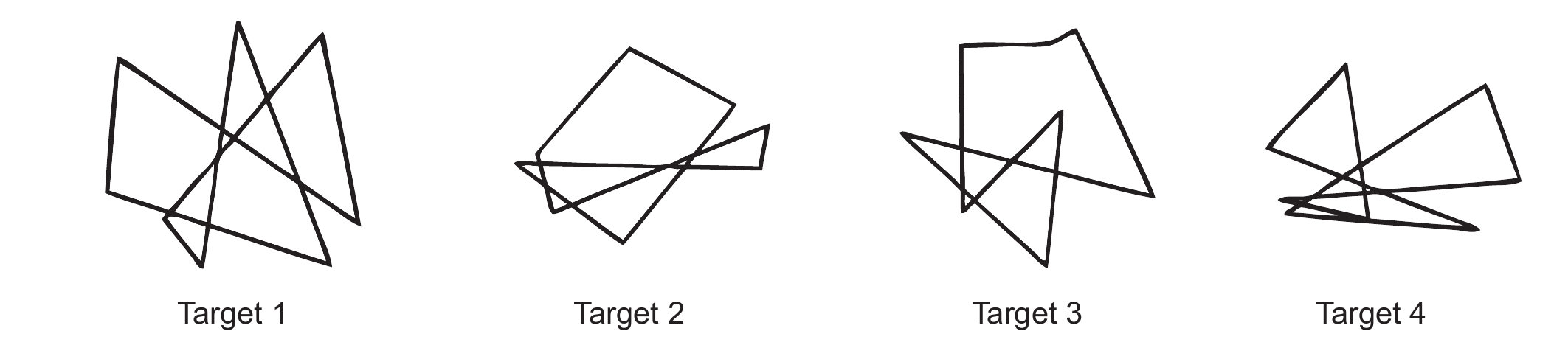} \\
\textbf{B} \\
\includegraphics[width=160mm]{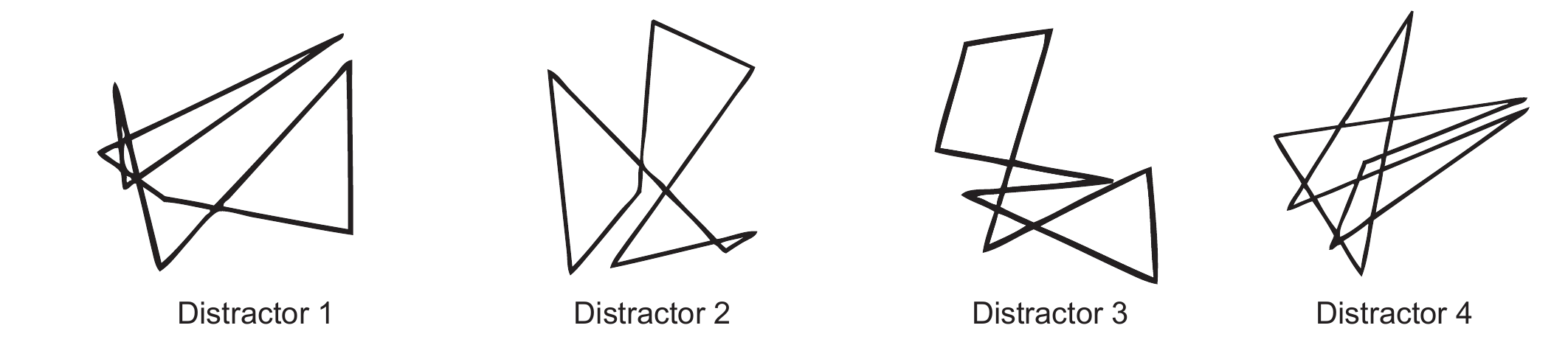} \\
\caption{Novel ``paper clip'' objects}{Four target (\textbf{A}) and four distractor object pairs (\textbf{B}) used in the experiment. See Chapter \ref{chap:pleast} Methods for additional information.}
\label{fig:bpleast_objs}
\end{figure}

Novel ``paper clip'' objects similar to those used in previous investigations of three-dimensional object recognition \cite{BulthoffEdelman92,EdelmanBulthoff92,LogothetisPaulsBulthoffEtAl94,LogothetisPaulsPoggio95,SinhaPoggio96} were used as stimuli (see Chapter \ref{chap:pleast} Methods). A total of eight objects were used -- four as targets and four as distractors. The four target objects were also used in the Chapter \ref{chap:pleast} experiment. Target and distractor objects were paired together for the purposes of the experiment. All objects are shown in Figure \ref{fig:bpleast_objs}.

% paperclip targets fig

\subsection{Procedure}
\begin{figure}[h!]
\includegraphics[width=160mm]{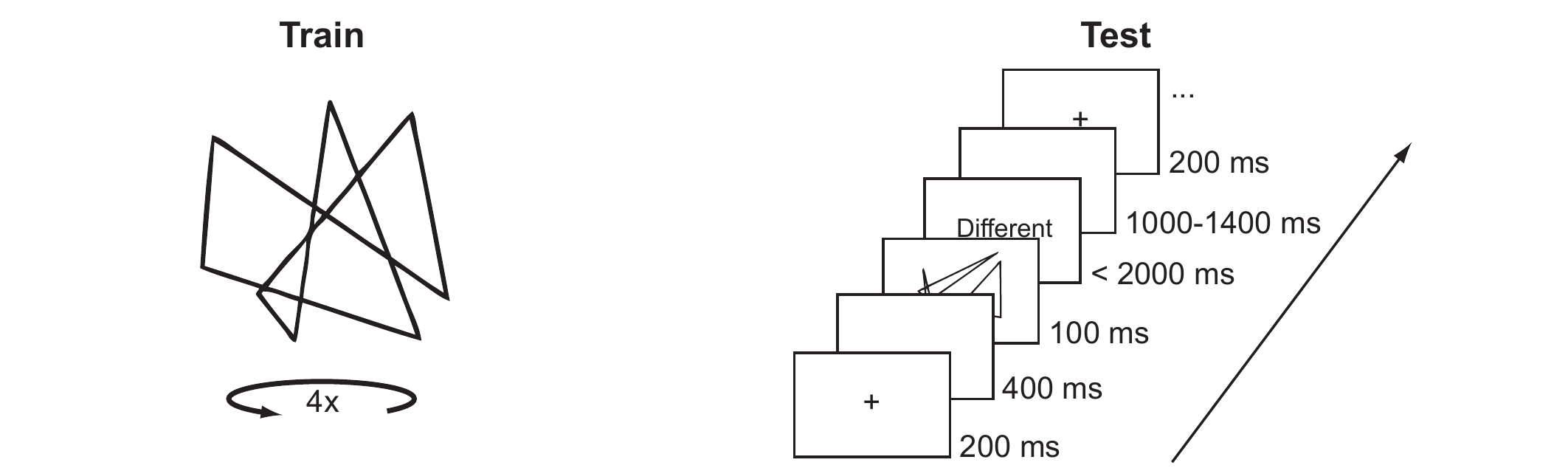} \\
\caption{Experimental procedure}{Experimental trials were composed of a training period followed by a testing period. The training period depicted a target object rotating a total of four times around its vertical axis. Rotation was either spatially and temporally predictable, spatially predictable or temporally predictable only, or completely unpredictable. The test period contained 30 trials that depicted either the training object or its paired distractor at 15 viewing angles each.}
\label{fig:bpleast_task}
\end{figure}

The experiment was divided into 16 blocks, each containing a training period followed by a series of test trials (Figure \ref{fig:bpleast_task}). During the training period of a given block, participants observed one of the target objects rotate about its y-axis. The object either rotated coherently (i.e., spatially predictable, S+ conditions) or in a random manner (S- conditions). Coherent rotation was composed of adjacent views spaced 12 degrees apart. The object made four complete rotations during the study period. All views of the object were still presented four times each in the random case. The presentation rate during the study period was either 10 Hz with a 50 ms on time and 50 ms off time (i.e., temporally predictable, T+ conditions) or variable with a 50 ms on time and off times ranging from 17-400 ms (T- conditions). 

The S+/- and T+/- conditions were crossed and each of the target-distractor object pairs was assigned to one of the four conditions. These assignments were approximately counterbalanced across participants (Assignment 1: \textit{N}=15; Assignment 2: \textit{N}=17; Assignment 3: \textit{N}=15; Assignment 4: \textit{N}=15). Each block condition with its target-distractor pairing was repeated for four blocks during the experiment. Block order was randomized for each participant.

During each block, participants were instructed to study the target object during the training period and then complete a series of 30 test trials. On each test trial, either the target object or its paired distractor was presented. Participants were instructed to respond ``same'' if they believed the object depicted the trained target object or ``different'' if they believed it depicted the distractor object. Half of the test trials contained 15 views of the target object spaced 24 degrees apart, and the other half contained 15 views of the distractor, also spaced 24 degrees apart. Test trials were shown in a random order and feedback was withheld to prevent participants from changing their response criteria over the course of a block. 

The experiment was displayed on an LCD monitor at native resolution operating at 60 Hz using the Psychophysics Toolbox Version 3 \cite{Brainard97,Pelli97}. All stimuli were presented at central fixation on an isoluminant 50\% gray background and subtended approximately 5 degrees of visual angle. Test trials began with a fixation cross (200 ms) followed by a blank (400 ms) followed by the probe stimulus (100 ms). Participants were required to respond within 2000 ms or the trial was aborted. Subsequent test trials were separated by a variable intertrial interval of 1000-1400 ms.

The experiment began with a practice block to ensure that participants understood the task that was later discarded from analysis. The training period during the practice block was always spatially and temporally predictable and used a reserved target object and distractor that were not further used in any of the experimental blocks. During the practice test trials, participants received feedback after responding according to whether they were correct or incorrect. After completing the practice block, participants were informed that future training periods could be presented in spatially and/or temporally unpredictable manners.

% paperclip targets fig

\section{Results}
\begin{figure}[h!]
\begin{center}
\begin{tabular}{ll}
\includegraphics[width=80mm]{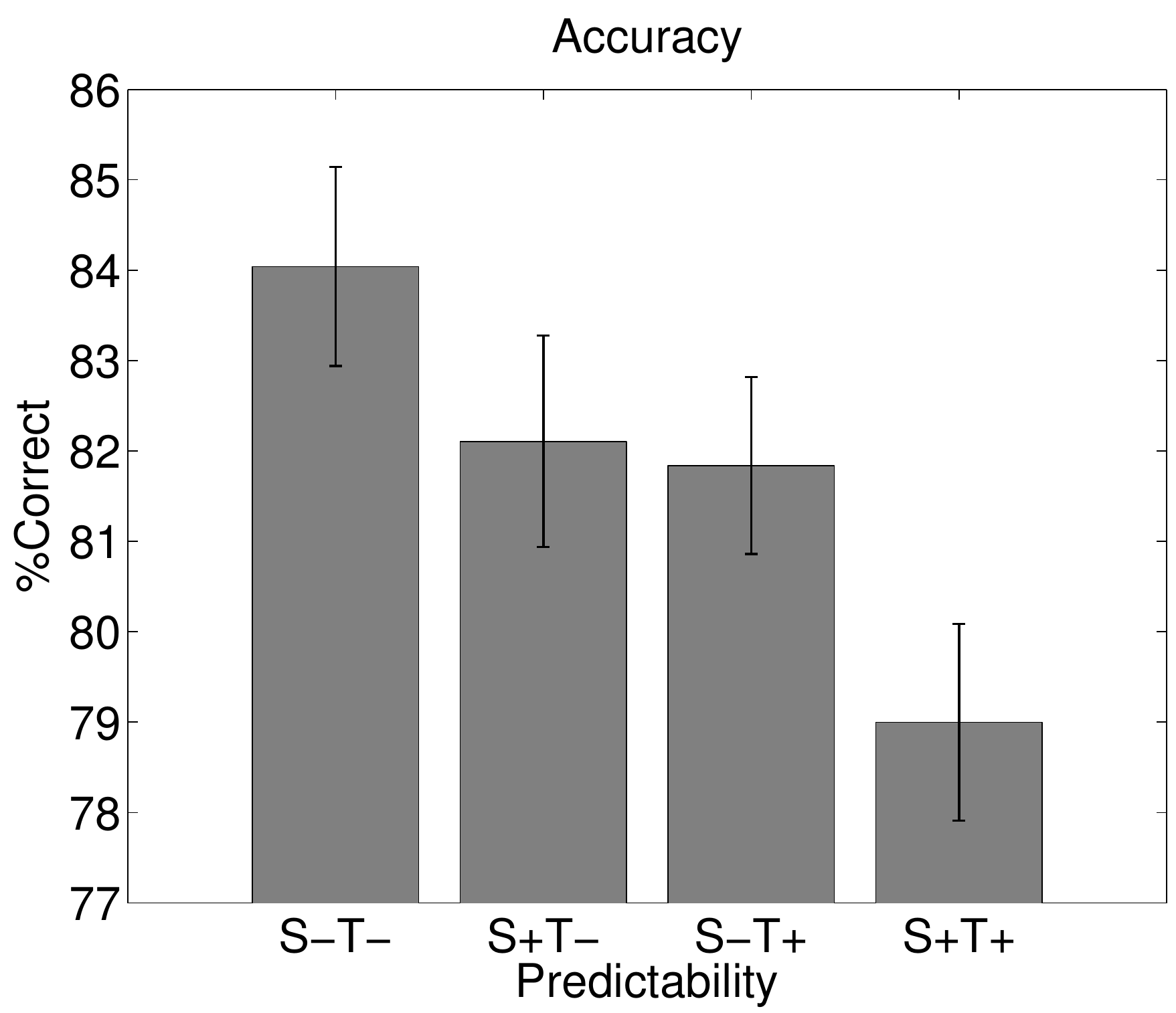} & 
\includegraphics[width=80mm]{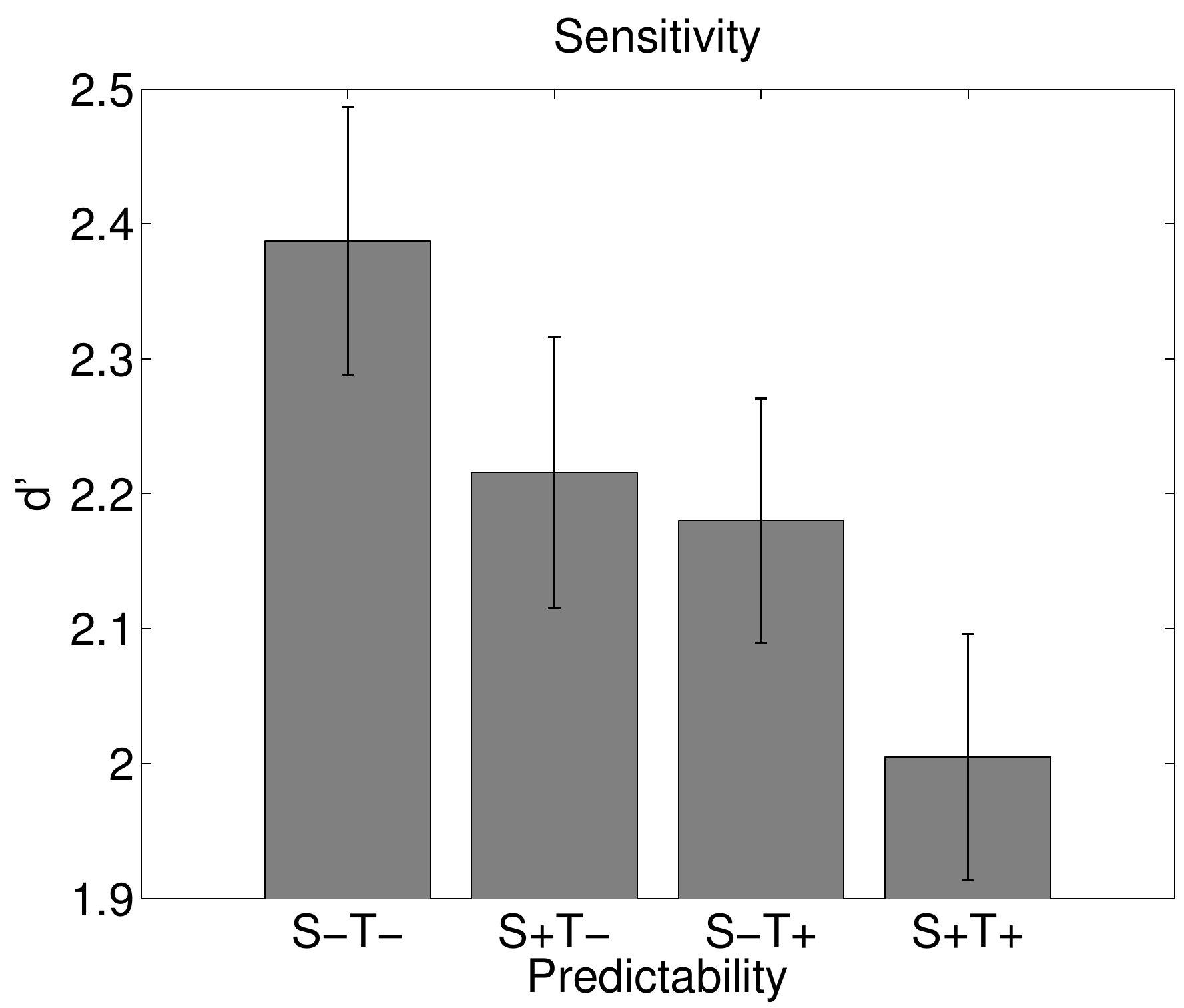} \\
\includegraphics[width=80mm]{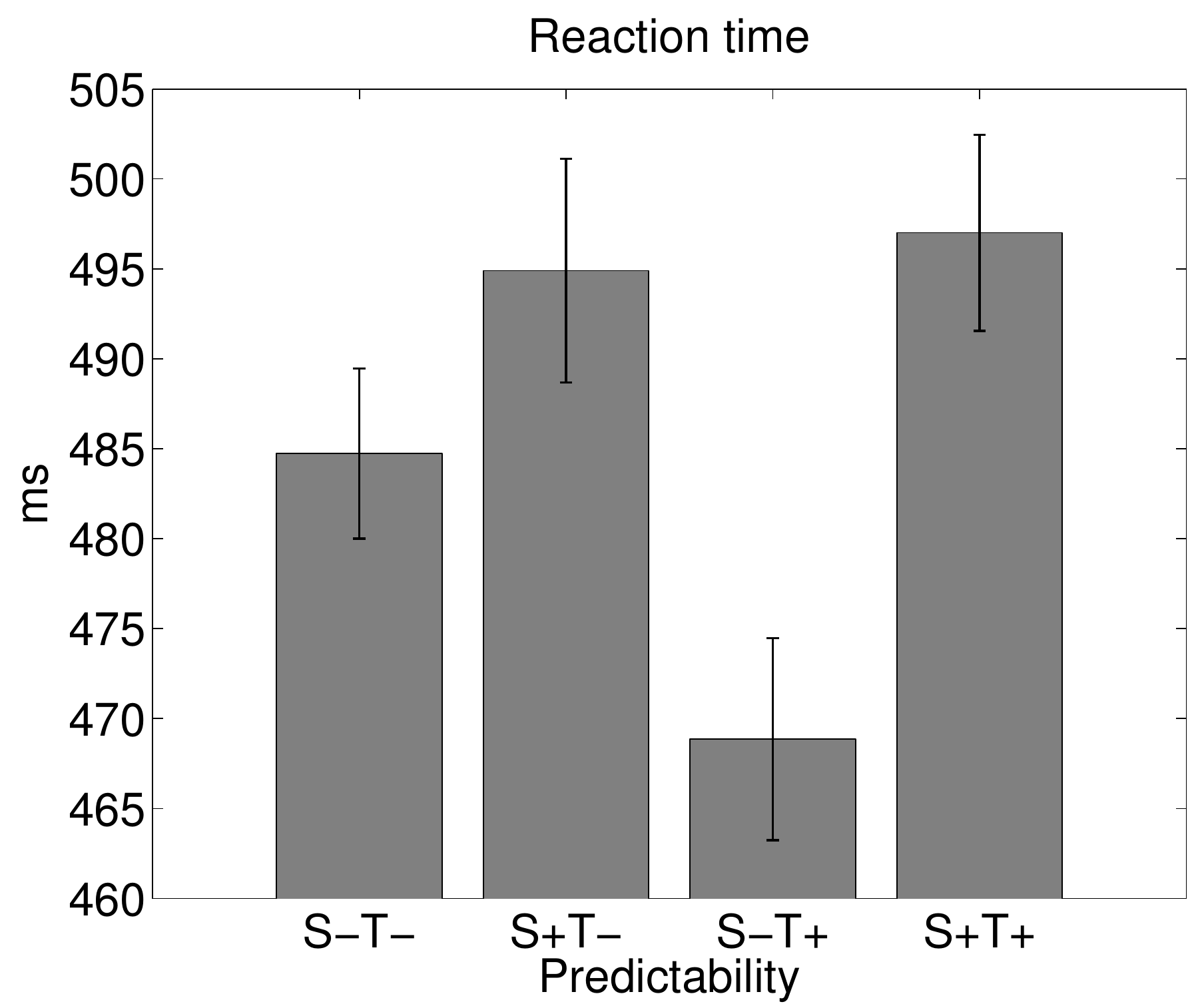} &
\includegraphics[width=80mm]{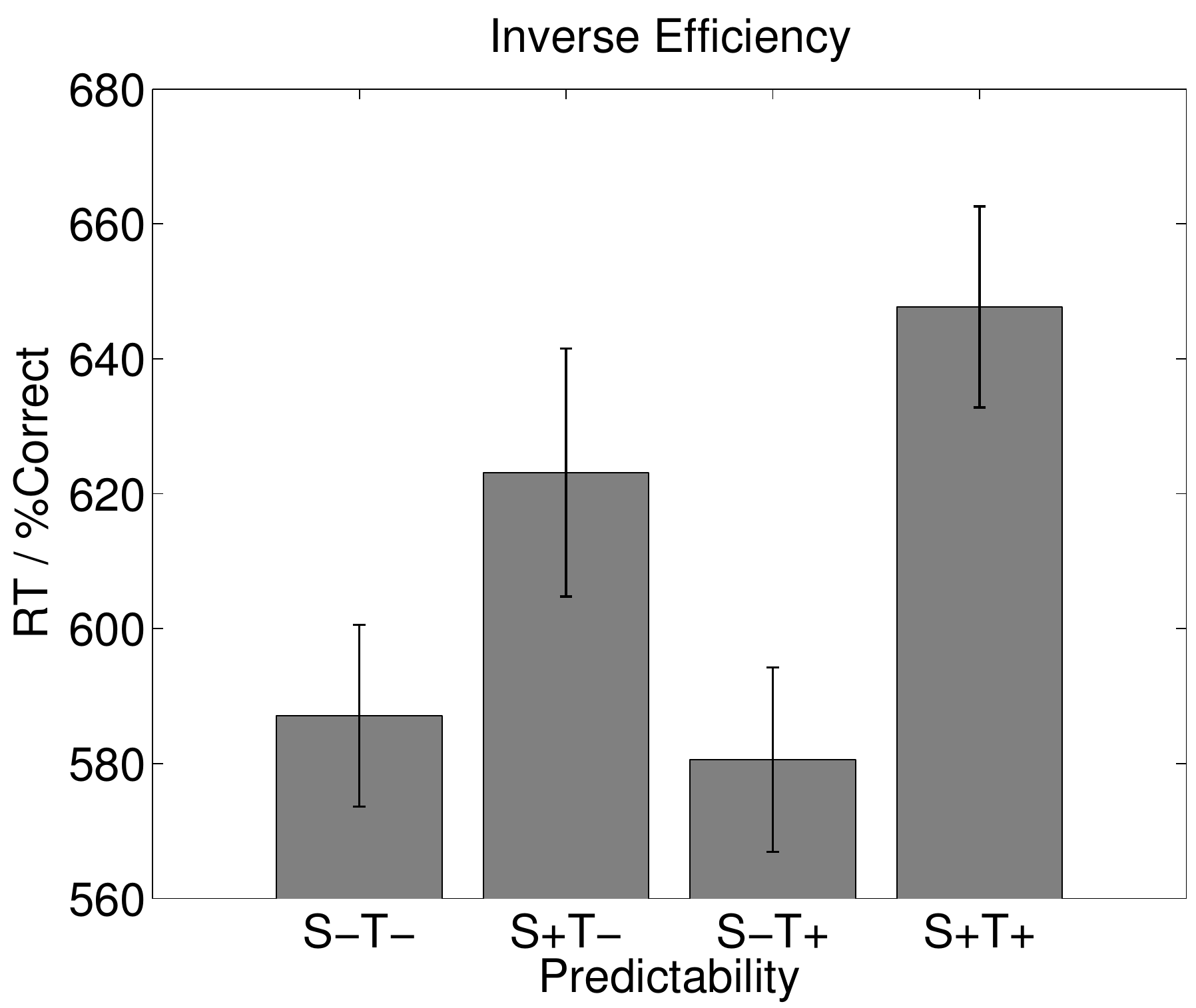} \\
\end{tabular}
\end{center}
\caption{Behavioral measures of spatial and temporal predictability}{Accuracy, \textit{d'} (sensitivity), reaction time, and inverse efficiency (reaction time divided by percent correct) as a function of predictability during the training period. S-/+ refers to spatially unpredictable and predictable, T-/+ to temporally unpredictable and predictable. Error bars depict within-subjects error using the method described in \protect\incite{Cousineau05} adapted for standard error.}
\label{fig:bpleast_behave}
\end{figure}

Three subjects were excluded from behavioral analysis for accuracy 2.7$\sigma$ (or further) below mean accuracy across subjects. All three excluded subjects were assigned condition-object 3 resulting in the final counterbalancing -- Assignment 1: \textit{N}=15; Assignment 2: \textit{N}=14; Assignment 3: \textit{N}=15; Assignment 4: \textit{N}=15. The resulting 59 subjects were submitted to a 2x2 ANOVA with spatial and temporal predictability as within-subjects factors and counterbalancing assignment as a between-subjects factor. Accuracy and reaction times were collected during the experiment and were used to compute \textit{d'}, a measure of sensitivity that takes into account response bias, and inverse efficiency, a measure that combines accuracy and reaction times \cite{TownshendAshby78,TownshendAshby83}. These behavioral measures are plotted in Figure \ref{fig:bpleast_behave}.

% oh, behave

% Acc
% Assignment: F(1, 57) = 0.156, p = 0.694
% Spatial: F(1, 57) = 4.496, p = 0.0383 * 
% Temporal: F(1, 57) = 4.149, p = 0.0463 * 
% Int: F(1, 57) = 0.197, p = 0.659

Overall, subjects were less accurate when the training period was spatially predictable (\textit{F}(1, 57) = 4.50, \textit{p} = 0.038) or temporally predictable (\textit{F}(1, 57) = 4.20, \textit{p} = 0.046). The interaction between spatial and temporal predicability failed to reach significance (\textit{F}(1, 57) = 0.20, \textit{p} = 0.659). Subjects were least accurate for the combined spatial and temporal predictability training context (denoted S+T+ in Figure \ref{fig:bpleast_behave}). This training context significantly differed from the completely unpredictable training context (S-T-) (\textit{t}(58) = -2.8587, \textit{p} = 0.001), and trended toward significance for conditions with only spatially or only temporally predictable training (S+T+ versus S+T-, \textit{t}(58) = -1.60, \textit{p} = 0.116; S-T- versus S+T+ versus S-T+, \textit{t}(58) = -1.77, \textit{p} = 0.082).

% d'
% Assignment: F(1, 57) = 0.762, p = 0.386
% Spatial: F(1, 57) = 3.066, p = 0.0853 .
% Temporal: F(1, 57) =  3.00, p = 0.0887 .
% Int: F(1, 57) = 0.00, p = 0.985  

When responses are transformed into \textit{d'}, effects of spatial predictability and temporal predictability during the training period trended toward significance (spatial, \textit{F}(1, 57) = 3.07, \textit{p} = 0.085; temporal, \textit{F}(1, 57) = 3.00, \textit{p} = 0.089). The interaction between spatial and temporal predicability failed to reach significance (\textit{F}(1, 57) = 0.00, \textit{p} = 0.985). The pattern of results as a function of predictability during the training period was the same as for accuracy, and thus this failure to reach critical significance likely reflects the loss of power when transforming responses into \textit{d'} due to discarding misses and correct rejections. 

% RT
% Assignment: F(1, 57) = 0.85, p = 0.36
% Spatial: F(1, 57) = 10.66, p = 0.002 *
% Temporal: F(1, 57) = 1.00, p = 0.321
% Int: F(1, 57) = 2.17, p = 0.146

% IE
% Assignment: F(1, 57) = 1.069, p = 0.306
% Spatial: F(1, 57) = 9.157, p = 0.00371 *
% Temporal: F(1, 57) = 0.244, p = 0.623
% Int: F(1, 57) = 1.219 , p = 0.274   

Reaction times for correct trials were slower when the training period was spatially predictable (\textit{F}(1, 57) = 10.66, \textit{p} = 0.002). A similar effect for temporal predictability failed to reach significance (\textit{F}(1, 57) = 1.00, \textit{p} = 0.321), nor did the interaction between spatial and temporal predictability (\textit{F}(1, 57) = 2.17, \textit{p} = 0.146). Effects on inverse efficiency (defined as reaction time divided by percent correct) were similar to those of reaction times. Inverse efficiency was highest when the training period was spatially predictable (\textit{F}(1, 57) = 9.12, \textit{p} = 0.004), but did not significantly differ as a function of temporal predictability (\textit{F}(1, 57) = 0.24, \textit{p} = 0.623), nor when considering the interaction between spatial and temporal predictability (\textit{F}(1, 57) = 1.22, \textit{p} = 0.274).

\begin{figure}[h!]
\begin{center}
\begin{tabular}{ll}
\includegraphics[width=80mm]{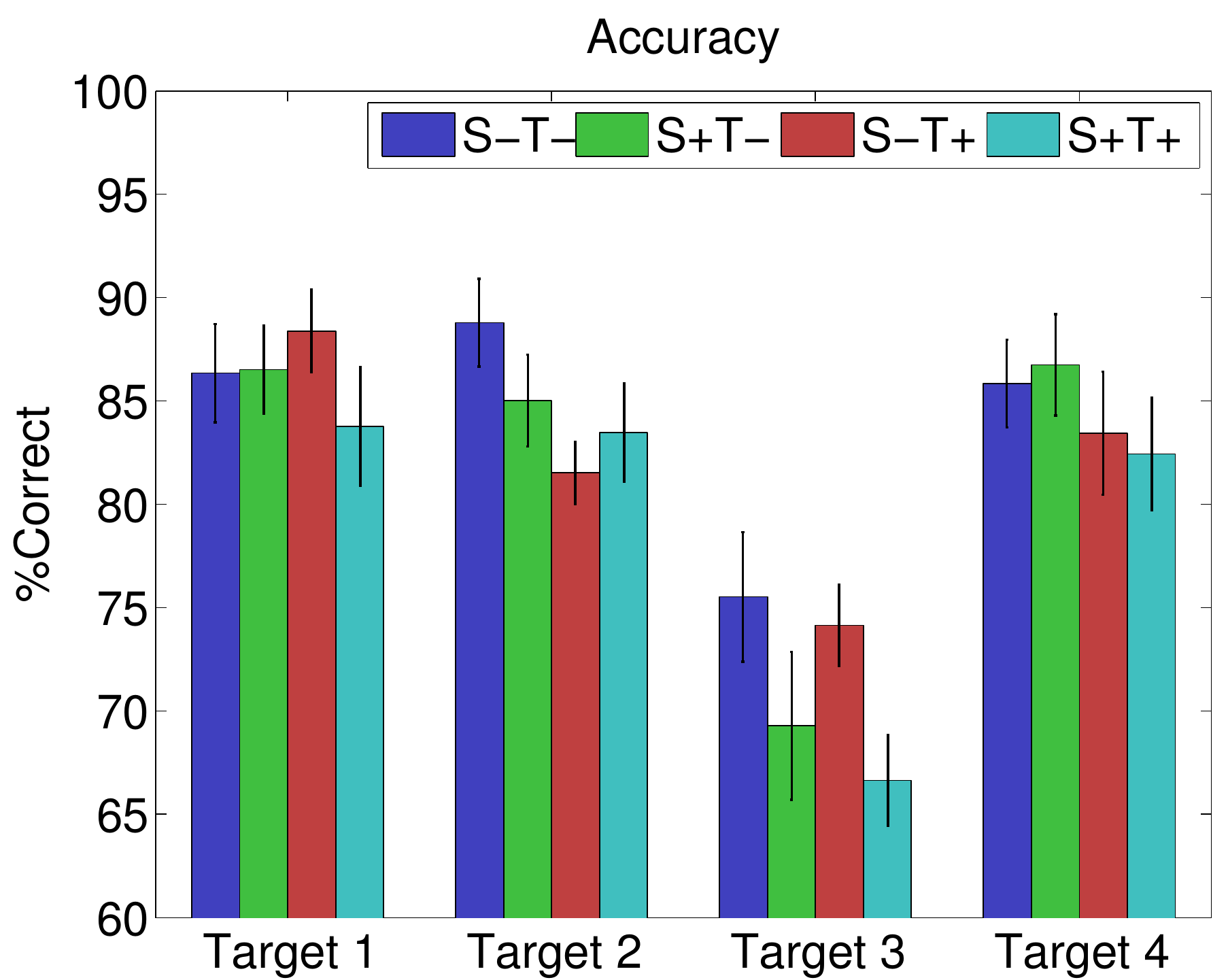} &
\includegraphics[width=80mm]{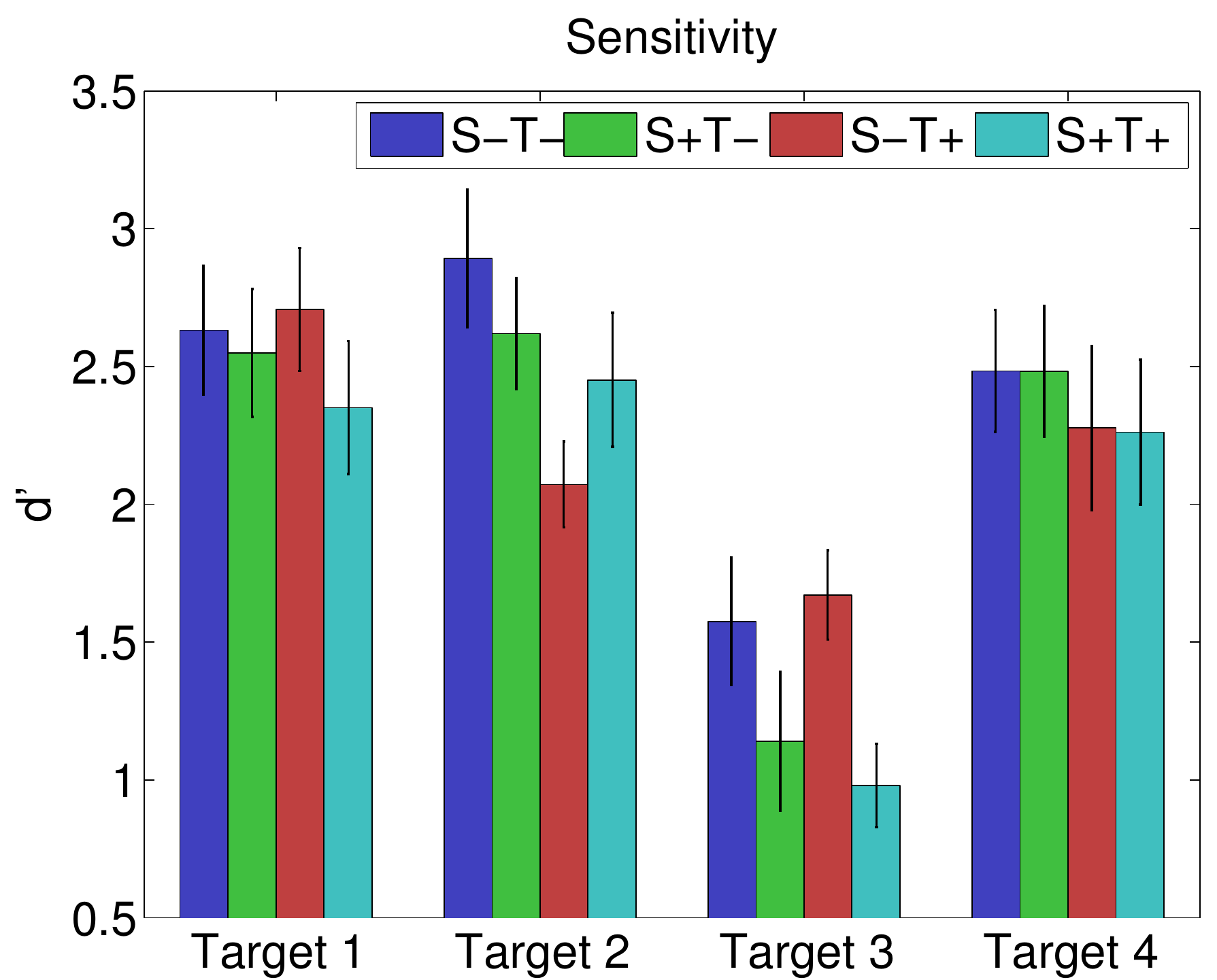} \\
\includegraphics[width=80mm]{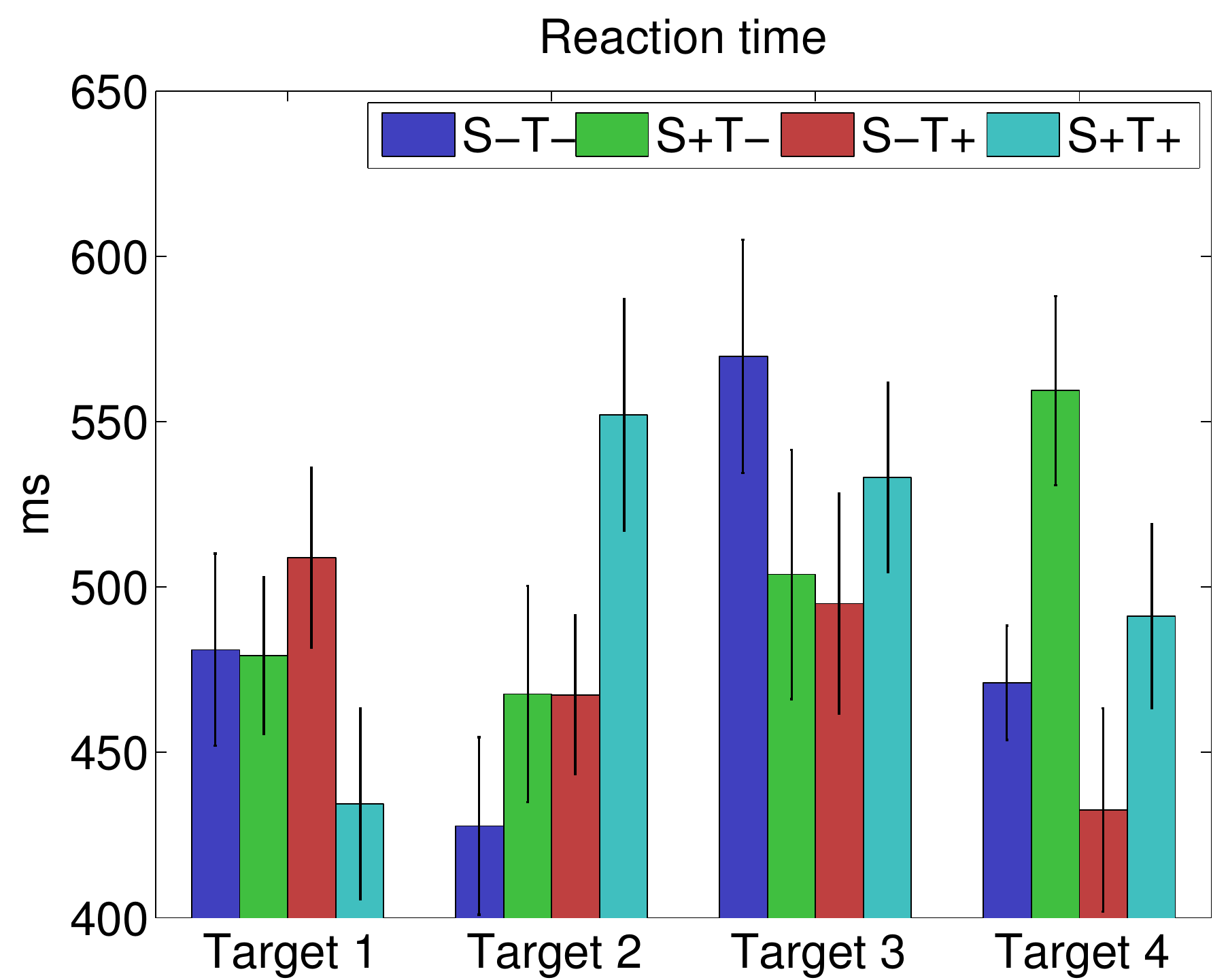} &
\includegraphics[width=80mm]{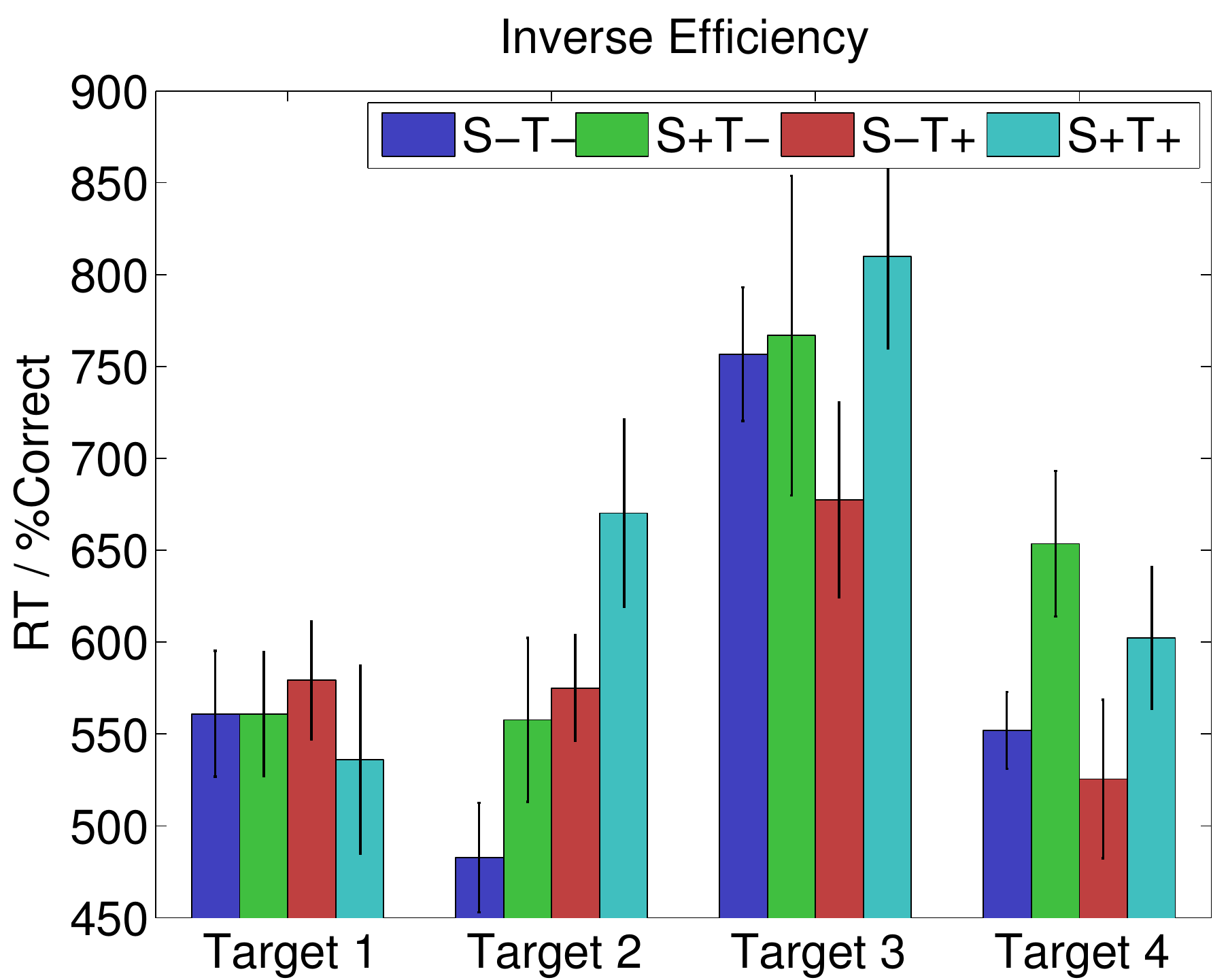} \\
\end{tabular}
\end{center}
\caption{Behavioral measures for each target object}{Accuracy, \textit{d'}, reaction time, and inverse efficiency for each target object. Horizontal axes denote target and colors predictability during the training period. Error bars depict between-subjects standard error.}
\label{fig:bpleast_behave_obj}
\end{figure}

Effects were highly variable across target objects (Figure \ref{fig:bpleast_behave_obj}). Target-condition assignment did not significantly affect any of the behavioral measures on average (all \textit{p}'s $>$ 0.05), but often interacted with predictability effects and their interactions. One reason for this variability regards the orthographic projection used to render the objects. Previous research has indicated that recognition accuracy fluctuates as a function of how well the two-dimensional projection of an object captures its full three-dimensional structure \cite{BalasSinha09b}. For example, when there is a large amount of foreshortening due to the projection, it could be difficult to infer the length of line segments that compose the object, impairing recognition. These degenerate projections are generally diametrically opposed around the object. 

% oh, behave, pt 2: objs

\begin{figure}[h!]
\begin{center}
\begin{tabular}{ll}
\includegraphics[width=80mm]{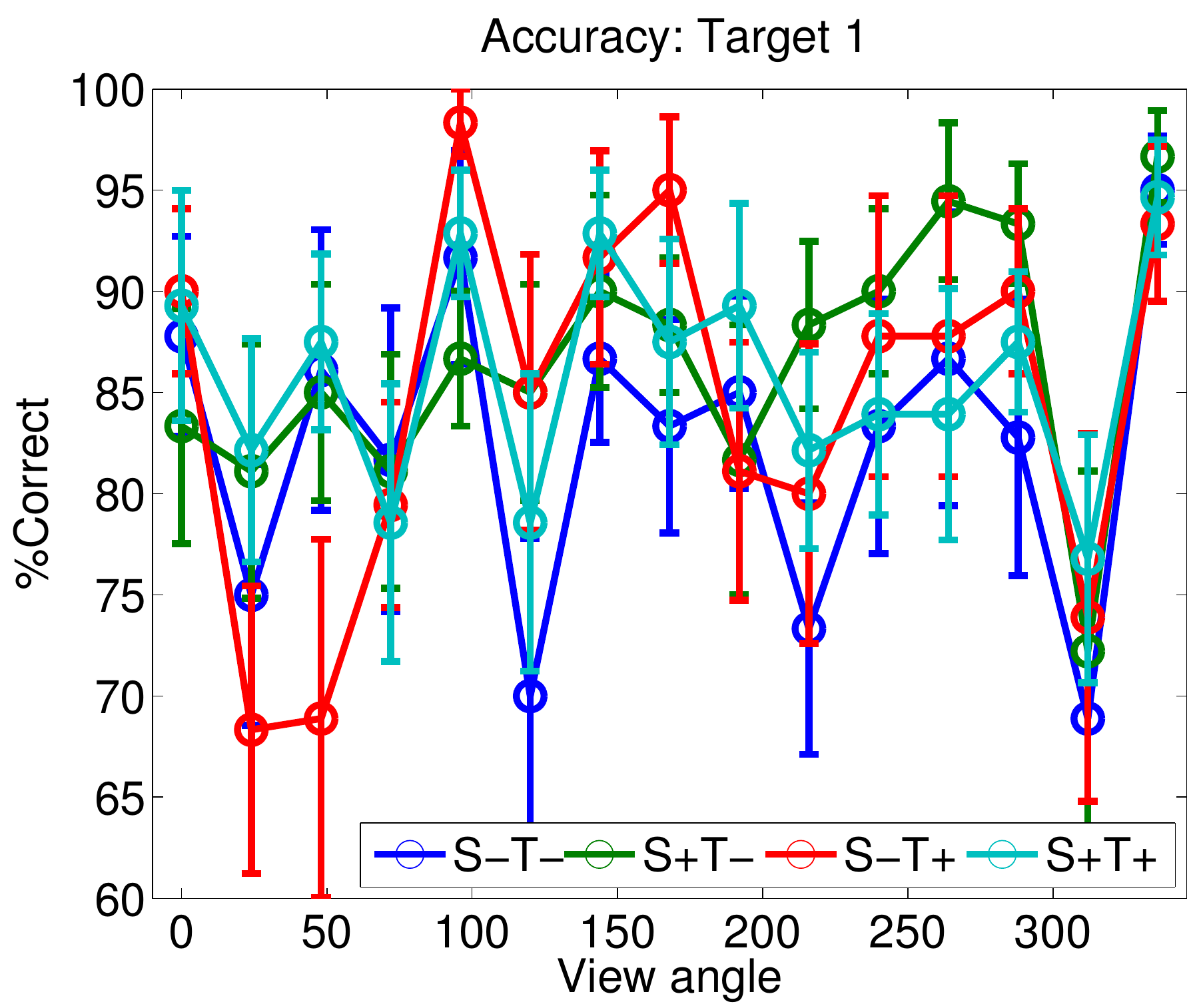} &
\includegraphics[width=80mm]{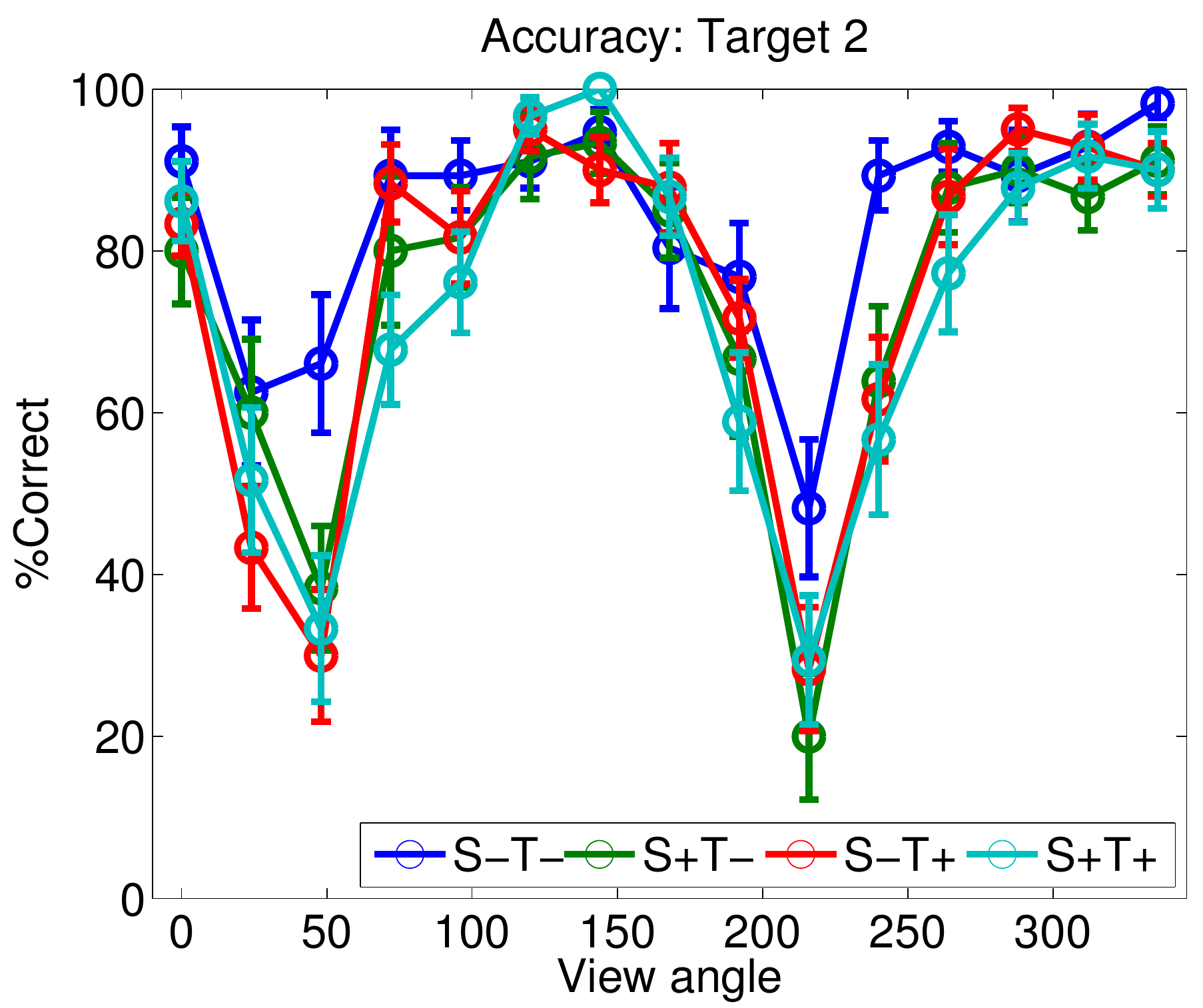} \\
\includegraphics[width=80mm]{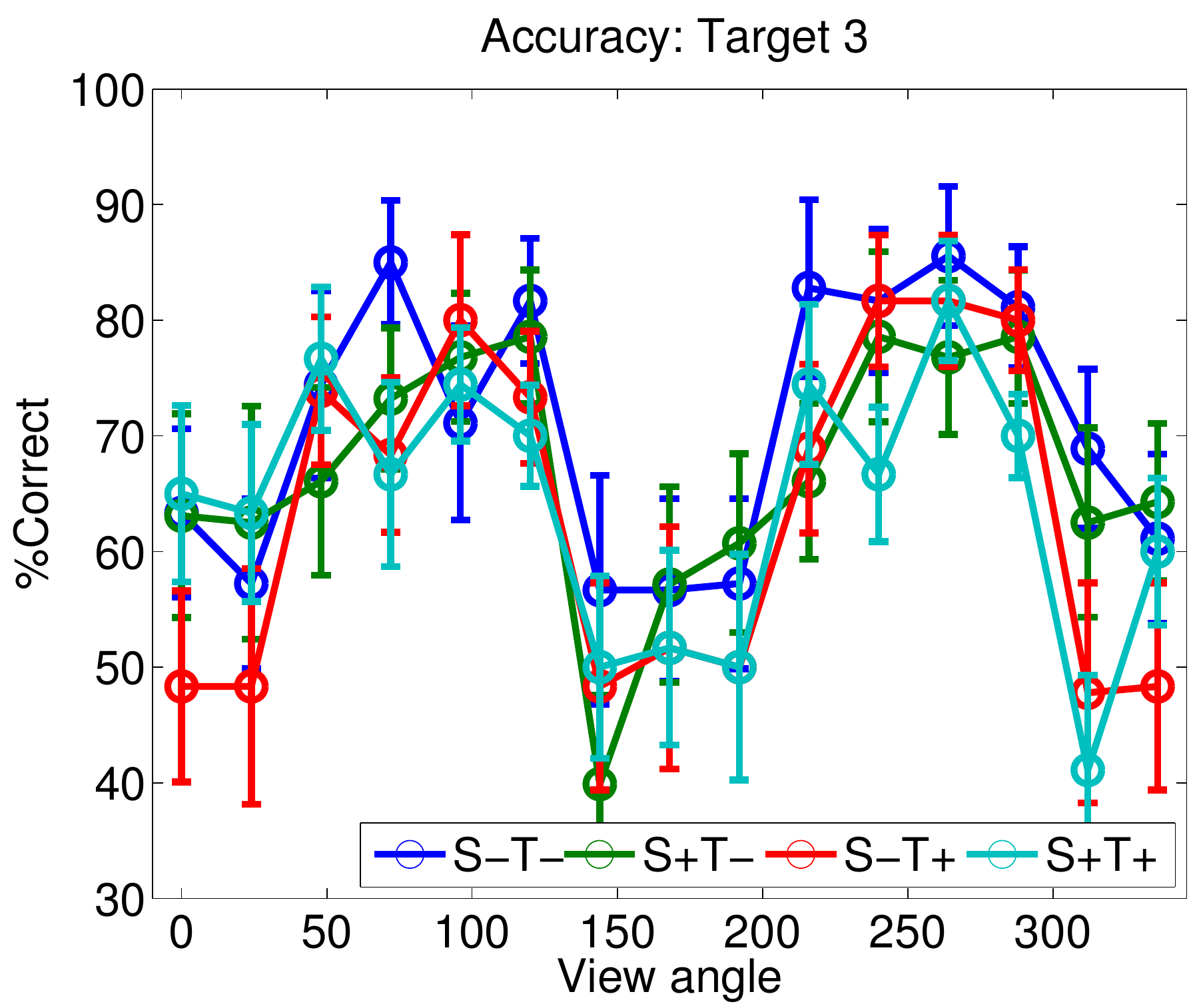} &
\includegraphics[width=80mm]{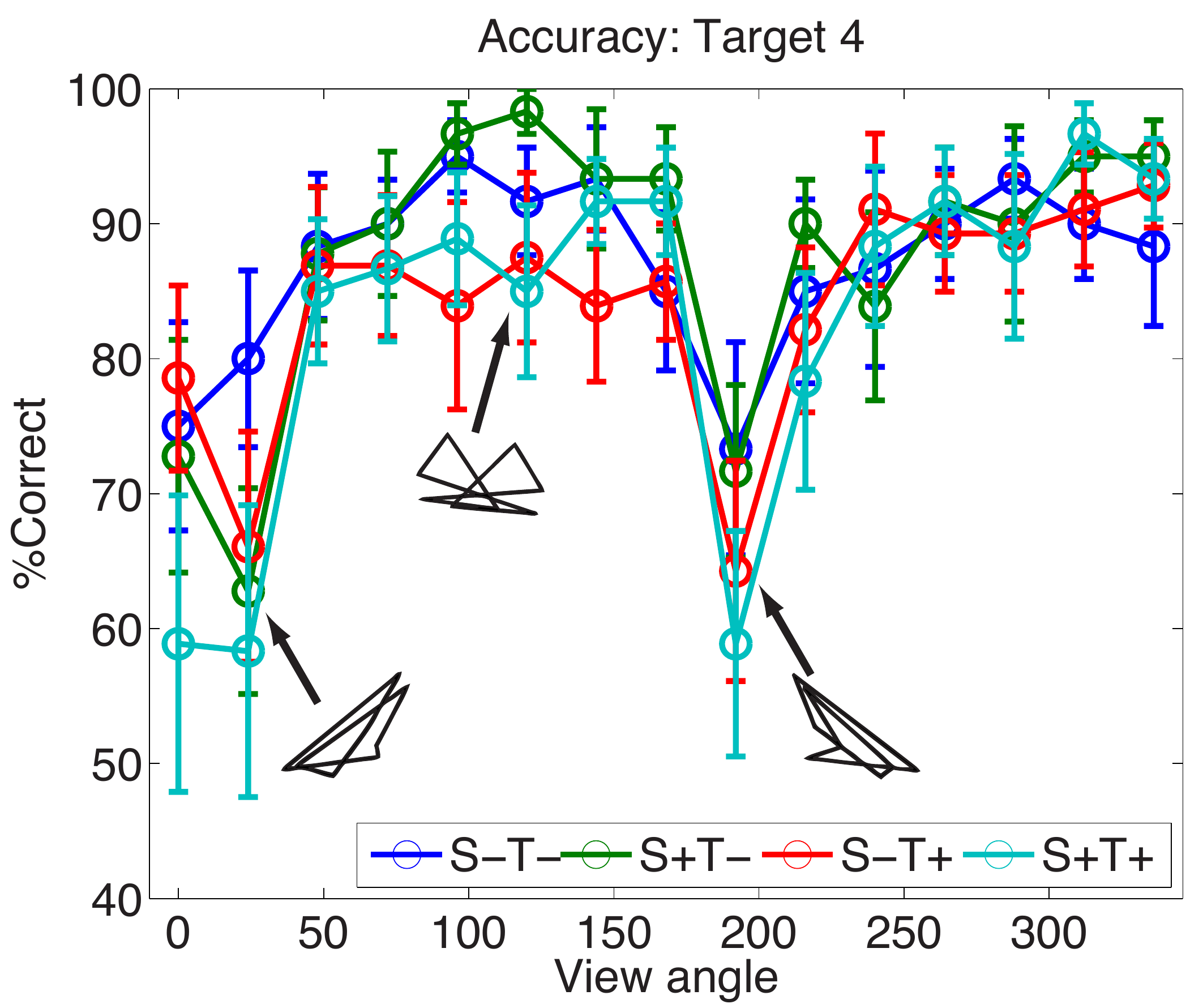} \\
\end{tabular}
\end{center}
\caption{Accuracy as a function of viewing angle for each target object}{Target accuracy at each viewing angle presented during the test periods. Horizontal axes denote viewing angle and colors predictability during the training period. Error bars depict between-subjects standard error. Diametrically opposed foreshortened views and one canonical view are shown for target object 4.}
\label{fig:bpleast_behave_rot}
\end{figure}

To investigate this hypothesis, accuracy was computed as a function of viewing angle for each target object to investigate whether it interacted with predictability during the training period (Figure \ref{fig:bpleast_behave_rot}). Only accuracy was considered in this analysis as each data point only corresponded to four trials per subject and thus transformation to \textit{d'} was not plausible. Test trials during which distractor objects were presented were also excluded from this analysis since there was no consistent relationship between the targets and distractors across viewing angles and thus they would only contribute noise. With the exception of target object 1, all objects demonstrated fluctuations in accuracy as a function of viewing angle with two diametrically opposed degenerate views. The most consistent differences in accuracy between training conditions appeared to be localized to the troughs of the accuracy function, corresponding to these degenerate views.

% oh, behave, pt 3: rots

\begin{figure}[h!]
\begin{center}
\begin{tabular}{ll}
\includegraphics[width=160mm]{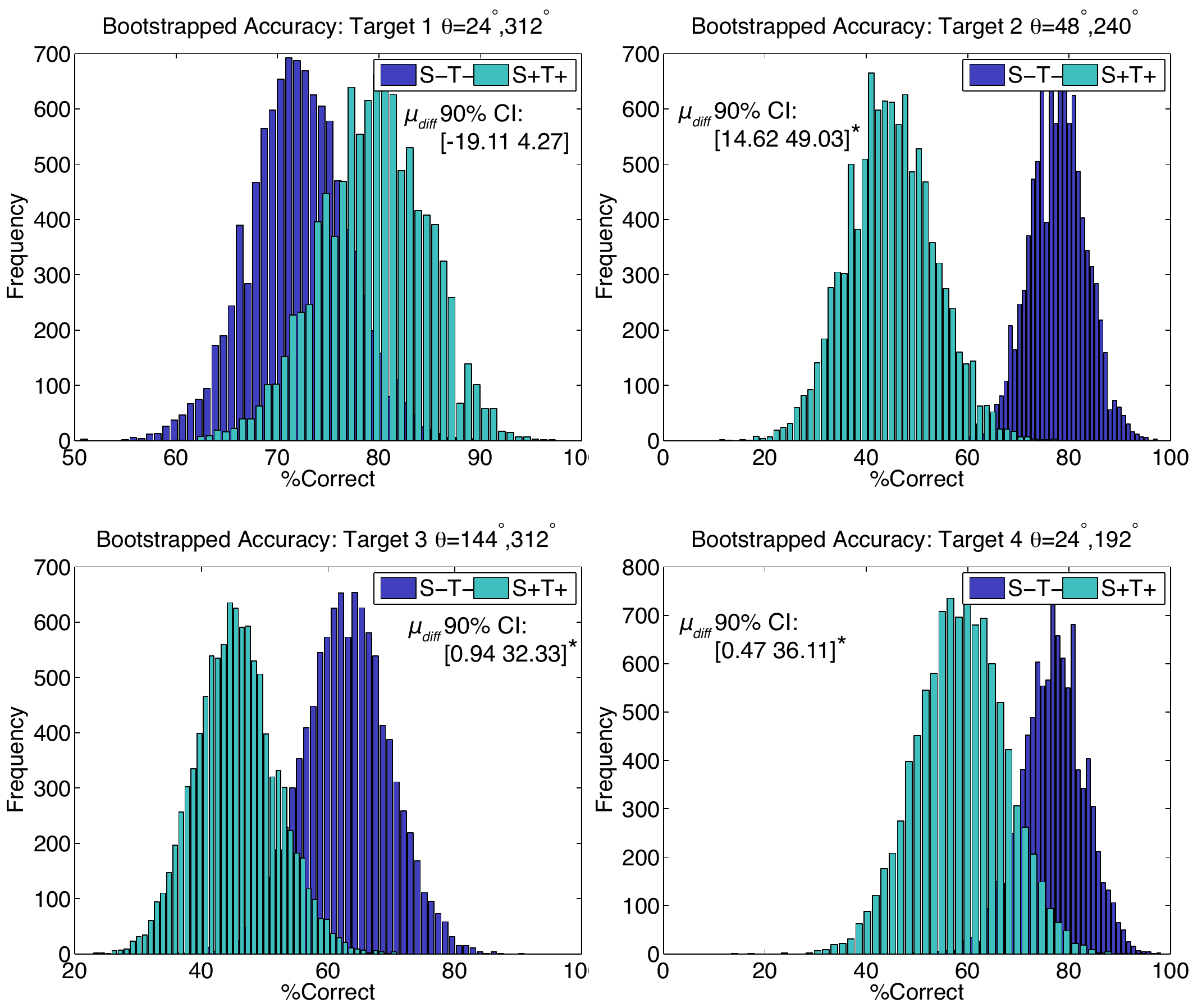}
\end{tabular}
\end{center}
\caption{Bootstrapped accuracy for degenerate views}{Average target accuracy for degenerate views resampled with replacement from the 59 subjects for 10000 iterations. Viewing angles for averaging is noted for each target object. Asterisks denote significant differences based on 90\% confidence intervals.}
\label{fig:bpleast_behave_bootstrap}
\end{figure}

Standard statistical tests did not have enough power to detect differences between conditions for degenerate views due to the low trial counts for each data point. To address this design limitation, a bootstrapping method was used to resample the available data in these cases. The completely unpredictable (S-T-) and combined spatial and temporal predictability (S+T+) were used to assess differences due to training context since these two conditions elicited the greatest difference in average accuracy in the full analysis. The accuracy function over viewing angles was collapsed across conditions and the two minima associated with degenerate views were identified for each object. For target object 1, the two views were at $\theta$ = (24$^\circ$, 312$^\circ$), object 2: $\theta$ = (48$^\circ$, 240$^\circ$), object 3: $\theta$ = (144$^\circ$, 312$^\circ$), and object 4: $\theta$ = (24$^\circ$, 192$^\circ$). S-T- and S+T+ accuracy was averaged at each object's degenerate views and resampled with replacement from the 59 subjects for 10000 iterations. This produced degenerate view accuracy distributions for spatiotemporally unpredictable and predictable training contexts (Figure \ref{fig:bpleast_behave_bootstrap}). Accuracy was lower for degenerate views studied under the spatiotemporally predictable context for all target objects except target 1, which didn't exhibit the patterned accuracy function that other targets did. The predictability difference in accuracy for degenerate views was significant at the 90\% alpha level (i.e., the confidence interval of the difference between means did not include zero) for all target objects except target 1.

% oh, behave, pt 4: bootstraps

\section{Discussion}\subsection{Summary of results}
The work described in this chapter investigated how predictability biased representations of novel objects over prolonged learning. The experimental paradigm used to address this question involved training participants to recognize novel objects while manipulating their spatial and temporal predictability. Somewhat surprisingly, accuracy was lowest when stimuli were learned in a combined spatially and temporally predictable context and highest when learned in a completely unpredictable context. Reaction times were also slower when objects were learned with spatial predictability. These findings were unexpected because the spatial structure of the physical world has been suggested to be a powerful learning mechanism when coupled with the putative temporal association rules of visual neurons \cite{StringerPerryRollsEtAl06,WallisBaddeley97,IsikLeiboPoggio12,SakaiMiyashita91,MeyerOlson11,CoxMeierOerteltEtAl05,LiDiCarlo08,LiDiCarlo10,LiDiCarlo12}.

Behavioral measures were highly variable across objects used in the experiment. There was some indication that the principal differences between predictability conditions during training were driven primarily by degenerate viewing angles caused by three-dimensional foreshortening in the objects used. In three out of four objects, accuracy was lower for degenerate views learned in a spatiotemporally predictable context compared to a completely unpredictable one.

\subsection{A behavioral disadvantage for spatial prediction during object learning}
Intuitively, spatial predictability should be advantageous for learning the three-dimensional structure of objects given that it is the learning context within which the visual system evolved. However, the literature contains a mixture of contradictory effects regarding the utility of spatially predictable information during object learning and recognition. Initial experiments described in \incite{LawsonHumphreysWatson94} with depth-rotated line drawings of familiar objects demonstrated the expected increase in recognition accuracy for spatially predictable sequences. A number of studies have found that studying depth-rotating sequences of novel objects under one ordering and then testing with a different ordering impairs recognition \cite{Stone98,VuongTarr04,ChuangVuongBulthoff12}, implying that learned spatial predictability about the sequence is used to encode the object \cite{BalasSinha09c}. The foreshortening model advanced in \cite{BalasSinha09b} also provided a better match to observers' data by incorporating spatial information (e.g., the first- and second-order derivatives of the foreshortening function over object views).

Some of the experiments described in \incite{LawsonHumphreysWatson94}, however, demonstrated better accuracy for sequences studied with weak spatial predictability (maximum of two consecutive spatially coherent frames in a sequence) than total spatial predictability. Experiments described in \incite{HarmanHumphrey99} failed to find any positive or negative effects of spatial predictability on accuracy. They did, however, increase in reaction time for objects learned in a spatially predictable context, similar to the one reported here. 

One possible reason for the behavioral disadvantage for objects learned with spatial predictability is that less attention is necessary in these conditions. A constantly changing sequence of views might require more attentive processing to encode whereas the relatively small amount of change between views in spatially predictable sequences is comparatively ``unsurprising'' such that some views might be overlooked during encoding \cite{TarrGauthier98}. However, there was some indication that the adverse effects of spatial predictability in the current experiment were driven primarily by the degenerate views of particular stimuli. A more focused experiment is clearly necessary to explicitly test the hypothesis that behavioral performance is impaired for degenerate views learned in a spatially predictable context but relatively intact for canonical views. 

\subsection{Building viewpoint invariance for three-dimensional objects}
Given the cumulative literature considered here, the most reasonable interpretation of the experimental results is as follows: Three-dimensional objects are typically represented in a viewpoint-dependent manner \cite{WallisBulthoff99,EdelmanBulthoff92,TarrGauthier98,LogothetisPaulsBulthoffEtAl94,LogothetisPaulsPoggio95}. Each of these views is associated with the given object's identity to the degree to which three-dimensional features are recoverable from the two-dimensional projection. This can lead to low accuracy for degenerate viewing angles caused by extreme three-dimensional foreshortening demonstrated here and in previous studies \cite{BalasSinha09b}. 

When three-dimensional objects are studied in a full spatiotemporally predictable context, temporal association mechanisms are invoked that build viewpoint invariance \cite{CoxMeierOerteltEtAl05,LiDiCarlo08,LiDiCarlo10,LiDiCarlo12}. This invariance accounts for variability due to foreshortening but is actually \textit{not optimal} for three-dimensional objects that need to be recognized from individual static views. If the full spatiotemporal sequence is available when the object needs recognized (as is typical in real world object recognition) accuracy is not impaired by degenerate views and activation of the newly acquired invariant features might even be facilitated \cite{BalasSinha09c}. However, if only static views are available during recognition, or perhaps worse, the full sequence is available but not in its spatially predictable order, it is not possible to activate these invariant features, leading to impaired recognition \cite{Stone98,VuongTarr04,ChuangVuongBulthoff12}.

The corollary of this interpretation is that when three-dimensional objects are studied in a completely unpredictable context, temporal association mechanisms are actually \textit{not invoked}. Further research is needed to determine if this assertion is plausible, and if so, precisely why these mechanisms would not be active. However, one possibility is that temporal associations operate over samples of an input sequence at regular intervals (e.g., 100 ms, see Chapter \ref{chap:leabrati}). When input sequences don't align to these sampling intervals, the associations between subsequent inputs cannot be formed. The result is that despite prolonged learning, the objects remain represented in a viewpoint-dependent manner, which facilitates recognition for static views.

%% file: chap_sims.tex
\chapter{Neural model of spatiotemporal prediction for object recognition}
\label{chap:sims}

\sloppy
\interfootnotelinepenalty=10000

\section{Introduction}
The work presented in this chapter describes a neural network model of the broader LeabraTI framework (Chapter \ref{chap:leabrati}) complete with the columnar substructure required for its temporally interleaved predictive learning. The specific implementation was used to investigate the role of spatiotemporal predictive learning in a visual object recognition task, analogous to the tasks implemented in the Chapter \ref{chap:pleast} and \ref{chap:bpleast} experiments. The principal behavioral results of these experiments are first reviewed before turning to the model implementation and simulations that reproduce these results.

The Chapter \ref{chap:pleast} experiment investigated the role of predictive processing during a novel object recognition task. The experiment made use of novel three-dimensional ``paper clip'' objects \cite{BulthoffEdelman92,EdelmanBulthoff92,LogothetisPaulsBulthoffEtAl94,LogothetisPaulsPoggio95,SinhaPoggio96} that required integration over multiple sequential views to extract their three-dimensional structure. Stimulus sequences were either presented in a spatially predictable or random order and either at a predictable 10 Hz rhythm or arrhythmically (these two factors were manipulated independently). The results of the experiment indicated that both the spatial and temporal predictability of an entraining sequence enhanced discriminability of a subsequently presented probe stimulus using a same-different judgement.

The Chapter \ref{chap:bpleast} experiment expanded on the previous chapter's experiment by investigating the role that spatial and temporal predictability played during prolonged learning about the same paper clip objects. The experiment involved an explicit training period during which observers studied the objects while they were rotated through their views followed by a series of test trials that required same-different judgements about static probe stimuli. Spatial and temporal predictability were manipulated independently manipulated during the training period. Somewhat surprisingly, the results of the experiment were an almost complete reversal of the previous chapter's experiment. Discriminability was lowest when stimuli were learned in a combined spatially and temporally predictable context and highest when learned in a completely unpredictable context.

The model described next was capable of producing the results of both experiments. LeabraTI predicts that spatially predictable sequences presented at a regular temporal interval aligned with the brain's endogenous 10 Hz prediction rate should maximally activate representations due to the multiple prediction-sensation events that successfully integrate visuospatial information at optimal temporal intervals. The result is a ``synergistic'' superadditivity effect for combined spatially and temporally predictable sequences, similar to findings that have been demonstrated in previous investigations of predictability on attentional allocation \cite{DohertyRaoMesulamEtAl05,RohenkohlGouldPessoaEtAl14}.

The Chapter \ref{chap:bpleast} discriminability reversal effect due to prolonged learning was able to be produced by increasing the scale of a single projection of synaptic weights in the model. Further analysis of the model's representational similarity (\nopcite{KriegeskorteMurBandettini08}; \abbrevnopcite{KriegeskorteMurRuffEtAl08}) suggested that this reversal was due to viewpoint invariance learned from spatiotemporal association that ``trickled down'' to retinotopic feature representations. This was problematic for certain viewpoints, causing them to be confused with other objects and leading to lower accuracy on average.

\section{Methods}

\subsection{Model architecture}

\begin{figure}[h!]
\begin{center}
\includegraphics[width=160mm]{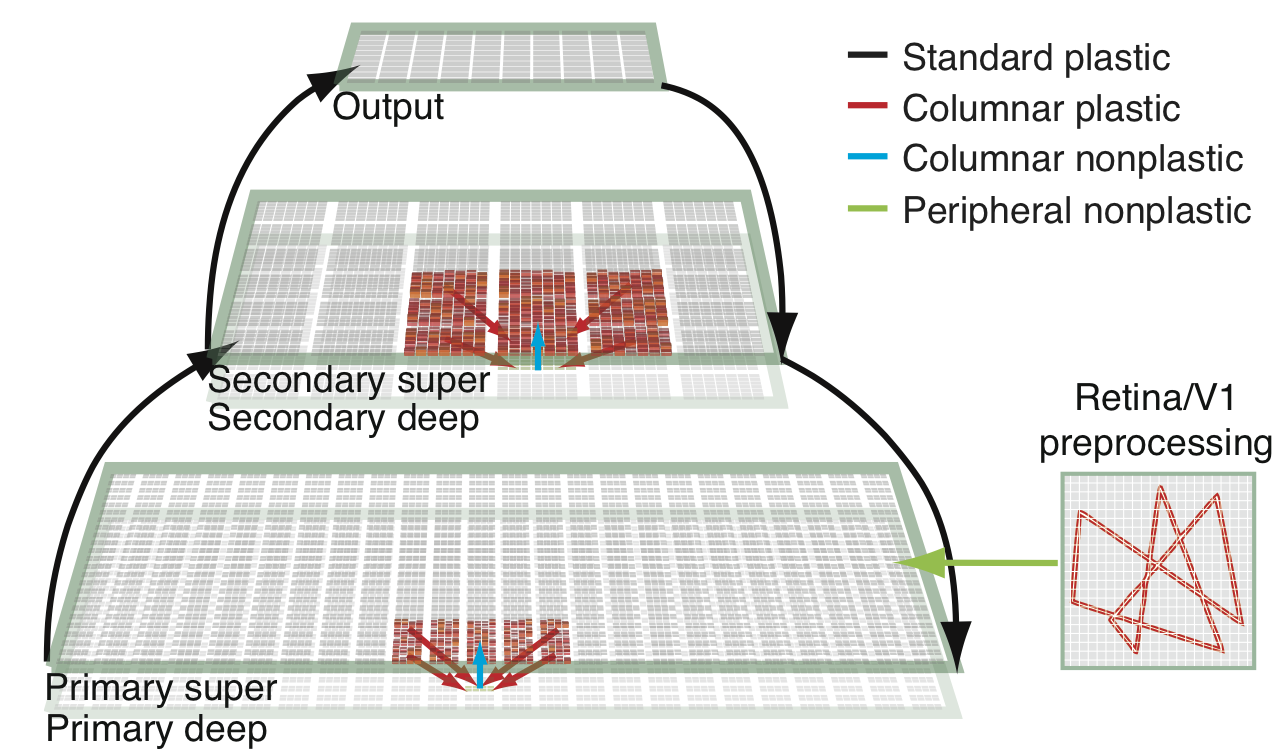}
\end{center}
\caption{Model architecture}{The model's layers and principal projections. Primary and secondary visual layers contained columnar substructure in which deep units integrated from 5x5 columns of superficial units in the primary case or 3x3 columns in the secondary case. Ascending synapses from deep to superficial units were nonplastic and connected in a one-to-one manner.}
\label{fig:v1_v2_output}
\end{figure}

The model architecture is illustrated in Figure \ref{fig:v1_v2_output}. The model consisted of three layers and one preprocessing stage whose parameters are described in detail in the following paragraphs. Two of the layers contained columnar substructure necessary for learning using the LeabraTI algorithm. To simplify the overall LeabraTI computation, only superficial (Layer 2/3) and deep (Layer 6) neuron subtypes were explicitly modeled. Projections between these neuron populations primarily corresponded to the descending Layer 5 $\rightarrow$ Layer 6 synapses in the brain, which are assumed to be plastic, and the ascending Layer 6 $\rightarrow$ Layer 4 transthalamic synapses which are assumed to be relatively stable and nonplastic. This simplification captures the core computational properties of the LeabraTI framework while reducing the overhead of simulating the full  lamination of the neocortex with detailed microcolumnar circuitry.

% model fig

\textbf{Retina and V1 preprocessing}: Input was provided to the model via a 24x24 retinotopic filter bank that preprocessed images offline from the model proper. This preprocessing step is consistent with a large class of biological models describing object recognition in cortex \cite[e.g.,]{RiesenhuberPoggio99,WallisRolls97,MasquelierThorpe07,OReillyWyatteHerdEtAl13} and in the case of the present model, represents visual processing from the level of the retina through V1 simple cells \cite{HubelWiesel62}. Grayscale bitmap images were scaled to 24x24 pixels and convolved with Gabor filters at four orientations (0$^\circ$, 45$^\circ$, 90$^\circ$, and 135$^\circ$) and two polarities (off-on and on-off) producing a 24x24x4x2 set of inputs. Each Gabor filter was implemented as 6x6 pixel kernel, with a wavelength $\lambda$ = 6 and Gaussian width terms of $\sigma_x$ = 1.8 and $\sigma_y$ = 1.2. A static nonlinearity was applied to the output of the filtering step in the form of a modified \textit{k}-Winners-Take-All (\textit{k}WTA) inhibitory competition function that reduced activation across the 4x2 filter bank to the equivalent of \textit{k} = 1 fully active units \cite[see][Supporting Information]{OReillyWyatteHerdEtAl13}.

\textbf{Primary visual layers}: 24x24 retinotopic layer arranged into groups of 4x2 units (4608 total units), decomposed into superficial and deep neuron subtypes. Each superficial unit received the output of the retina/V1 preprocessing step. \textit{k}WTA inhibition for superficial units was set to 60\% of the average of the top \textit{k} active units compared to the average of all other superficial units with each 4x2 unit group using a value of \textit{k} = 2. Deep units received from 5x5 columns of superficial units (200 inputs per deep unit) integrated into a single value that was used to drive the context input channel for each superficial unit in a one-to-one manner.

\textbf{Secondary visual layers}: 6x6 retinotopic layer arranged into groups of 7x7 units (1764 total units), also decomposed into superficial and deep neuron subtypes. Each superficial unit received from 8x8 topographical neighborhoods of early visual columns (512 afferents per unit) and sent back reciprocal connections with the same topography. \textit{k}WTA for superficial units was set to 60\% of the average of the top \textit{k} active units compared to the average of all other superficial units with 15\% activity within each unit group. Deep units received from 3x3 columns of superficial units (441 inputs per deep unit) integrated into a single value that was used to drive the context input channel for each superficial unit in a one-to-one manner.

\textbf{Output layer}: 10x10 layer (100 total units) without unit group or columnar substructure. Each unit received a full projection from secondary visual columns (1764 afferents per unit) and fully projected back to all units in all columns. A scale of 10\% was used to limit the influence of the output units on secondary visual columns during the training period, preventing ``hallucinatory'' representations that can become disconnected from bottom-up inputs. A \textit{k}WTA value of \textit{k} = 1 was used to enforce a localist representation. The localist representation is a computational simplification that allowed an invariant identity representation similar to what inferior temporal (IT) neurons encode using a population code \cite{HungKreimanPoggioEtAl05,LiCoxZoccolanEtAl09}. 

\subsection{LeabraTI learning algorithm}

LeabraTI was implemented as an extension of the standard Leabra algorithm which is described in detail in \incite{OReillyMunakata00} and \incite{OReillyMunakataFrankEtAl12}. Standard Leabra learning operates across two phases: a \textit{minus} phase that represents the system's expectation for a given input and a \textit{plus} phase, representing observation of the outcome. The difference between the minus and plus phases, along with additional self-organizing mechanisms, is used in computing the synaptic weight update function at the end of each plus phase. 

LeabraTI extends standard Leabra learning by interleaving its minus and plus phases over temporally contiguous input sequences. In standard Leabra, the minus phase depends on clamped inputs from the sensory periphery to drive the expectation while the plus phase additionally makes use of clamped outputs from other neural systems to drive the outcome. In LeabraTI, the minus phase expectation is not driven by the sensory periphery, but instead by lagged context represented by deep (Layer 6) neurons. During the plus phase, driving potential shifts back to the sensory periphery. Deep neurons' context is also updated after each plus phase.

LeabraTI was only used to update the synaptic weights between superficial and deep neurons. Inter-areal feedforward and feedback projections bifurcate from the local column, directly synapsing disparate populations of superficial neurons and thus weight updates in these cases were handled by standard Leabra equations. In computing the weight update, the standard Leabra delta rule \cite{OReilly96} uses the difference in rate between the plus and minus phases of receiving units (\textit{y}) in proportion to the rate of sending units (\textit{x}) in the minus phase:
\begin{align*}
\Delta_{leabra} w_{ij} &= x^-(y^+ - y^-)
\end{align*}

In the LeabraTI framework, deep neurons are considered to be the receiving units as they are the terminus of the descending columnar synapses. However, deep units are proposed to only be active during the minus phase when they drive the prediction, and thus cannot be used to compute an error signal.  To address this issue, we invert the LeabraTI delta rule:
\begin{align*}
\Delta_{leabrati} w_{ij} &= super^-(deep^+ - deep^-) \\
			  &\approx deep^-(super^+ - super^-)
\end{align*}

Additionally, the temporally extended nature of the algorithm requires that the receiving units represent the current state (time \textit{t}) and sending units the previous moment's state (time \textit{t} - 1). While conceptualized as the previous equation, the actual implementation is as follows:
\begin{align*}
\Delta_{leabrati} w_{ij} &= super_{t-1}^+(super_{t}^+ - super_{t}^-)
\end{align*}

This formulation allows the driving potential of deep neurons to be computed just once using the previous plus phase state of superficial neurons (multiplied by the superficial $\rightarrow$ deep learned weights) and held constant as a sustained input to superficial neurons during the minus phase. This is a gross simplification of the actual biological process of deep neurons, but is vastly more computationally efficient than explicit modeling by computing an additional rate for each deep neuron at each time step. This formulation is also equivalent to the simple recurrent network (SRN) \cite{Elman90,Servan-SchreiberCleeremansMcClelland91}, thus providing a realistic biological substrate for its computational function. 

One limitation of LeabraTI's interleaving of minus and plus phases over time is that the initial minus phase in an input sequence does not have access to the previous moment's  context. Even if there was lagged context available, it would represent information from a prior, possibly unrelated input sequence. To address this, weight updates are disabled for the first minus-plus phase pair, and enabled thereafter. In the brain, this process might be facilitated by a neural mechanism that is sensitive to the repetition of inputs over time such as acetylcholine \cite{ThielHensonMorrisEtAl01,ThielHensonDolan02}.

\subsection{Training and testing environment}
LeabraTI requires training to establish the spatial associations over subsequent time steps. In human development, this is expected to be facilitated by coarse transformations of retinal inputs due to environmental or self motion. This initial learning stage develops generic features that capture how inputs change from moment-to-moment (100 ms periods in LeabraTI). The actual inputs are not critical except that they accurately reflect the average statistics of the environment. In training the model, a simplified  ``paper clip'' environment was assumed, using the four objects from the Chapter \ref{chap:bpleast} experiment.

% plus phase mixture parameter, but not used (?)
During training, an input sequence depicted one of the four objects rotating coherently through all 30 view renderings (adjacent views spaced 12 degrees apart). During the minus phase, the model made a prediction about the upcoming view of the object and during the plus phase, the view was processed by the retina/V1 filter banks and clamped as an input to the model. The output unit corresponding to each object was also clamped during the plus phase to bias views belonging to the same object toward similar lower-level feature representations. The minus phase lasted 50 cycles whereas the plus phase lasted 20 cycles, consistent with the idea that internally generated predictions are sustained whereas sensory events are transmitted in rapid ``burst'' packets from the sensory periphery.  Training proceeded for 20 epochs of 10 randomly selected input sequences each. The learning rate on all plastic synapses started at 1.0 and was halved every 8 epochs.

Training efficacy was evaluated by computing the average cosine (normalized dot product) between the minus and plus phase for the primary and secondary visual layers:
\begin{align*}
cos \theta = \frac{1}{n}\sum_{k=1}^{n}\frac{layer_k^- \cdot{} layer_k^+}{||layer_k^-||||layer_k^+||}
\end{align*}

\begin{figure}[h!]
\begin{center}
\begin{tabular}{ll}
\includegraphics[width=80mm]{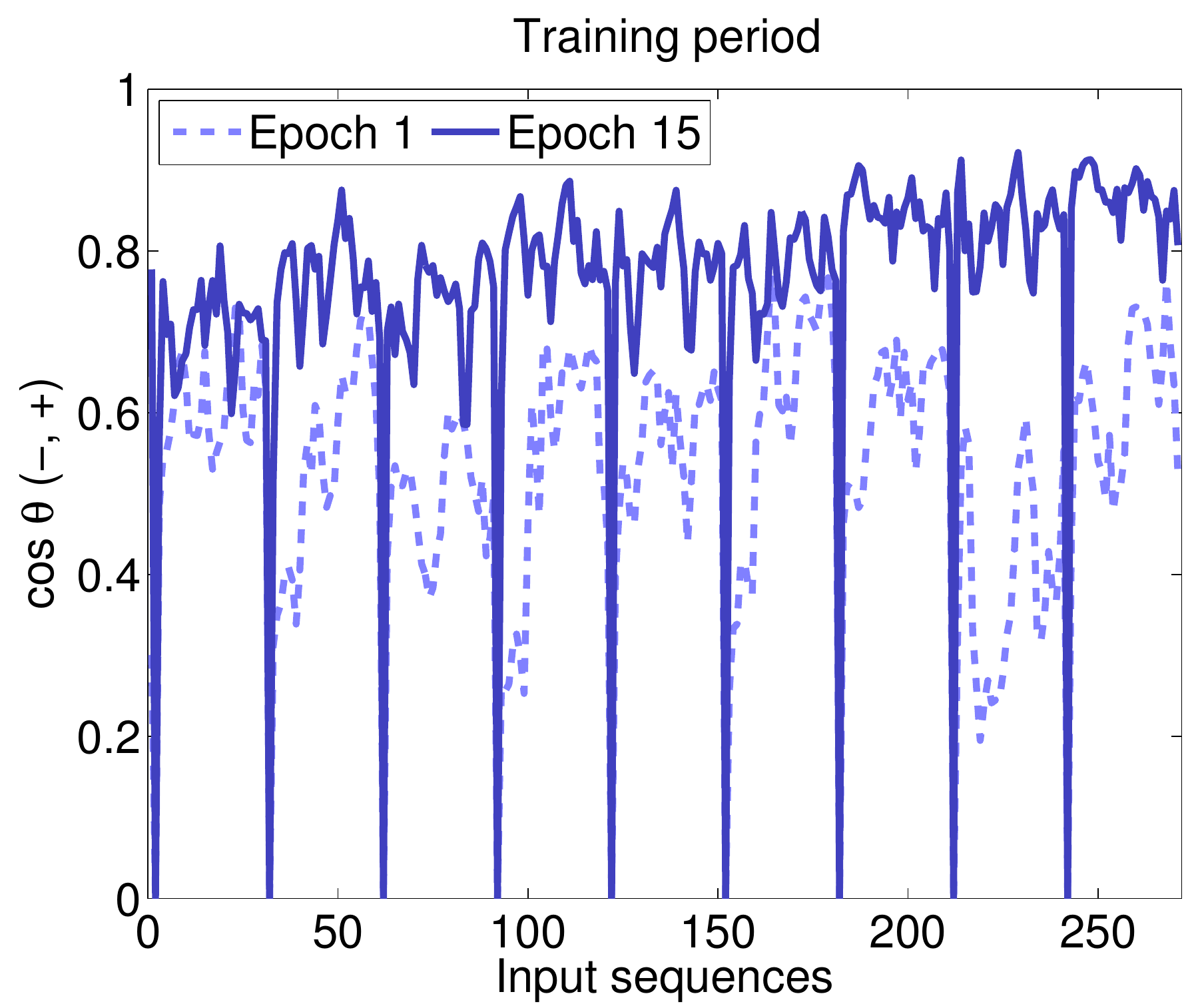}
\includegraphics[width=80mm]{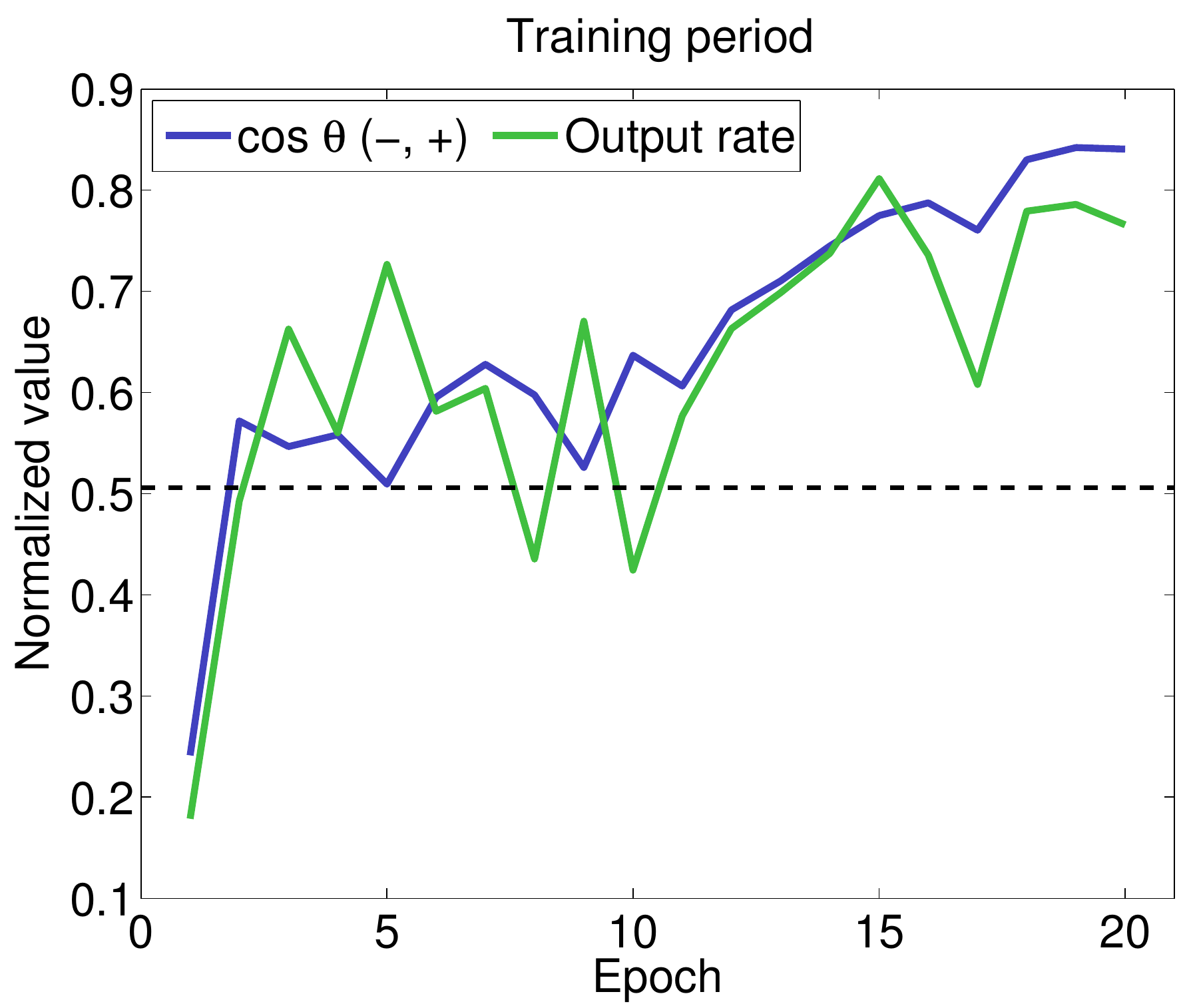}
\end{tabular}
\end{center}
\caption{Model training}{Average cosine between minus and plus phase for layers with columnar substructure and output response rate over the course of training. Sharp drops in the cosine to zero indicate the start of a new input sequence and are unlearnable. The lower bound for the cosine was computed as the reproduction of the plus phase from the previous moment's (\textit{t} - 1) state and the overall average is indicated by the dotted line. A cosine greater than this level indicates positive prediction.}
\label{fig:sims_train}
\end{figure}

The cosine varies between 0 and 1 and expresses the degree of similarity of LeabraTI's prediction to the actual outcome in layers with columnar substructure (Figure \ref{fig:sims_train}). A value of zero indicates the minus phase prediction is completely orthogonal to the plus phase sensation and a value of 1 indicates complete overlap. The lower-bound on the cosine is not likely to be zero as it would require spurious activations in retinotopic regions that do not contain any features. A better approximation of the lower bound is the case when the system simply reproduces the plus phase from the previous moment's (\textit{t} - 1) state. This can be thought of as the amount of perceptual overlap between adjacent views of the stimuli, and thus any additional features that contribute to a higher cosine value indicate positive prediction.

% training fig

Typically, after the initial feature training phase, neural models are trained to classify stimulus-response pairs (\nopcite{RiesenhuberPoggio99}; although, see also \nopcite{OReillyWyatteHerdEtAl13}). As such, human learning studies indicate that stimulus predictability and response mappings can be learned independently \cite{WyartNobreSummerfield12,KokRahnevJeheeEtAl12,HorschigJensenVanSchouwenburgEtAl13}. The present model, however, was compact enough and input environment simple enough that the initial features and response mappings could be learned jointly without decoupling between these two levels of encoding. The response rate of the target output unit was used to evaluate the efficacy of the learned response mappings. 

Consistency between lower-level features and learned responses was ensured by using two sets of synapses with different update intervals that contribute a weighted mixture to the input of each receiving unit. The first ``standard'' set of synapses was updated after every plus phase, whereas a second ``stable'' set of synapses was updated at the end of each epoch. In the present model, a 80\% stable to 20\% standard synaptic mixture was used. This allowed the model to more slowly integrate learning across an entire epoch's worth of input sequences without runaway representations caused by being exposed to the same rotating object's features over multiple time steps while still maintaing the moment-to-moment spatiotemporal predictive learning central to LeabraTI.

Testing involved presenting input sequences accordant with each of the four predictability conditions used in the Chapter \ref{chap:pleast} and \ref{chap:bpleast} experiments. In the spatially unpredictable conditions (S-), random views were selected for each plus phase and used to compute deep neurons' updated context. To model the effect of temporal unpredictability (T-), a variable number of time steps (up to four) separated each context update. Each time the context update was skipped, a decay factor of 50\% was applied to superficial neurons' context input channel. The default scale of this channel was 100\% and thus four time steps without a context update decayed the scale to 12.5\%. The net effect of temporal unpredictability was a weakening of the prediction at each time step until the next view was actually presented and the updated context could be computed.

The completely unpredictable condition (S-T-) utilized both the variable update interval and decay whereas the combined spatial and temporal predictability condition (S+T+) was identical to the training procedure (i.e., a coherently rotating object with constant update interval). In all cases, predictions about each upcoming view were made during each minus phase given the current context state. Weight updates that normally occurred at the end of each plus phase during training were disabled on all plastic synapses during testing.

\section{Results and Discussion}
\begin{figure}[h!]
\begin{center}
\begin{tabular}{ll}
\includegraphics[width=80mm]{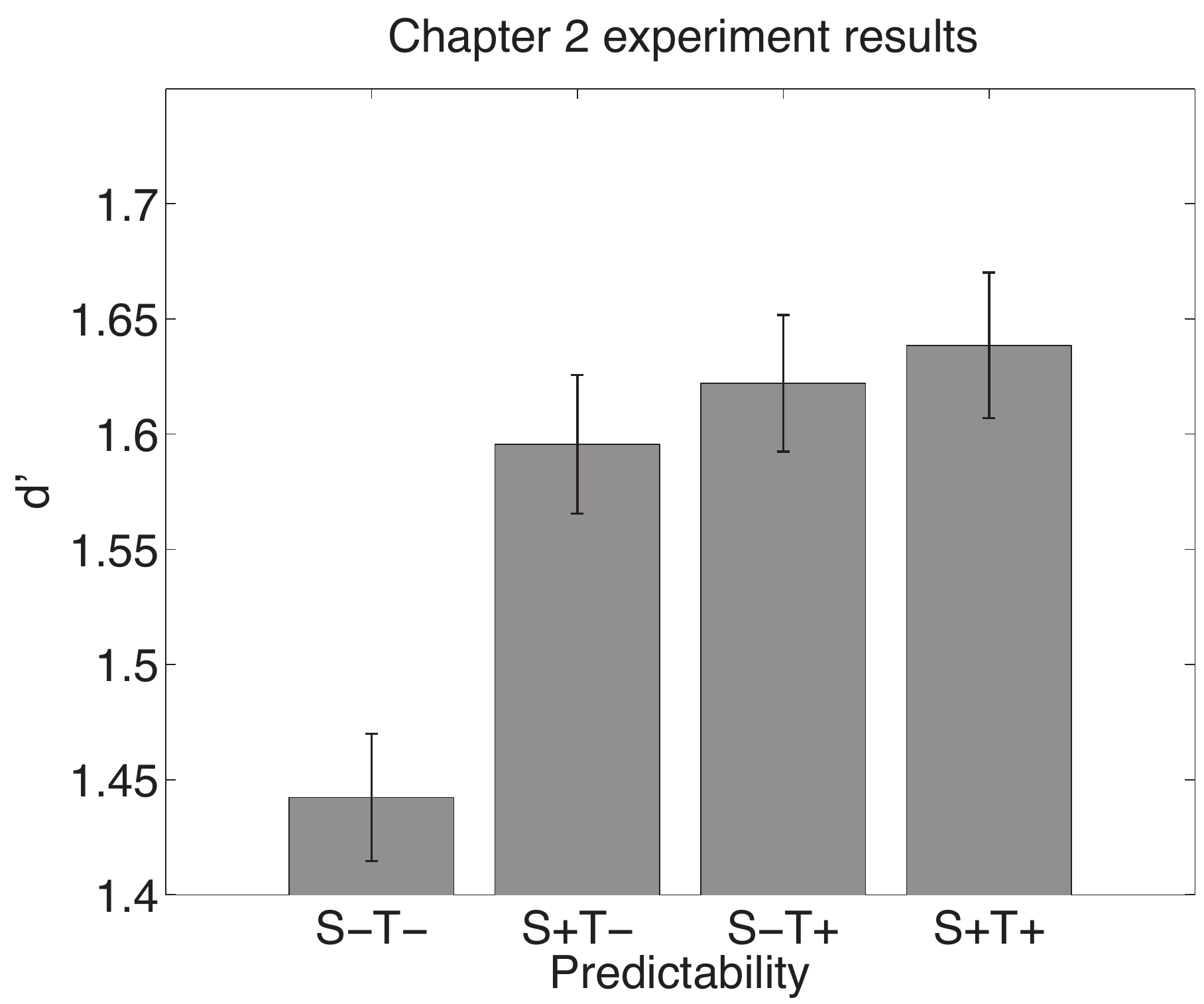} &
\includegraphics[width=80mm]{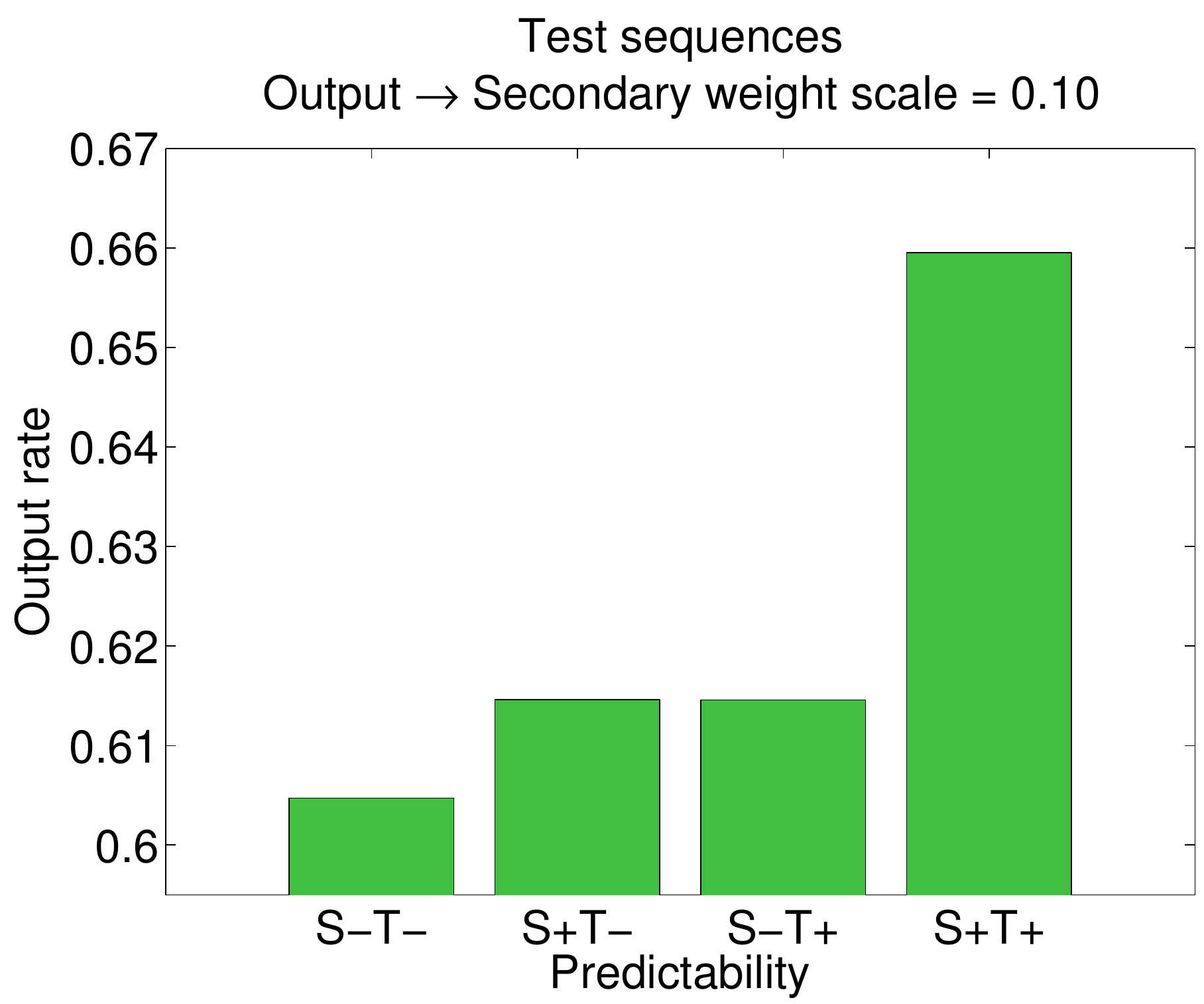} \\
\includegraphics[width=80mm]{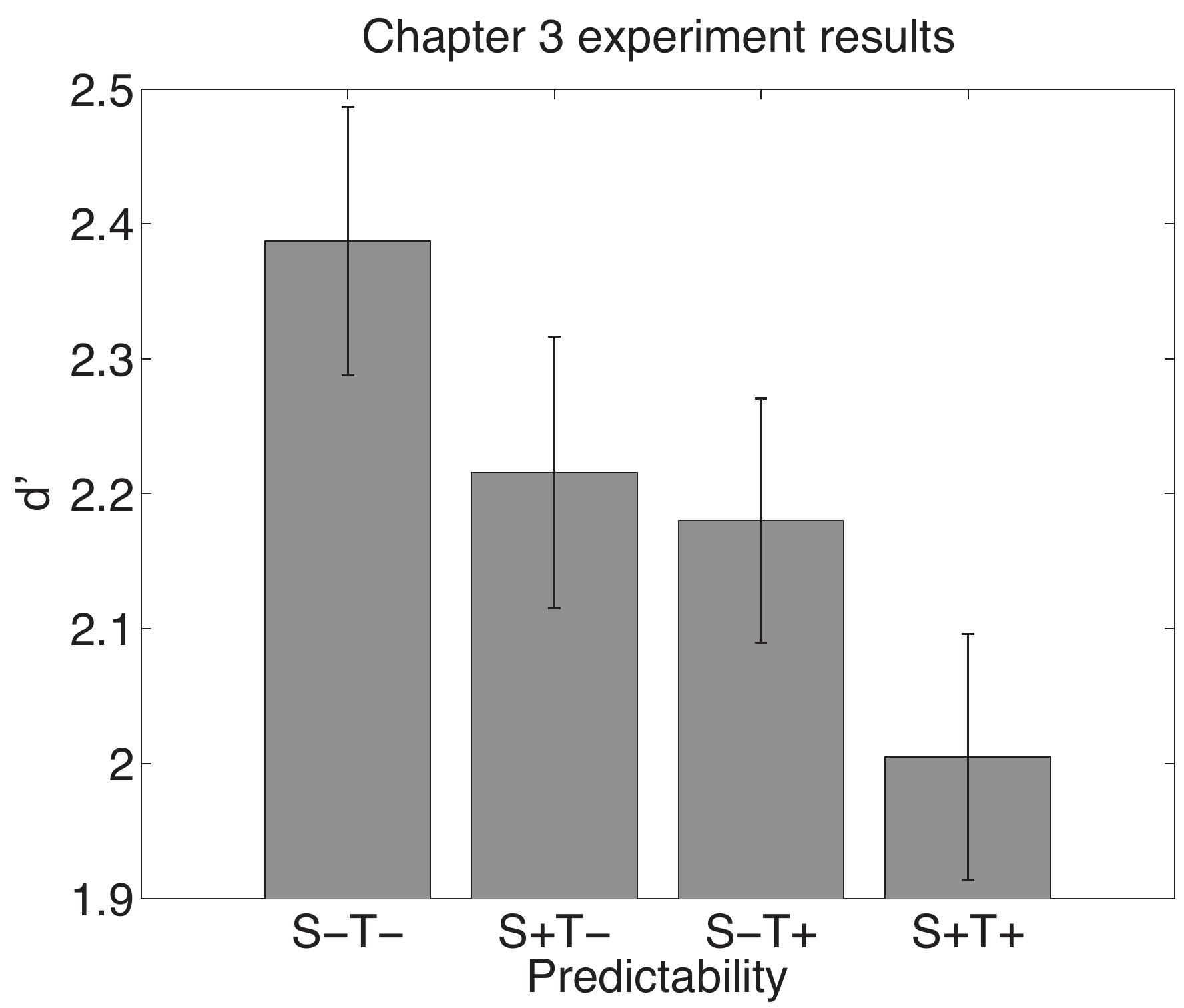} &
\includegraphics[width=80mm]{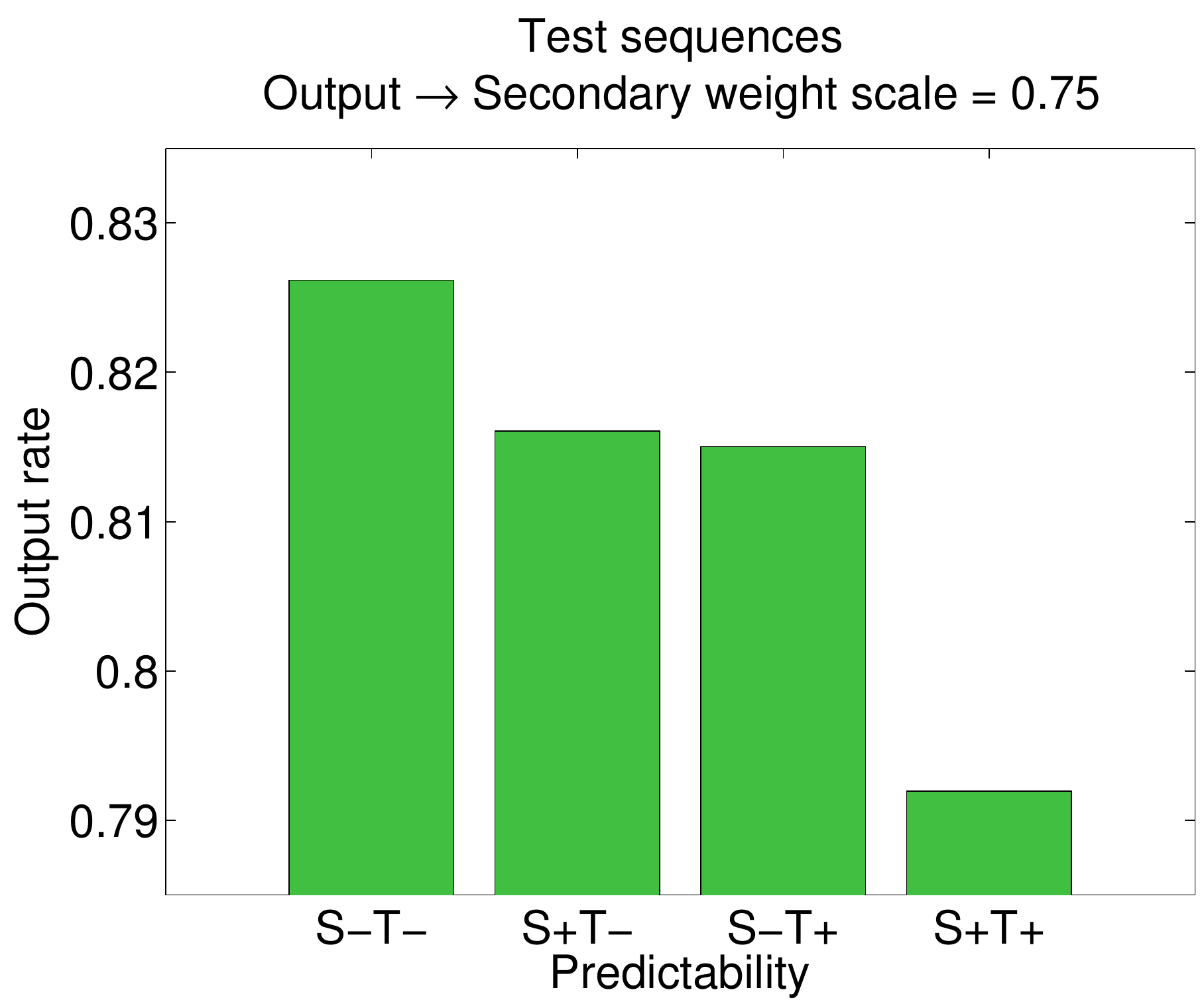} \\
\end{tabular}
\end{center}
\caption{Experiment and modeling results}{Chapter \ref{chap:pleast} (\textbf{top}) and \ref{chap:bpleast} (\textbf{bottom}) experiment and model results. \textit{d'} (sensitivity) was used as the common behavioral measure across experiments and the response rate of the models' target output unit was used in comparison.}
\label{fig:sims_test}
\end{figure}

%\subsection{Modeling prediction during novel object recognition and after prolonged learning}
The results from the Chapter \ref{chap:pleast} and \ref{chap:bpleast} experiments along with the results of the model test sequences are plotted in Figure \ref{fig:sims_test}. \textit{d'} (sensitivity) was used as the common behavioral measure across experiments due to the issues with response bias in raw accuracy found in the Chapter \ref{chap:pleast} experiment. The response rate of the target output unit was used to compare the model with the experimental results. All model results reflect the weights learned after 15 training epochs, as this was the point when the output rate was maximal, allowing for the largest potential differences between conditions during testing. This epoch choice also mitigated overfitting issues given the relatively simple training environment.

% testing fig

The Chapter \ref{chap:pleast} experiment tested subjects' ability to differentiate objects that were presented after a short series of spatiotemporally predictable or unpredictable entraining views. Subjects only ever saw 168 degrees of an object spread across 8 views on any given trial. Although feedback was given following the response on each trial, the relatively short exposure to disparate object views combined with the relatively large set of 16 possible target objects likely discouraged substantial learning. The trained model without modifications was capable of producing these results. Output unit rate was super-additive in the combined spatial and temporal predictability case as the testing sequence in this case perfectly matched the training environment in terms of spatial and temporal properties and thus maximally activated both superficial and deep units. This is a fundamental prediction of the LeabraTI model that was simply additive in the behavioral data (although a synergistic effect was found in EEG data). Synergistic effects of combined spatial and temporal predictability have also been demonstrated in previous investigations of predictability on attentional allocation \cite{DohertyRaoMesulamEtAl05,RohenkohlGouldPessoaEtAl14}.

The Chapter \ref{chap:bpleast} experiment produced an almost complete reversal of the results from the Chapter \ref{chap:pleast} experiment. This second experiment differed from the first a number of meaningful ways. First, a smaller set of only four target objects was used. Subjects also observed each of the objects rotate completely through each of its views four times and were explicitly instructed to study the object as it rotated. No feedback was given during test trials, but each object was seen four separate times during the experiment and many subjects reported being aware of the fact that there were a total of four unique objects. It is reasonable to conclude that these differences encouraged overtraining of the objects and that spatial and temporal predictability interact with this overtraining in different ways.

LeabraTI is predicated on spatiotemporal regularity and is thus somewhat inappropriate for evaluating learning under spatially and temporally unpredictable contexts. Furthermore, the interpretation of the Chapter \ref{chap:bpleast} experiment results was that such predictive learning mechanisms were \textit{not invoked} in those learning contexts. To account for these results without further predictive learning, a simple proxy was used for overtraining the stimuli in which the scale of the weights on the Output $\rightarrow$ Secondary visual synapses was increased. Typically, a relative scale of 10\% is used on feedback projections so that feedforward inputs drive the majority of weight changes with feedback playing a more modulatory role \cite{CrickKoch98,ShermanGuillery98}. This is crucial for the training period to prevent ``hallucinatory'' representations that can become disconnected from bottom-up inputs and produces the best testing results since the model adapts its weights to the strength of inputs for each layer. 

\begin{figure}[h!]
\hspace{2mm} \textbf{A} \hspace{56mm} \textbf{B}
\begin{center}
\includegraphics[width=160mm]{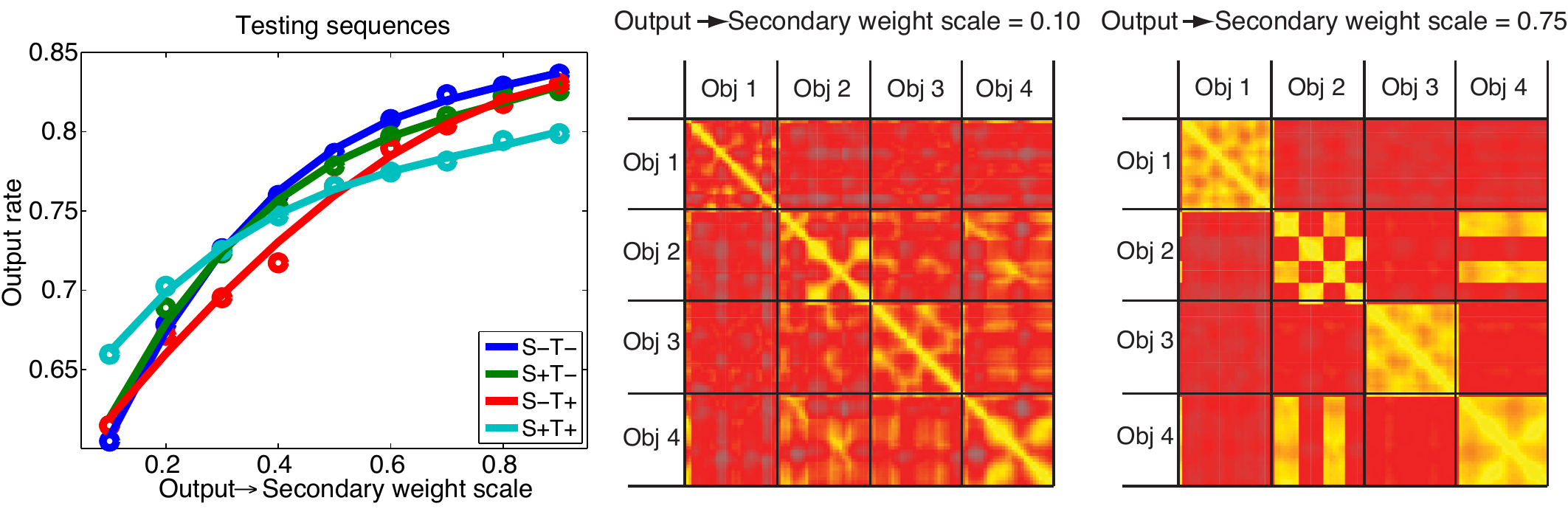}
\end{center}
\caption{Effect of prolonged learning and representational similarity}{\textbf{A}: Target output rate as a function of Output $\rightarrow$ Secondary visual synaptic weight scale. Lines indicate best fit third order polynomials. \textbf{B}: Pairwise cosine over secondary visual unit minus phase activations across all views of all objects. Yellow indicates greater similarity. Results shown for both 10\% and 75\% Output $\rightarrow$ Secondary visual weight scales.}
\label{fig:sims_ramp_distmatrix}
\end{figure}

Increasing the scale of the weights on the Output $\rightarrow$ Secondary visual synapses to 75\% produced the same reversal observed in the Chapter \ref{chap:bpleast} results in which training in the combined spatial and temporal predictability context impaired recognition relative to the completely unpredictable case. Synaptic weight scaling is one of the many effects of learning, especially when considering the long timescale self-organizing mechanisms presumed by Leabra that reinforce the most active units \cite{OReillyMunakata00,OReillyMunakataFrankEtAl12}. The full range of the reversal effect when increasing Output $\rightarrow$ Secondary visual synaptic weight scale is plotted in Figure \ref{fig:sims_ramp_distmatrix}A. Overall, the effect is graded and thus varying the amount of exposure observers have with the stimuli would probably modulate the degree of the reversal effect. 

To determine the effect of learning on the representation of the objects, the cosine was used to compute a pairwise similarity metric over secondary visual unit minus phase activations across all views of all objects (i.e., representational similarity, \nopcite{KriegeskorteMurBandettini08}; \abbrevnopcite{KriegeskorteMurRuffEtAl08}). LeabraTI training produced a representation that captures some similarity across sequential views but each view remained relatively distinct, as would be expected of intermediate visual representations \cite{KobatakeTanaka94,UllmanVidal-NaquetSali02,HayworthBiederman06,FreemanSimoncelli11}. The proxy for learning used here strengthens the synapses between secondary visual units and higher-level areas that code increasingly invariant representation. In the model, this higher-level area was a localist output layer which can be considered to be coding the same invariant representation that IT cortex does using a population code \cite{HungKreimanPoggioEtAl05,LiCoxZoccolanEtAl09}. 

The representational similarity suggests that prolonged learning under a spatiotemporally predictable context builds viewpoint invariance, consistent with previous suggestions \cite{StringerPerryRollsEtAl06,WallisBaddeley97,IsikLeiboPoggio12,WallisBulthoff01,WallisBackusLangerEtAl09}. Synaptic weight scaling might cause this invariance to ``trickle down'' to lower-level retinotopic feature representations. This is problematic for objects that suffer from severely degenerate views, in particular Object 2.\footnote{In the Chapter \ref{chap:bpleast} experiment, accuracy for Object 2 suffered the most of all objects for degenerate views, falling from ceiling to below chance levels.}For Object 2, two distinct views were represented, split by the degenerate view. However, one of these views was represented similarly to Object 4. This object confusion was less of an issue when the objects were recently acquired (10\% Output $\rightarrow$ Secondary visual weight scale) and might account for the comparatively lower performance of objects studied for prolonged periods with spatiotemporal predictability.

%% file: chap_discuss.tex
\chapter{General Discussion}
\label{chap:discuss}

\sloppy

\section{Summary of principal results}
The work described within this thesis has centered around how prediction is used in sensory processes such as object recognition and prolonged object learning. The work is heavily motivated by the LeabraTI (TI: Temporal Integration) framework (Chapter \ref{chap:leabrati}) which leverages the laminocolumnar structure of the neocortex \cite{Mountcastle97,BuxhoevedenCasanova02,HortonAdams05} to learn to predict temporally structured sensory inputs. Predictive learning in the LeabraTI framework is made possible by temporally interleaving predictions and sensory processing across the same populations of neurons so that powerful error-driven learning mechanisms \cite{OReillyMunakata00,OReillyMunakataFrankEtAl12} can be used to compute a prediction error that can be learned against to minimize the difference between predictions and sensory events over time.

LeabraTI relies on a 10 Hz prediction-sensation period as its core ``clock cycle'', suggested to correspond to the widely studied alpha rhythm observable across posterior cortex using scalp EEG \cite{PalvaPalva07,HanslmayrGrossKlimeschEtAl11,VanRullenBuschDrewesEtAl11}. Chapter \ref{chap:pleast} investigated the role of the alpha rhythm in prediction by using an entrainment paradigm \cite{SchroederLakatosKajikawaEtAl08,CalderoneLakatosButlerEtAl14} in which stimuli were presented rhythmically at 10 Hz so that predictions and sensory information could be interleaved regularly at the optimal rate proposed by LeabraTI. The experiment made use of three-dimensional objects that required integration over multiple sequential views to extract their three-dimensional structure. Thus, relatively rapid predictive learning mechanisms that operate over subsequent 100 ms periods could be leveraged to optimally encode the the objects. The spatial coherence between views and temporal onset of each view were independently manipulated to determine their effect on stimulus encoding quality and the putative role of the alpha rhythm in predictive processing.

The results of the Chapter \ref{chap:pleast} experiment indicated that spatial coherence and predictable temporal onset of each stimulus in an entraining sequence enhanced discriminability of a subsequently presented probe stimulus. Oscillatory analyses indicated strong bilateral alpha power and phase coherence modulation as a function of stimulus predictability. Specifically, spatial predictability of entraining stimuli suppressed alpha power with a lower degree of phase alignment relative to unpredictable stimuli. Temporally predictable entraining stimuli had the opposite effect, with increased alpha power and phase alignment, indicating successful entrainment. Importantly, phase alignment due to temporal predictability remained elevated compared to temporally unpredictable stimuli during a 200 ms blank period between the entraining sequence and probe, indicating that the effects of temporal predictability could persist without exogenous entrainment. In addition to these bilateral main effects, right hemisphere sites exhibited synergistic effects of combined spatial and temporal probe predictability on EEG amplitude and 10 Hz phase coherence approximately 200 ms after probe onset. 

Overall, the results of the Chapter \ref{chap:pleast} experiment support the basic claims put forward by the LeabraTI framework. The predictable 10 Hz presentation rate of the entraining sequence improved encoding of the target object, enhancing discriminability for the subsequent probe stimulus. This finding was accompanied by increased alpha phase alignment that remained elevated until the onset of the probe, which was necessary for ensuring that the probe event was processed precisely when the brain was expecting sensory information and not when it was generating a prediction. 

% revisi—n todo -- talk about effect of spatial predictability on alpha -- this is also important for LeabraTI support

Given this basic support for the LeabraTI framework, the Chapter \ref{chap:bpleast} experiment was designed to investigate the role of prolonged predictive learning of dynamic stimuli. Previous work has suggested that a temporal association rule similar to the one central to LeabraTI might be leveraged for constructing stable representations of spatially coherent visual inputs \cite{StringerPerryRollsEtAl06,WallisBaddeley97,IsikLeiboPoggio12} and indeed, a line of experiments by DiCarlo and colleagues demonstrated that predictable object transformation sequences build transformation invariance crucial for robust object recognition \cite{CoxMeierOerteltEtAl05,LiDiCarlo08,LiDiCarlo10,LiDiCarlo12}. Given these results, one might hypothesize that combined spatiotemporal predictability would be optimal for prolonged learning of three-dimensional objects. 

The Chapter \ref{chap:bpleast} experiment used a subset of the object stimuli from the previous chapter's experiment along with an explicit training period during which observers studied the objects while they were rotated through their views. The study period was followed by a series of test trials that required same-different judgements about static probe stimuli. Somewhat surprisingly, the results of the experiment were an almost complete reversal of the previous chapter's experiment. Discriminability was lowest when stimuli were learned in a combined spatiotemporally predictable context and highest when learned in a completely unpredictable context. Furthermore, there was some indication that the principal differences between predictability conditions during training were driven primarily by degenerate viewing angles caused by three-dimensional foreshortening in the objects used \cite{BalasSinha09b}. In three out of four of the objects used in the experiment, accuracy was lower for degenerate views learned in a spatiotemporally predictable context compared to a completely unpredictable one.

Chapter \ref{chap:sims} revisited the LeabraTI framework and described a neural network model that implemented the columnar substructure necessary for predictive learning. The model was trained to recognize the same three-dimensional objects used in the Chapter \ref{chap:pleast} and \ref{chap:bpleast} experiments with the goal of being able to reproduce the conflicting behavioral results of the experiments. LeabraTI predicts that spatially predictable sequences presented at a regular temporal interval should elicit a synergistic effect on behavioral measures due to the multiple prediction-sensation cycles that successfully integrate visuospatial information at optimal temporal intervals \cite[see also]{DohertyRaoMesulamEtAl05,RohenkohlGouldPessoaEtAl14}. Such a synergistic effect was demonstrated in the Chapter \ref{chap:pleast} EEG results, but was simply additive for behavioral measures. Still, the model provided a reasonable account of these data. Furthermore, the model was able to produce the reversal effect observed in Chapter \ref{chap:bpleast} by increasing the scale of a single projection of synaptic weights as a simple proxy for prolonged learning. 

The model further indicated that the synaptic weight scaling that accompanies prolonged learning promoted viewpoint invariance that has been suggested to be formed by spatiotemporal associations \cite{StringerPerryRollsEtAl06,WallisBaddeley97,IsikLeiboPoggio12,WallisBulthoff01,WallisBackusLangerEtAl09}. However, this invariance ``trickled down'' to lower-level retinotopic feature representations. This was problematic for the objects used, since some of them suffered from extreme foreshortening, causing severely degenerate views. This caused confusion between objects, potentially accounting for the overall reversal observed between the Chapter \ref{chap:pleast} and \ref{chap:bpleast} experiments.

\section{Open questions}

\subsection{Does the brain predict at 10 Hz or 5 Hz?}
The results of the Chapter \ref{chap:pleast} experiment indicated that delta-theta band oscillations centered around 5 Hz indexed predictability leading into the probe judgement, in addition to the 10 Hz effects of primary interest. Specifically, 5 Hz power was suppressed due to spatial predictability during the 200 ms blank period between the entraining sequence and probe. 5 Hz power was also suppressed due to temporal predictability of the entraining sequence but then increased at the onset of the probe and remained elevated for over 250 ms after its presentation. 5 Hz phase angle alignment was elevated due to temporal predictability throughout the probe judgement and furthermore and exhibited synergistic enhancement in alignment when spatial prediction was also possible. Altogether, these results suggest delta-theta band oscillations, might also play a role in predictive processing in addition to alpha oscillations as suggested by the LeabraTI framework.

There are two potential explanations for these delta-theta band predictability effects. First, the 5 Hz effects could simply be due to their being s subharmonic frequency of the fundamental effects of predictability observed at 10 Hz. Steady state visual evoked potentials (SSVEP) from exogenous rhythmic stimulation are known to cause power increases at harmonic frequencies, in addition to the fundamental frequency of stimulation. It is unclear, however, whether increases in harmonic power are simply due to sharing the same zero crossing as the fundamental frequency waveform (and thus contributing power) or whether they actually might serve a functionally distinct role from the fundamental frequency \cite[e.g.,]{Herrmann01,KimGraboweckyPallerEtAl11}. 

The second potential explanation of the observed delta-theta band effects is that sensory predictions occur at both 10 Hz and 5 Hz. Two recent reviews have suggested that sensory prediction is at least partially subserved by delta-theta oscillations \cite{ArnalGiraud12,GiraudPoeppel12}. These theories were formulated to explain prediction during speech recognition, with a focus on multiple windows of integration that are necessary for robust comprehension of speech's hierarchy of time varying features (e.g., phonemes, syllables, words). Delta-theta band predictions support integration over periods of 150 ms or longer which can be used to predict the overall speech envelope \cite{AikenPicton08} or phrasal structure of sentences. These theories are supported by recent empirical evidence of the importance of delta-theta band oscillations in the speech \cite{ArnalWyartGiraud11} and general audition domains \abbrevcite{StefanicsHangyaHernadiEtAl10,ArnalDoellingPoeppel14}. Others have noted, however, that auditory processing does not exhibit the ubiquitous perceptual discretization or strong alpha-band modulations that visual processing does \cite{VanRullenZoefelIlhan14,WeiszHartmannMullerEtAl11}, which is the fundamental mechanism by which LeabraTI interleaves predictions with sensory events. This does not necessarily mean that audition is not a predictive process. It could suggest, however, that auditory prediction happens at a more abstract level after feature extraction, consistent with the slower overall prediction rate, opposed to the sensory level for visual prediction and as suggested by LeabraTI. 

In the Chapter \ref{chap:pleast} experiment, 5 Hz power and phase coherence were sensitive to the spatial predictability of the entraining sequence, and so it seems unlikely that they were simply a subharmonic side effect of 10 Hz rhythmic stimulation. Thus, the most reasonable explanation of the 5 Hz predictability effects observed in the Chapter \ref{chap:pleast} experiment results is that they reflected a slower, more abstract visual prediction process such as anticipation of the appearance of the probe and whether it might depict the same object as the entraining sequence or a distractor.

\subsection{Are ``paper clip'' objects somehow special?}
The ``paper clip'' objects used throughout the current work have a long history of use in studies of three-dimensional object recognition in human observers \cite{BulthoffEdelman92,EdelmanBulthoff92,SinhaPoggio96} as well as monkey physiology studies \cite{LogothetisPaulsBulthoffEtAl94,LogothetisPaulsPoggio95}. The objects are easy to generate systematically and thus can be combined with a staircase procedure to titrate difficulty or can be generated \textit{en masse} to find the parameters that elicit maximal responses during neural recordings. Various effects with the objects have been replicated using computational models of object recognition with identical stimuli \cite{RiesenhuberPoggio99} and geometric properties of the objects are known to capture a large amount of variability in behavior \cite{BalasSinha09b}. Thus, it can be reasonably concluded that paper clip objects are a useful class of stimuli for studying three-dimensional object recognition.

However, other work brings under question the ecological validity of paper clip objects. The objects are constructed from thin line segments separated by empty space and thus self-occlusion of features is less of a problem than for three-dimensional volumetric objects with surfaces. This might imply that viewpoint invariance is not actually necessary to represent the full three-dimensional structure of paper clip objects, since the majority of features can be extracted from a single static view. Accordingly, studies comparing three-dimensional objects composed of line segment with volumetric objects found that the line segment objects were not represented in a viewpoint invariant manner \cite{FarahRochlinKlen94,PizloStevenson99}.

Thus, it is is possible that a spatiotemporally predictable training context is simply not optimal for learning to represent paper clip objects. Temporal association mechanisms \cite{StringerPerryRollsEtAl06,WallisBaddeley97,IsikLeiboPoggio12,WallisBulthoff01,WallisBackusLangerEtAl09} might bias the development of viewpoint invariance by forming  associations between canonical and degenerate views that promote viewpoint robustness by accounting for feature variability, but lower overall accuracy levels and slow reaction times. Another way of stating this idea is in terms of the nature and complexity of the representation at each stage of visual processing. At early stages of vision, objects are represented in terms of spatially localized oriented edges \cite{HubelWiesel62}, which are an optimal feature for encoding paper clip objects. At later stages, such as intermediate visual areas and inferior temporal (IT) cortex, neurons encode stimuli in terms of surfaces and complex volumetric features (\abbrevnopcite{CoxSchmidPetersEtAl13}; \nopcite{HayworthBiederman06,KourtziConnor10}). Thus, it may be the case that the spatiotemporal associations between entire views learned by IT neurons \cite{SakaiMiyashita91,MeyerOlson11,CoxMeierOerteltEtAl05,LiDiCarlo08,LiDiCarlo10,LiDiCarlo12} are actually surface-based encodings, which cannot effectively be used to represent paper clip objects since they are composed of line segments separated by empty space. This idea might also account for the conflicting effects in the literature regarding whether sequence predictability is actually advantageous for object recognition, as some investigations used line drawing stimuli \cite{LawsonHumphreysWatson94} whereas others used surfaced volumetric stimuli \cite{HarmanHumphrey99} (see Chapter \ref{chap:bpleast} Discussion).

\subsection{Is synaptic weight scaling a good proxy for prolonged learning?} 
% also seems weird to examine invariance at level of V2, although see \cite{KobatakeTanaka94}
The model described in Chapter \ref{chap:sims} made use of a relatively simple proxy for prolonged learning in order to account Chapter \ref{chap:bpleast} experiment results. This proxy was necessary because it was somewhat inappropriate to use the LeabraTI algorithm for learning under spatially and temporally unpredictable contexts, especially considering the interpretation of the Chapter \ref{chap:bpleast} experiment results was that such predictive learning mechanisms were not invoked in those learning contexts. The proxy for learning involved training the model with spatiotemporally predictable input sequences using the LeabraTI algorithm and then increasing the scale of the synaptic weights on one of the principal projections before presenting spatially and temporally unpredictable input sequences with synaptic plasticity disabled. Synaptic weight scaling was suggested to be a reasonable first order approximation of prolonged learning since it is one of the many effects of learning, especially when considering the long timescale self-organizing mechanisms presumed by Leabra that reinforce the most active units \cite{OReillyMunakata00,OReillyMunakataFrankEtAl12}.

An alternative to synaptic weight scaling that would also be more biologically plausible would be to explicitly model other visual areas that do not require LeabraTI's predictive learning. For example, IT cortex has been suggested to have a different alpha rhythm properties than earlier visual areas (i.e., V1, V2, and V4) both in terms of its physiological generators and its behavioral correlates \cite{BollimuntaChenSchroederEtAl08}. Under this view, IT neurons don't actively predict their own inputs, but do code the spatiotemporal associations \cite{CoxMeierOerteltEtAl05,LiDiCarlo08,LiDiCarlo10,LiDiCarlo12} of lower-level neurons that actively perform prediction. Thus, modeling IT cortex would not necessarily require predictive learning and could be used to represent and operate on the three-dimensional object stimuli with more standard learning mechanisms. 

To implement such a heterogenous model would require the following more complex approach: First, layers that learn via the LeabraTI algorithm would need to be pre-trained on a set of unrelated input sequences (e.g., natural images) to establish how inputs are capable of predictably transforming from moment-to-moment (100 ms periods in LeabraTI). This pre-training would discover features that capture the principal variance across input transformations but are still relatively task-independent. After pre-training, the model would need to be expanded to include IT cortex and Output layers, which would then be trained on a more specific object recognition task using the standard Leabra algorithm.  %These standard learning mechanisms would thus operate selectively on the Intermediate Visual $\rightarrow$ IT and IT $\rightarrow$ Output synapses and their respective feedback projections. 
Such a multistage training regime has been successfully used to train other deep neural network architectures without suffering from error signal dilution \cite{HintonSalakhutdinov06} and has showed promise for combining predictive feature learning with other task-specific architectures  \cite{OReillyWyatteRohrlichEtAlInPrep}.

The synaptic weights produced by the pre-training step along with a relatively short amount of task-specific training on the paper clip objects would probably be sufficient to produce the results of the Chapter \ref{chap:pleast} experiment in which objects were still psychologically novel and likely to be represented by combinations of task-independent features. In the prolonged task-specific training required to produce the results of the Chapter \ref{chap:bpleast} experiment, the combined spatially and temporally predictable training context would maximally activate early and intermediate visual features due to a close match with the environmental statistics of pre-training, leading to \textit{smaller} error signals that are used to train the later IT and Output stages. The completely unpredictable training context, in contrast, would only weakly activate early and intermediate visual features due to the unexpected spatiotemporal irregularity across presentations, causing larger error signals in later stages. Effectively, IT neurons might learn more robust representations from an unpredictable training context since associations between views of the objects are constantly psychologically novel. This more robust representation would probably benefit recognition of the objects from static views without full sequence information, and might also be considered more viewpoint-centric (compared to the viewpoint invariance learned from spatially and temporally predictable training), although this would need to be quantified via the representation similarity methods used in Chapter \ref{chap:sims} (\nopcite{KriegeskorteMurBandettini08}; \abbrevnopcite{KriegeskorteMurRuffEtAl08}).

%%%
% save for general discussion
%%%
% TODO: importance of ``surprise'' and randy's theory of bottom-up alpha suppression 

\section{Conclusion}
This thesis has advanced a comprehensive theory of neocortical predictive processing of sensory input sequences. The cumulative work has spanned the detailed analysis of the underlying biological neural circuitry in accommodating predictive learning, experimental support for the theory's core testable predictions, and a neural network model that was constrained by biological details yet capable of producing the experimental results. Some of the idiosyncratic details that differentiate between various models of sensory prediction remain to be resolved and refinements on experimental methods and more complex neural models would also be illuminating to this end. Overall though, the current work represents a significant contribution to our understanding of the core computations that make the brain a ``prediction machine.''